\documentclass[12pt]{book}
\usepackage{graphicx}
\usepackage{amsmath}
\usepackage{amsfonts}
\usepackage{amssymb}
\newcommand{\be}{\begin{equation}}
\newcommand{\ee}{\end{equation}}
\newcommand{\ba}{\begin{eqnarray}}
\newcommand{\ea}{\end{eqnarray}}
\begin{document}

\begin{titlepage}
\[\] {\Large
\hspace{0.5cm}
\begin{minipage}[15cm]{15cm}
\begin{center} {\bf {\Huge Relic gravitational
waves\newline\newline in the expanding Universe}}\newline\newline
\end{center}
\end{minipage}
\hspace{-5cm}
\begin{minipage}[15cm]{15cm}
\centerline{PhD Thesis of} \centerline {{\Large  Germ\'an
Izquierdo S\'aez}}
 \centerline{to obtain the degree of}
 \centerline{Doctor of Physics}
 \[\]
 \centerline{Supervisor: Dr. Diego Pav\'on}

\end{minipage}
}
\begin{figure}
\hspace{6cm}
\includegraphics{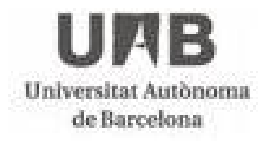}
\end{figure}

\[\]\[\]\[\]\[\]\[\]\[\]

\hspace{7cm}
\begin{minipage}[10cm]{10cm}
\begin{flushright}
{\it Departamento de F\'isica \newline  (F\'isica Estad\'istica)
\newline Universidad Aut\'onoma de Barcelona
\newline June 23, 2005 \newline}
\end{flushright}
\end{minipage}

\end{titlepage}

\setcounter{chapter}{0}
\pagenumbering{roman}

\thispagestyle{empty}
\[\]\[\]\[\]\[\]\[\]\[\]
\begin{flushright} \textbf{To my parents and my sister,
\\I owe them so many things...}
\end{flushright}
\newpage

\thispagestyle{empty}
\begin{center}{\Large
Acknowledgments}
\end{center}

This thesis had been developed during three years of quasi-full
time work. I do not consider it to be ``The Work of my Life" or
``My Little Contribution to  Human Knowledge"; but I do consider
it an extremely important part of them (more than $70\%$ of the
total, for sure). So many people have added to it in one way or
other than I surely forgot to acknowledge someone. This is why I
firstly would like to thank these ones.

Then, I thank my family, for showing so much understanding toward
me; for standing up the evil Germ\'{a}n (whom, fortunately, no one
knows very well), and for all the support they had provided me
from the outset. I hope I will be able, some day,  to tell them
exactly what my research is about.

Next, I thank my thesis advisor, Diego, for his advise and
guidance. He had transformed a recently graduated lazy student in
a much-less-lazy PhD student (great effort was needed to achieve
this). In these years we have shared bliss and deceptions in
supporting the Real Madrid. I hope we will share fresh joys for a
long period.

Time for my friends. Thanks to all my friends of the university:
Dani (so many years together, so many things we have learnt, so
many hilarious situations we have lived through), Jorge and Santi
(two flatmates, and bad joke lovers as me), Ester (another old,
old friend), Alvaro (the roommate that everybody would like to
have), Vicente, Raul, Aitor, Gabriel, ... Also, thanks to  my
friends overseas: Hong, Mariam, Jose Antonio, Andrea... And thanks
to my friends beyond the academic world (for keeping me in touch
with the ``outside"): Agust\'in, Carlos, Jordi, Jose, Ricardo...

I wish to thank everybody of the Statistical Physics group of the
UAB, for the good research atmosphere as well as the  help they
did provide me with during  all these years. Especially to Javier,
the first person I turned to when having a computer problem. Also,
I thank the people I met at the ICG of the University of
Portsmouth, for the hospitality they gave me during my two stays.

Finally, I would express my gratitude to the ``Programa de
Formaci\'{o} d'Investigadors de la UAB" for funding me for the
whole period of my PhD.\\
\\

And now, something completely different...

\tableofcontents
\newpage

\mainmatter
\chapter{Introduction}
\section{Cosmology and gravitational waves}
It can be safely said that the advent of general relativity
signaled the commencement of modern cosmology. The latter aims at
a global description of the observable universe (i.e., the
Universe). Based on observations at very large scales and some or
other theory of gravitation cosmologists try to propose simple and
plausible world models (i.e., models whose mathematical complexity
is kept at a reasonable minimum and do not run into conflict with
observational data). At present there are four main sources of
observational data: the light from the faraway galaxies, whose
redshifts seem to indicate that the Universe is expanding in an
accelerated fashion; the abundance of light elements at cosmic
scales and clusters thereof; the mass distribution of galaxies and
clusters; and the anisotropies of the cosmic microwave background
radiation (CMB). These are usually interpreted in favor of the
$\Lambda$CDM model, a flat Friedman-Robertson-Walker universe that
began expanding from an initial phase of very high density and
temperature (Big Bang), and today $70\%$ of its energy would be in
the form of a cosmological constant, $\Lambda$, and the rest in
the form of cold (non-relativistic) matter.

The future detection of gravitational waves (GWs henceforth) is
expected to provide us with invaluable information about the
instant of their decoupling from other fields, i.e., about
$10^{-43}$ seconds after the Big Bang. Likewise, GWs are a direct
prediction of the Einstein's field equations whence it would
entail a very strong evidence in favor of general relativity.
Regrettably, the weak interaction of the gravitational field with
matter -that renders GWs so valuable from a cosmological
viewpoint- hinders their detection.

GWs can be found originated at single sources or in a background
similar to the CMB. Of the first kind are, for instance, the GWs
from supernovae burst and the periodic signals from spherically
asymmetric neutron stars or quark stars; of the second kind are
those GWs originated at the decay of cosmic strings and the relic
GWs.

This thesis is mainly concerned with  relic GWs\footnote{From
chapter \ref{amplimech} on, we refer to relic GWs just as GWs for
the sake of simplicity.}, i.e., the GWs generated by  parametric
amplification of the quantum vacuum during the expansion of the
Universe which form a background whose spectrum depends on the
scale factor. The CMB is a remnant of the radiation that once
dominated the expansion of the Universe and brings us information
of the decoupling between radiation and non-relativistic matter
(i.e., dust). In the same way the spectrum of the relic GWs can
provide us with prime information about the evolution of the scale
factor, i.e., about the history of the Universe.

Even in the absence of any observed spectrum, some valuable
information can be extracted from the existing bounds over it
\cite{allen96}. The presence of the relic GWs at the instant of
the radiation--dust decoupling produces temperature fluctuations
via the Sachs-Wolfe effect and, consequently, temperature
anisotropies in the CMB. As these anisotriopies have been
measured, a bound over the spectrum of relic GWs present at this
instant follows. GWs passing by us transversely to our line of
sight to a pulsar of well measured period will cause the arrival
times of the pulses to shift. Many years of observation of the
pulses arriving from a number of stable millisecond pulsars lead
to another bound over the spectrum of the relic GWs. The theory of
primordial nucleosyntesis predicts rather successfully the cosmic
abundance of light elements. If at nucleosyntesis time, the
contribution of relic GWs to the total energy were too large, the
neutrons would have been more avaliable resulting in an
overproduction of helium. These restrictions over the GWs spectrum
translates into constraints on the free parameters of the universe
models (see top panel of Fig. \ref{gwdetection}).

We assume throughout that GWs are negligibly damped by the cosmic
medium; this is in keeping with common lore and well supported by
different studies \cite{damped}.

\section{Gravitational waves detection}

Because of insufficient technological means no direct detection of
GWs has been possible so far. Nevertheless, some indirect
evidences of their existence have been found. Binary systems
(i.e., two stars orbiting their common center of mass) have an
energy loss due to the emission of gravitational radiation that
can be modelled by means of the general relativity. The
observational data of the variation of the rotational period of
the binary pulsar $PSR1913+16$ confirmed the predictions of the
theory \cite{Hulse, Will}. Recently, a similar but more precise
indirect proof of the existence of GWs has been obtained from the
data of the variation of the orbital period of the double pulsar
$J0737-3039$ \cite{kramer}.

The gravitational waves detection is a very topical subject and
great effort is under way in that direction. There are two common
types of detectors: resonant mass detectors and laser
interferometers \cite{Rowan}.

The first kind of detectors operates as the GWs produces a stress
over a resonant bar (stretching or compressing it) and the
vibrational amplitude or phase of the antenna experiments changes
that can be measured. By making the antenna's quadrupole modes
resonant at the wave's frequency, the detector keeps a ``memory"
of the excitation, allowing extra time to detect the signal. The
sensitivity of this kind of antenna can be improved by increasing
its mass (thus, increasing the force the GWs exert on it), by
making it equally sensitive in all directions and polarizations
and by lowering the thermal motion of molecules in the detector
(thermal noise). Today there are seven cryogenic resonant-bar
detectors in operation and three proposed resonant spheres.
Analyzing the experimental results of the resonant bars EXPLORER
and NAUTILUS, Astone \textit{et al.} showed that the number of
coincident detections is greatest when both of them are pointing
into the center of our galaxy \cite{astone}. This conclusion, if
not a direct detection of the GWs, can be considered a further
indirect evidence of their existence.

The laser interferometer detectors use test masses which are
widely separated and freely suspended (as pendulums reduce the
effects of thermal noise); laser interferometry provides a means
of sensing the motion of these masses produced as they interact
with a gravitational wave. This technique is based on the
Michelson interferometer and is particularly suited to the
detection of gravitational waves as they have a quadrupole nature.
Waves propagating perpendicular to the plane of the interferometer
will result in one arm of the interferometer being increased in
length while the other arm is decreased and vice versa. The
induced change in the length of the interferometer arms results in
a small change in the intensity of the light observed at the
interferometer output \cite{Rowan}. Nowadays there are four laser
interferometers taking data at their level of sensitivity: TAMA300
\cite{tama300}, GEO600 \cite{geo}, VIRGO \cite{virgo} and LIGO
\cite{ligo} (whose arms have a length of $300$, $600$, $3000$ and
$4000$ meters, respectively).

Ground based detectors have a minimum frequency of detection
bounded by seismic noise and atmospheric effects. Ground-based
interferometers are also obviously limited in length to a few
kilometers, restricting their coverage to events such as supernova
core collapses and binary neutron star mergers. As a partial
solution to these limitations, LISA - Laser Interferometer Space
Antenna aimed at a launch in 2014 \cite{Lisa}- has been proposed
by an American/European team; it consists of an array of three
drag free spacecraft at the vertices of an equilateral triangle of
length of side $5\times10^6$ km. Proof masses inside the
spacecraft (two in each spacecraft) form the end points of three
separate but not independent interferometers. Each single two-arm
Michelson type interferometer is formed from a vertex (actually
consisting of the proof masses in a `central' spacecraft), and the
masses in two remote spacecraft. In the low-frequency band of
LISA, sources are well known and signals are stable over long
periods (many months to thousands of years). The primary sources
for LISA are expected to be compact galactic binaries,
supermassive black holes binaries, and extreme mass ratio
inspirals. Other space based laser interferometers with a better
sensitivity than LISA are being proposed: the Advanced Laser
Interferometer Antenna (ALIA) and the Big Bang Observer (BBO)
\cite{crowder}.

\begin{figure}[tbp]
\includegraphics*[scale=0.6]{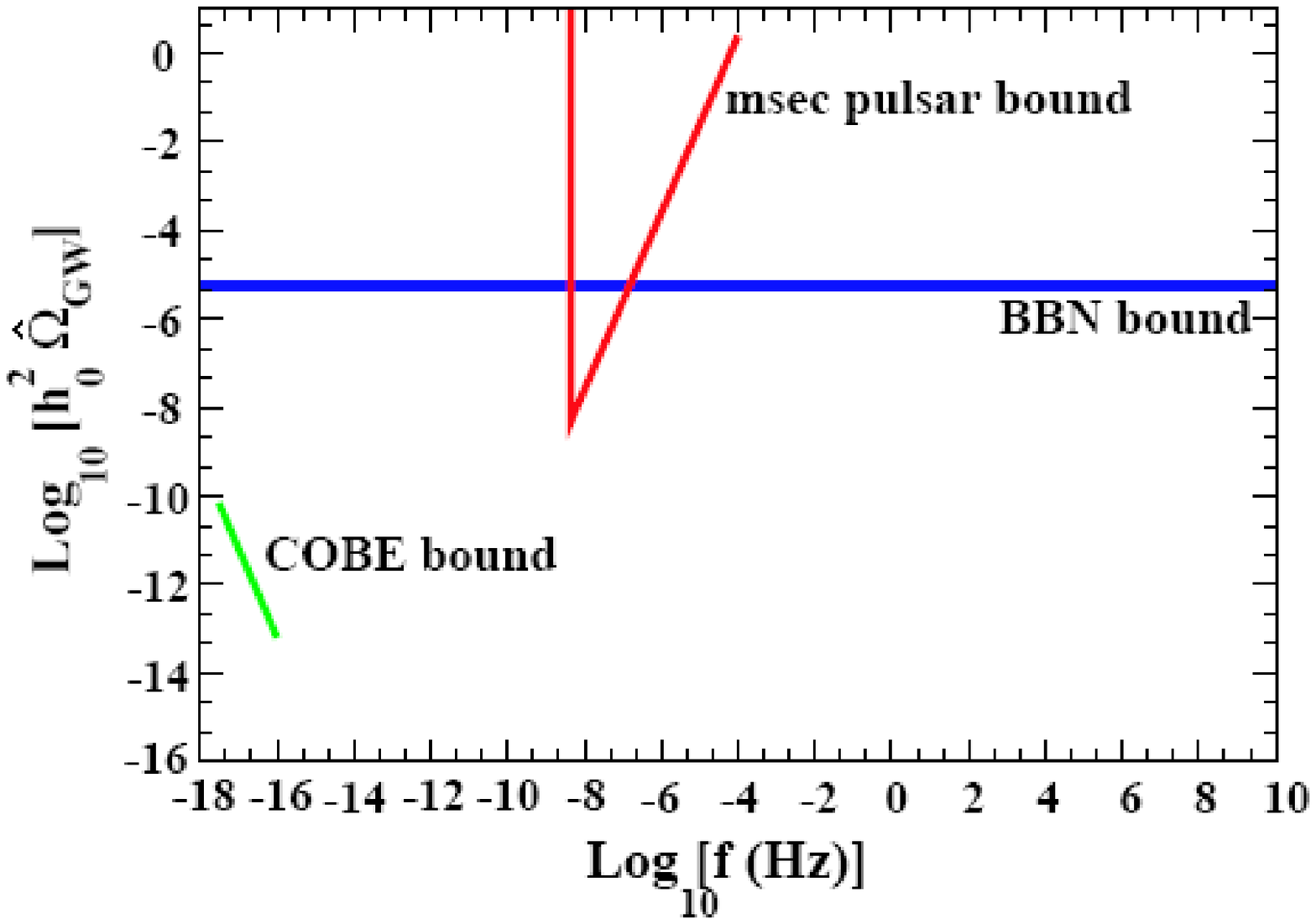}
\includegraphics*[scale=0.5]{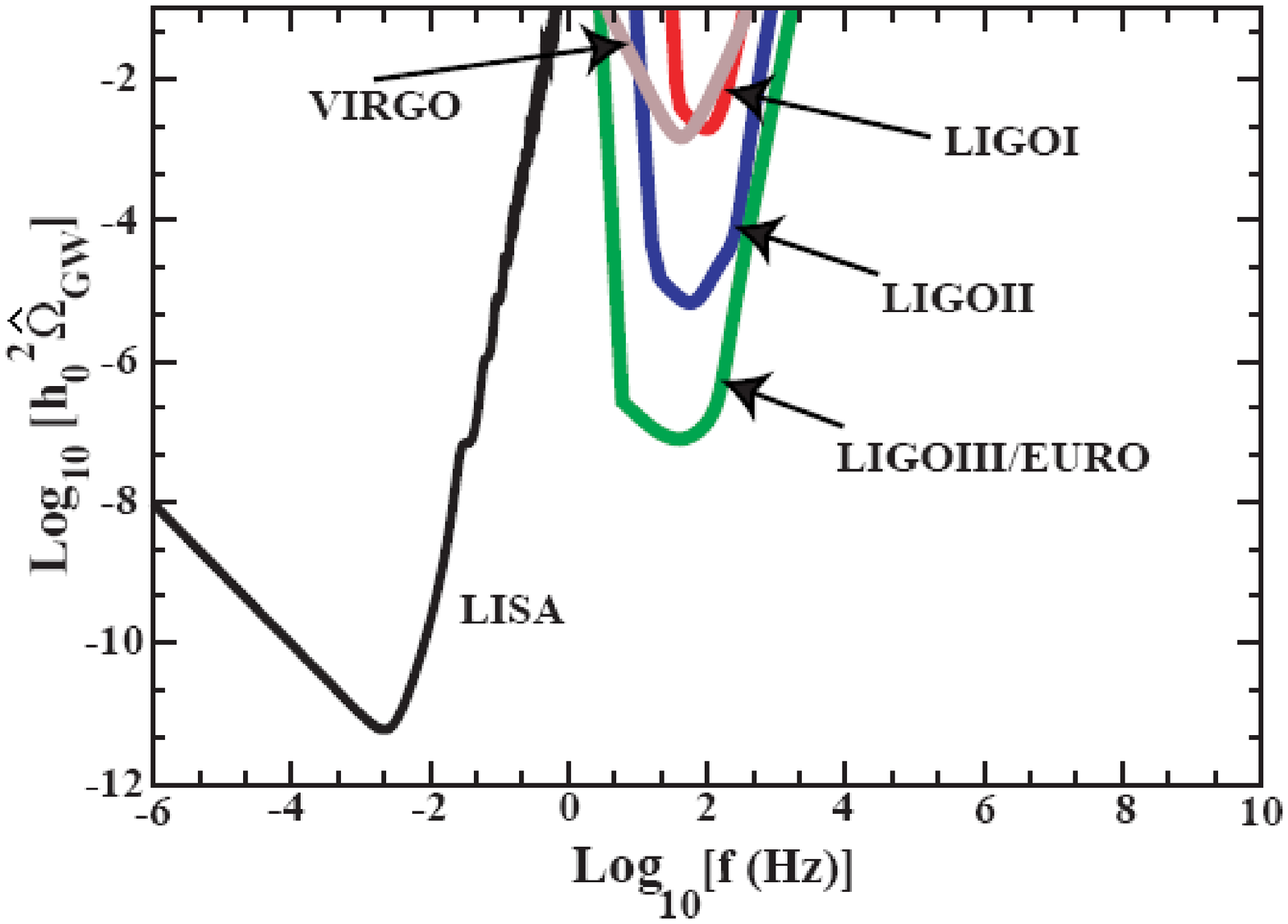}

\caption{Top panel: Bounds over ${\hat{\Omega}}_{GW}$ from the CMB
anisotropy (COBE), from the millisecond pulsar timing and from the
Big Bang nucleosyntesis (BBN). Bottom panel: sensitivity versus
frequency of LISA and ground-based detectors of first, second and
third generations. The parameter $h_0$ is defined from
$H_0=h_0\times 100km/sec/Mpc$ and represents the existing
experimental uncertainty in evaluating $H_0$. (Adapted from Figs.
1 and 2 of Ref. \cite{buon})} \label{gwdetection}
\end{figure}

It is quite common, when tackling the subject of the GWs
detection, to define the dimensionless parameter
${\hat{\Omega}}_{GW} (f)$ as%
\be%
{\hat{\Omega}}_{GW} (f)=f\frac{d\rho_g/df}{\rho_{c}},
\ee%
where $f$ denotes the frequency of the wave, $\rho_g$ is the
energy density of the GWs and $\rho_c=3H^{2}/(8\pi\, G)$. This
parameter gives a measure of the intensity of the GWs signal and
can be compared with the sensitivity of the GWs detectors (see
bottom panel of Fig. \ref{gwdetection}).

\section{Outline of the thesis}

In chapter \ref{amplimech}, we review the mechanism of
amplification of the relic GWs and their power spectrum. In
chapter \ref{mbh&gwspect}, we obtain the relic GWs spectrum in a
universe scenario consisting in a De Sitter stage of expansion
followed by a radiation dominated stage and a dust dominated stage
(three-stage model). Later, we obtain the spectrum in a four-stage
model in which and additional era (dominated by a mixture of
radiation and mini black holes) between the De Sitter and the
radiation one. We compare both models and found bounds on the free
parameters of the four-stage scenario. In chapter
\ref{presentaccel}, the spectrum of GWs is obtained in a scenario
with an accelerated era of expansion dominated by dark energy
right after the De Sitter-radiation-dust dominated eras. Although
the spectrum is formally equal to those of the three--stage model
of chapter \ref{mbh&gwspect}, it evolves differently. The
possibility of the existence of a second dust era if the
accelerated era comes to an end in some future time is also
considered. In chapter \ref{GWsentropy}, working in the De
Sitter-radiation-dust-dark energy scenario and assuming the GWs
entropy is proportional to the number of GWs within the event
horizon (being $A$ the proportionality constant), we test the
cosmological generalized second law of thermodynamics. An upper
bound to the constant $A$ is found. In chapter \ref{GSLDE}, we
leave the consideration of GWs to extend the study of the
generalized second law of thermodynamics of the previous chapter
to a universe dominated by dark energy. It turns out that the
generalized second law is fulfilled in both phantom and
non--phantom dark energy models provided the dark energy fluids
have an entropy given by Gibbs' equation. Finally, the Summary
discusses and sum up the main results of this work. Except where
otherwise stated, we use the units such that $\hbar=G=c=k_B=1$.

Before proceeding into the main body of the thesis it is sobering
and expedient to recall the viewpoint of Willem de Sitter
regarding cosmology research \cite{desitter}:
\begin{quote}
``It should not be forgotten that all this talk about the Universe
involves a tremendous extrapolation, which is a very dangerous
operation".
\end{quote}

\chapter{The amplification mechanism of gravitational waves \label{amplimech}}
\markboth{CHAPTER \thechapter. THE AMPLIFICATION MECHANISM OF
GW}{}

\section{The wave equation}
The existence of wave-like solutions of the linearized vacuum
field equations, i.e., GWs, was first predicted by Einstein in
1916 when he realized of the propagation effects inherent in the
gravitational field equations \cite{ein}. The GWs equation to
linear order was obtained by Lifshitz who considered perturbations
that not interfered with their own propagation (as they carry
little energy and momentum) \cite{lif}. He proceeded by
introducing a small perturbation in the background FRW metric,
$g_{ij}$,
\begin{equation*}
\overline{g}_{ij}=g_{ij}+h_{ij}(x_i)\qquad(i,j=0,1,2,3),
\end{equation*}
where $\left| h_{ij}\right| \ll \left| g_{ij}\right|$, and by
evaluating the Einstein equations for the perturbation to linear
order in $h_{ij}$ in a general coordinates system for which
\begin{equation}
h_{00}=h_{0\alpha}=0 \qquad(\alpha=1,2,3).  \label{coord}
\end{equation}
Three kinds of solutions to the equations for the perturbation
follow: scalar solutions, which represent density perturbations;
vectorial solutions, which represent rotational perturbations; and
tensorial solutions, which represent sourceless weak GWs.

We focus our attention on the latter kind by considering a flat
FRW universe with background metric
\begin{equation}
ds^{2}=-dt^{2}+a(t)^{2}\left[ dr^{2}+r^{2}d\Omega ^{2}\right]
=a(\eta )^{2}[-d\eta ^{2}+dr^{2}+r^{2}d\Omega ^{2}],  \label{gij}
\end{equation}%
where $t$ and $\eta $, the cosmic and conformal time,
respectively, are related through the scale factor by $a(\eta
)d\eta ={dt}$.

Additionally to the conditions set by equation (\ref{coord}), we
impose the gauge conditions
\begin{equation}
{h^{ij}}_{;j}=0,  \label{gauge}
\end{equation}%
where the semi-colon denotes covariant derivative with respect to
the background metric (\ref{gij}). From them, it is possible to
show that the perturbation has two independent components only
(i.e., the wave has just two polarizations) and that the wave
equation simplifies to \be
{\overset{..}{{h}}_{\alpha}}^{\beta}+3\frac{\overset{.}{a}}{a}
{\overset{.}{{h}}_{\alpha}}^{\beta}+g^{\gamma\delta}{{h_{\alpha}}^{\beta}}_{,\gamma,\delta}=0,
\label{eghij} \ee where the upper dot and the commas indicate
partial derivatives respect to time and spatial coordinates,
respectively.

Introducing the ansatz%
 \be
{h_{\alpha}}^\beta(t,\textbf{x})=h(t){G_{\alpha}}^\beta(\textbf{k},\textbf{x}),
\ee%
where ${G_\alpha}^\beta$ is a combination of plane-waves solutions
$exp(\pm i \textbf{k}·\textbf{x})$ that contain the two
polarizations of the wave mentioned above and fulfill%
 \be
{G_{\alpha}^{\beta}}^{;\gamma}_{;\gamma}=-k^2G_{\alpha}^{\beta},
\qquad {G^{\beta}_{\alpha}}_{;\beta}=G^{\alpha}_{\alpha}=0,
\label{eqG}\ee%
into equation (\ref{eghij}) leads to the evolution equation for
the
temporal part of the wave%
\be {\overset{..}{h}}(k,t)+3\frac{\overset{.}{a}}{a}
{\overset{.}{h}(k,t)}+\frac{k^2}{a^2}{h}(k,t)=0. \label{eqht}\ee%
Or, in terms of the conformal time%
\be h^{\prime \prime}(k,\eta)+2\frac{a^{\prime}}{a}
h^{\prime}(k,\eta)+k^2h(k,\eta)=0, \label{eqheta}\ee%
where the prime indicates derivative with respect to $\eta$.

 In the equations of above we have used the
comoving wave-number $k=|\textbf{k}|$, which is related with the
wave-length and the frequency of the wave by \be k=\frac{2\pi
a}{\lambda}=a\omega.\ee

Equation (\ref{eqheta}) can be suitably simplified by using the
ansatz $h(k,\eta)=\mu(k,\eta)/a(\eta)$. This yields the so called
Lifshitz's equation

\be \mu ^{\prime \prime }(\eta )+\left( k^{2}-\frac{a^{\prime
\prime }(\eta )}{ a(\eta )}\right) \mu (\eta )=0.  \label{eqmu}\ee
It parallels the time independent Schr\"{o}dinger equation where
the terms $k^2$ and $a^{\prime \prime}/a$ play the role of the
energy and the potential, respectively. It can be also interpreted
as the equation of an harmonic oscillator parametrically excited
by the term $a^{\prime \prime }/a$. When $k^{2}\gg \frac{a^{\prime
\prime }}{a}$, i.e., for high frequency waves, expression
(\ref{eqmu}) becomes the equation of an harmonic oscillator whose
solution is a free wave, and consequently the amplitude of
$h_{\alpha \beta }(\eta ,\mathbf{x})$ decreases adiabatically as
$a^{-1}$ in an expanding universe. In the opposite regime,
$k^{2}\ll \frac{a^{\prime \prime }}{a}$, the general solution to (\ref%
{eqmu}) is a linear combination of the pair of solutions%
\be \mu _{1}= a(\eta ),{\qquad} {\qquad}\mu _{2}= a(\eta )\int
d\eta \ a^{-2}(\eta ).
\ee%
If the universe is expanding, $\mu _{1}$ will grow faster than
$\mu _{2}$ and will soon dominate. In this case, the amplitude of
$h_{\alpha \beta
}(\eta ,\mathbf{x})$ will remain constant as long as $%
k^{2}\ll \frac{a^{\prime \prime }}{a}$. If at some future time
this condition is no longer satisfied, the wave will have an
amplitude larger than it would in accordance with the adiabatic
behavior. This phenomenon is called \textquotedblleft
superadiabatic\textquotedblright\ or \textquotedblleft
parametric\textquotedblright\ amplification of gravitational waves
\cite{grish93}.

GWs that fulfill the condition $k^{2}\ll a^{\prime \prime }/a$
($k^{2}\gg a^{\prime \prime }/a$) are considered to be well inside
(outside) of the Hubble radius \cite{maia94}, i.e., their wave
lengths are shorter (longer) than $\lambda_H=H^{-1}(\eta)$, where
$H=a^{\prime}/a^2$ is the Hubble
function, and%
\be k<2\pi a H(\eta)\qquad \left(k>2\pi a H(\eta)\right).
\ee%
This can be traced to the fact that the ratio between $(2\pi a
H(\eta))^2$ and the term $a^{\prime \prime }/a$ is always constant
for a perfect fluid with an equation of state of barotropic type,
$p=(\gamma-1)\rho$, and tends to zero as soon as $\gamma
\rightarrow 4/3$.

\section{Gravitational waves creation}

In the previous subsection we have seen how the GWs may undergo
``parametric" amplification. This approach is based on the idea
that the amplitude of each GW is enlarged during the expansion of
the universe. Usually the initial amplitude is assumed to be the
vacuum one, $ A(k)\sim k$ \cite{grish77}. The rationale behind
this  is the following. One assimilates the quantum zero-point
fluctuations of vacuum with classical waves of certain amplitudes
and arbitrary phases; consequently it is permissible to equalize
$\omega/2$ with the energy density of the gravitational waves,
$h^{2}/\lambda^{2}$ times $\lambda^{3}$. Thereby the initial
vacuum amplitude of GWs is given  by $h \propto A(k) \propto k$
\cite{grish93}.

A different approach can be developed by realizing that Lifshitz's
equation (\ref{eqmu}) is not invariant under conformal
transformations of
the metric \cite{grish74, Birrell}%
\be g_{ij}\rightarrow\widehat{g}_{ij}=\Omega^2(t,\textbf{x})g_{ij},\ee %
where $\Omega^2(t,\textbf{x})$ is a real, continuous, finite and
non vanishing function, except when $a^{\prime \prime}=0$. With
this in mind, the ``parametric'' amplification may be interpreted
as  spontaneous GWs creation due to the action of the expansion of
the Universe over the initial vacuum.

\subsection{A simple example: the quantum pendulum}

The ideal pendulum, a weight hanging from a string of length $l$,
pedagogically illustrates the phenomena of particle creation from
the initial vacuum. The pendulum oscillates around its equilibrium
position with frequency $\sqrt{g/l}$. If the string is wound
around a reel, the length of the string can vary as well as the
oscillation frequency.

Being the pendulum initially at the minimum energy configuration,
we increase the length of the string by lowering the bob, in a
time interval $\tau=t_f-t_i$, from an initial value $l_i$ to a
final value $l_f$. Classically, the pendulum (initially at rest)
will be also at rest in the final state and the change of
potential energy will equal the friction work.

From a quantum mechanic point of view, the initial minimum energy
configuration of the system corresponds to the state with energy
$E_i=\omega_i/2$ and number of quanta $N_i=0$. If the time spent
in lowering the bob is much longer than the initial and final
periods of oscillation, i.e., $\tau \gg T_f=2\pi/\omega_f$ (the
adiabatic case), the bob undergoes more than one complete
oscillation during the process. In this case the final energy is
$E_f=\omega_f/2$ and the number of quanta in the final state is
$N_f=0$. The potential energy obtained is partly dissipated by the
reel, as in the classical approach, but also by the change of the
energy from the initial to the final state.

By contrast, if the time spent in lowering the bob is much shorter
than the initial and final periods of oscillation, i.e.,  $\tau
\ll T_i=2\pi/\omega_i$, things fare differently. The final energy
of the oscillator will be half the energy of the initial state
$E_f=E_i/2$ and the final vacuum energy will be $\omega_f/2$ (see
Ref \cite{allen96} for details). Consequently, the final number of
quanta $N_f=\frac{1}{4}\frac{\omega_i}{\omega_f}$ will differ from
zero. Note that these quanta are created because of the sudden
change in the vacuum state.

In the case of the GWs, we must replace the pendulum equation for
equation (\ref{eqmu}). The role of the variable length $l$ is now
played by the term $a^{\prime\prime}(\eta)/a(\eta)$. If
$a^{\prime\prime}(\eta)/a(\eta)$ varies slowly compared with the
wave number $k$, the initial vacuum state will evolve into the
final vacuum state and no GW will be generated. But if
$a^{\prime\prime}(\eta)/a(\eta)$ evolves fast enough, the final
state will not be the vacuum one and GWs will be spontaneously
generated.

\subsection{Bogoliubov coefficients}
In Minkowski spacetime, a scalar field $\phi$ of
mass $m$ obeys the Klein-Gordon equation%
\be \left( \eta ^{ij}\partial _{i}\partial _{j
}+m^{2}\right) \phi =0, \label{eqphimik}\ee%
where $\eta^{ij}$ represents the metric. The field $\phi$ can be
expanded as%
\be
\phi (x)=\sum_{\mathbf{k}}[A_{\mathbf{k}}\nu
_{\mathbf{k}}(x)+A_{\mathbf{k}}^{\dagger }\nu
_{\mathbf{k}}^{\ast }(x)],{\ } \label{eqf0} \ee%
in terms of the family of modes%
\be \nu_{\mathbf{k}}=\frac{1}{4\pi \sqrt{\pi k}}\,
e^{i(\mathbf{k\cdot x}-k t)},
\ee%
which are positive-frequency defined with respect to $t$%
 \be \frac{\partial
\nu_{\mathbf{k}}}{\partial t}=-i k
\nu_{\mathbf{k}}{\qquad }(k >0). \ee%
This set of modes forms a privileged family: it defines creation
and annihilation operators $A_{\mathbf{k}}$ and
$A_{\mathbf{k}}^{\dagger}$, respectively, representing real
particles for all inertial observers, as the quantum vacuum
defined by the modes is invariant under Poincar\'e
transformations.

In curved spacetime, the equation obeyed by the scalar field reads%
\be g^{ij}\phi_{;i;j}+\left(m^2+\xi R\right)\phi=0 \label{eqphicurv}\ee%
where $\xi$ is a dimensionless parameter and $R$ is the Ricci
scalar. The term $\xi R$ accounts for the coupling between the
scalar and the gravitational field.

When $m=0$ and $\xi=1/6$, equation (\ref{eqphicurv}) is
conformally invariant. In this case if the spacetime considered
can be transformed conformally in the Minkowski one, it is
possible to define a new field that obeys equation
(\ref{eqphimik}) and privileged modes can be chosen in order to
define a vacuum state. However, generally, there is no privileged
family of modes as the Poincar\'e group is no longer a symmetry.

In general, given a complete collection of modes $\nu _{m}(x)$
the field can be expanded as%
\be \phi (x)=\sum_{m}[A_{m}\nu _{m}(x)+A_{m}^{\dagger }\nu
_{m}^{\ast }(x)],{\ } \label{eqf} \ee%
where $A_m$ and $A_m^{\dagger}$ are, respectively, the
annihilation and creation operators associated to the quantum
vacuum state of the family, $|0\rangle$. The field can also be expanded as%

\be \phi (x)=\sum_{n}\left[ \overline{A}_{n}\overline{\nu }_{n}(x)+\overline{A}%
_{n}^{\dagger }\overline{\nu }_{n}^{\ast }(x)\right] \label{eqfbar}, \ee%
in terms of a different complete collection of modes,
$\overline{\nu }_{n}(x)$, with a different vacuum state,
$|\overline{0}\rangle$.

The Bogoliubov transformation%
\be \overline{\nu }_{n}(x)=\sum_{m}\left[ \alpha _{nm}\nu
_{m}(x)+\beta _{nm}\nu _{m}^{\ast }(x)\right] ,  \label{coefBog}
\ee%
where $\alpha_{nm}$ and $\beta_{nm}$ are the coefficients of
Bogoliubov, relates both families of modes. From equations
(\ref{eqf})-(\ref{coefBog}), it is readily seen that%
\ba
A_{m} &=&\sum_{n}(\alpha _{nm}\overline{A}_{n}+\beta _{nm}\overline{A}%
_{n}^{\dagger }), \\
\overline{A}_{n} &=&\sum_{m}(\alpha _{nm}A_{m}+\beta
_{nm}A_{m}^{\dagger }). \label{transop}
\ea%

As the creation/annihilation operators satisfy the commutation
relation
$[A_m,A_m^{\dagger}]=[\overline{A}_n,\overline{A}_n^{\dagger}]=1$,
the Bogoliubov coefficients must fulfill %
\be |\alpha_{nm}|^2-|\beta_{nm}|^2=1. \ee

The operator number of particles  of the first family,
$N_{m}=A_{m}^{\dagger }A_{m}$, acts upon the quantum vacuum
state of the second family according to%
\be
\langle \overline{0}|N_{m}|\overline{0}\rangle =\langle \overline{0}%
|A_{m}^{\dagger }A_{m}|\overline{0}\rangle =\sum_{n}|\beta
_{nm}|^{2}. \ee%
Therefore, the quantum vacuum state of the second family contains
$\sum_{n}|\beta _{nm}|^{2}$ particles of the first family. An
entirely similar relation holds for the first family vacuum state
and the second family number of particles operator.

The GWs equation (\ref{eqheta}) formally coincides with the
equation of a massless decoupled field in a spatially flat FRW
spacetime. Its solutions are%
\be h_{ij}(\eta ,\mathbf{x})=\int \left( A_{(k)}h_{ij}^{(k)}(\mathbf{k},x)+%
A_{(k)}^{\dagger}{h_{ij}^{(k)}}^{*}(\mathbf{k},x)\right) d^{3}k. \label{fieldh}\ee%
The modes $h_{ij}^{(k)}(\mathbf{k},x)$ can be written as%
\be h_{ij}^{(k)}(\mathbf{k},x)=\frac{\sqrt{8\pi }}{(2\pi )^{3/2}}e_{ij}(\mathbf{k%
})\frac{\mu _{(k)}(\eta )}{a(\eta )}e^{i\mathbf{k\cdot x}}\mathbf{,}\label{modeshk}\ee%
where $e_{ij}(\mathbf{k})$ contains both polarizations of the
wave, and the functions $\mu _{(k)}(\eta )$ are solutions to
(\ref{eqmu}). As the family of modes that quantify the field is
complete, the modes
$h_{ij}^{(k)}(\mathbf{k},x)$ must be orthogonal, i.e.,%
\be -i\int \left(h_{ij}^{(k)}(\mathbf{k},x){h_{ij}^{(k^{\prime
})}}^{*\prime }(\mathbf{k},x) -{h_{ij}^{(k)}}^{\prime
}(\mathbf{k},x){h_{ij}^{(k^{\prime })}}^{*}(\mathbf{k},x)\right)
d^{n-1}x=\delta (\mathbf{k}-\mathbf{k}^{\prime }),\ee%
and, therefore, $\mu_{(k)}$ obeys the further condition %
\be \mu_{(k)}\mu^{*\prime}_{(k)}-\mu_{(k)}^*\mu_{(k)}^{\prime}=i.
\label{condcuan} \ee%

If the spatially flat FRW universe contains a perfect fluid with
equation of
state%
\be p=(\gamma-1)\rho, \ee%
with $\gamma$ constant, the Einstein equations%
\be \left(\frac{a^{\prime}}{a^{2}}\right)^2=\frac{8\pi}{3} \rho , \ee %
\be \rho^{\prime}+3\frac{a^{\prime}}{a}(\rho+p)=0,
\ee%
can be solved to%
\be a(\eta)=a_i\left(\frac{a_iH_i}{l}\right)^{l}
\left(\eta-\eta_i+\frac{l}{a_iH_i}\right)^l,\label{scalef}\ee%
where $a_i$, $H_i$ and $\eta_i$ are, respectively, the initial
values of the scale factor, the Hubble function and the conformal
time and $l=2/(3\gamma-2)$. For simplicity, we define
$\eta_l=\eta-\eta_i+l/(a_iH_i)$ and Lifshitz's equation reduces to
the Bessel equation%
\be \mu^{\prime
\prime}(\eta)+\left[k^2-\frac{l(l-1)}{(\eta_l)^2}\right]\mu(\eta)=0,
\ee%
whose solutions can be written in terms of Hankel functions as %
\be \mu _{l}(\eta)=\frac{\sqrt{\pi }}{2}k^{-1/2}x_{l}^{1/2}\left[
D_{1l}H_{l-\frac{1}{2}}^{(1)}(x_{l})+D_{2l}H_{l-\frac{1}{2}}^{(2)}%
(x_{l})\right],\label{mul}\ee%
where $x_{l}=k\eta_l$, and $D_{1l}$ and $D_{2l}$ are integration
constants.

Henceforward, we shall work in the Heisenberg picture, the quantum
states are functions of time meanwhile the operators are time
independent. Initially the universe will be in some state from
which the number of GWs operator will be defined. As the scale
factor evolves, the quantum state evolves too and the constant
operator acts upon this state which is no longer the initial one.

In this scenario, it seems problematic to choose the appropriate
vacuum state or to define real particles. A possible solution to
this problem rests on the concept of adiabatic vacuum.

\subsection{The adiabatic vacuum approximation}

This approximation rests on the idea that the creation of
particles by a slow change in the initial state is minimal
\cite{Birrell}.

Solutions to the equation%
\be \chi ^{\prime \prime }+\omega ^{2}(\eta )\chi =0,  \label{vacioad}\ee%
with $\omega\in C^{\infty}$ are expressed as a linear
combination of basis functions which can be approximated by%
\be \chi ^{\pm }(\eta )=\frac{1}{\sqrt{2W(\eta )}}\exp \left( \mp
i\int^{\eta }W(\eta ^{\prime })d\eta ^{\prime }\right)
+\mathcal{O}(\omega ^{-N}), \label{xi+}\ee %
with%
\be W^{2}(\eta )=\omega ^{2}(\eta)\left[ 1+\chi _{2}(\eta )\omega
^{-2}+\chi _{4}(\eta )\omega
^{-4}+...\right] , \ee%
where the $\chi _{n}(\eta )$ are functions of $\omega (\eta )$ and
its derivatives up to the derivative of order $n$, which is
bounded as $\omega \rightarrow \infty$ \cite{fulling}. If we
introduce the adiabatic parameter $\Theta$ by replacing $\Theta\,
\omega (\eta)$ for $\omega (\eta )$ (at the end we can let $\Theta
=1$), the adiabatic behavior (slow expansion limit) of equation
(\ref{vacioad}) can be examined when $\Theta \rightarrow \infty$,
and, in this limit, the above expression can be considered a
power-series expansion in $\Theta^{-1}$ to order $N$.

Equations (\ref{eqmu}) and (\ref{vacioad}) share the same form.
Thus, we can define the adiabatic vacuum of order $N$ by taking
$\mu$ in equation (\ref{modeshk}) as an exact solution to
(\ref{eqmu}) whose WKB approximation is precisely $\chi^{+}$ in
(\ref{xi+}). To this order, the operators $A_{(k)}$ in
(\ref{fieldh}) correspond exactly to physical particles when
$\Theta \rightarrow \infty$ or when $k\rightarrow \infty$.

Note that in this limit, the modes $h_{ij}^{(k)}$ tend to the
positive-frequency Minkowski modes. Hence, the constants in
equation (\ref{mul}) turn out to be $D_{1l}=0$ and $D_{2l}=1$, thereby%
\be \mu _{l}=(\sqrt{\pi }/2)e^{i\psi
_{l}}k^{-1/2}x_{l}^{1/2}H_{l-\frac{1}{2}}^{(2)}(x_{l}). \label{mucuant} \ee%

\subsection{The creation mechanism}

Let us assume that the spatially flat FRW universe considered
above goes through a succession of stages, and that, in a generic
stage $r$, the Universe is dominated by a perfect fluid of
barotropic index $\gamma_r$. The transitions between the different
stages can be considered instantaneous or, more accurately, much
shorter than the duration of each stage.

The scale factor, in this scenario, is given by equation
(\ref{scalef}) with a different $l_r$  in each era and the
subindex $i$ denoting the beginning of the $r$-stage (i.e., the
end of the ($r-1$)-stage). Note that the scale factor must be
continuous at each transition $\eta_i$ as no discontinuities must
be present in the FRW metric. In each stage the solution to
Lifshitz's equation (\ref{eqmu}) is (\ref{mucuant}) with $l$
replaced by $l_r$.

The modes of two consecutive stages (namely the ($r-1$)-stage and
the $r$-stage) are related by a Bogoliubov transformation

\be \mu _{l_{r-1}}(\eta _{i})=\alpha_i \mu _{l_r}(\eta
_{i})+\beta_i \mu _{l_r}^{\ast }(\eta _{i}). \ee

Since the functions $\mu_{l_r}(\eta)$ and
$\mu^{\prime}_{l_r}(\eta)$ must be also continuous at $\eta_i$
(further details about the continuity of $\mu$ and the validity of
the sudden transition approximation are given in the Appendix), it
is possible to obtain the expression of the Bogoliubov
coefficients in terms of known quantities \cite{Maia93,MaiaTh}

\be \alpha_i =i[\mu _{l_{r-1}}^{\prime }(\eta _{i})\mu
_{l_r}^{*}(\eta _{i})-\mu _{l_{r-1}}(\eta _{i}){\mu
_{l_r}^{*}}^{\prime }(\eta _{i})],
\label{A-coef}\ee%

\be \beta_i =i[\mu _{l_{r-1}}(\eta _{i})\mu _{l_r}^{\prime }(\eta
_{i})-\mu _{l_{r-1}}^{\prime }(\eta _{i})\mu _{l_r}(\eta _{i})].
\label{B-coef} \ee%

The adiabatic approximation sets a bound over the GWs that can be
created in this transition. The limit of slow expansion,
$\omega\rightarrow \infty$, can be relaxed by assuming that there
is no creation of waves for $\omega(\eta_i) \gg\omega_i$. This
bound can be fixed as the frequency associated to the
characteristic time in which takes place the transition. Modes
whose period at the instant $\eta_i$ are shorter than this
transition time experiment an adiabatic expansion thereby they
still represent the quantum vacuum state, i.e., for them,
$\alpha_i=1$ and $\beta_i=0$. Meanwhile modes whose period at the
transition $\eta_i$ is larger than this characteristic time do not
represent the quantum vacuum state any longer, and $\beta_i\neq
0$. This characteristic time is usually chosen as the inverse of
the Hubble parameter at the transition, $H_i^{-1}$. Then, the
adiabatic bound for the frequencies at $\eta>\eta_i$ reads%
\be \omega_{i} (\eta)=2\pi \frac{a_i}{a(\eta)}H_i, \ee%
where we have taken into account the redshift.

Assuming that each created GW has an energy $2\, \omega (\eta )$
(the factor $2$ comes from the two polarization states), it is
possible to express the energy density of GWs created at the
transition $\eta_i$ with frequencies in the range $\left[ \omega
(\eta ),\omega (\eta )+d\omega (\eta
)\right] $ as%
\be d\rho _{g}(\eta )=2 \omega (\eta )\left[ \frac{\omega
^{2}(\eta )}{2\pi ^{2}}d\omega (\eta )\right] |\beta_i|^2=P(\omega
(\eta ))d\omega (\eta ),
\label{densquan} \ee%
 where $P(\omega (\eta ))=\left( \omega
^{3}(\eta )/\pi ^{2}\right) |\beta_i|^2 $ denotes the power
spectrum. As the energy density is a locally defined quantity,
$\rho _{g}$ loses its meaning for metric perturbations whose wave
length $\lambda =2\pi /\omega (\eta )$ exceed  the Hubble radius
$H^{-1}(\eta )$. Integrating over the frequency, we get the total
energy density

\be \rho _{g}(\eta )= \int^{\omega_i(\eta)}_{2\pi H(\eta)}P(\omega
(\eta ))d\omega (\eta ).
\ee%
The adiabatic cutoff amounts to a renormalization of the quantum
field theory. This upper bound ensures that the energy density
does not diverge at high frequencies. The adiabatic bound plays
the same role that the extraction of the energy of the vacuum in
the quantum field theories in Minkowski spacetime.

To relate the $\mu_{l}$ functions of two non-consecutive stages,
we must make use of the total Bogoliubov
coefficients which can be found recursively from%
\be \alpha _{T_{i}}=\alpha _{i}\alpha _{T_{(i-1)}}+\beta
_{i}^{\ast }\beta _{T_{(i-1)}}, \label{alphaT}\ee%

\be \beta _{T_{i}}=\beta _{i}\alpha _{T_{(i-1)}}+\alpha
_{i}^{*}\beta _{T_{(i-1)}}, \label{betaT}\ee%
where the subindex $(i-1)$ denotes the previous transition to the
$i$-th one \cite{Maia93}.

Equations (\ref{alphaT}) and (\ref{betaT}) readily follow from the
Bogoliubov transformation%

\begin{equation}
\left(
\begin{array}{l}
\mu _{l_{r-1}} \\
\mu _{l_{r-1}}^{\ast }%
\end{array}%
\right) =\left(
\begin{array}{ll}
\alpha _{i} & \beta _{i} \\
\beta _{i}^{\ast } & \alpha _{i}^{\ast }%
\end{array}%
\right) \left(
\begin{array}{l}
\mu _{l_r} \\
\mu _{l_{r}}^{\ast }%
\end{array}%
\right) .  \label{MatrizCoef}
\end{equation}

Thus, the modes of the $r-n$ stage, $\mu _{l_{r-n}}$, are related
with $\mu _{l_{r}}$

\be \left(
\begin{array}{l}
\mu _{l_{r-n}} \\
\mu _{l_{r-n}}^{*}%
\end{array}
\right) =\left(
\begin{array}{ll}
\alpha _{i-(n-1)} & \beta _{i-(n-1)} \\
\beta _{i-(n-1)}^{*} & \alpha _{i-(n-1)}^{*}%
\end{array}
\right) ...\left(
\begin{array}{ll}
\alpha _{i-1} & \beta _{i-1} \\
\beta _{i-1}^{*} & \alpha _{i-1}^{*}%
\end{array}
\right) \left(
\begin{array}{ll}
\alpha _{i} & \beta _{i} \\
\beta _{i}^{*} & \alpha _{i}^{*}%
\end{array}
\right) \left(
\begin{array}{l}
\mu _{l_r} \\
\mu _{l_r}^{*}%
\end{array}
\right) , \ee%
(note there is $n-1$ transitions between both eras). After
simplifying we get

\be \left(
\begin{array}{l}
\mu _{l_{r-n}} \\
\mu _{l_{r-n}}^{*}%
\end{array}
\right) =\left(
\begin{array}{ll}
\alpha _{T_{i}} & \beta _{T_{i}} \\
\beta _{T_{i}}^{*} & \alpha _{T_{i}}^{*}%
\end{array}
\right) \left(
\begin{array}{l}
\mu _{l_r} \\
\mu _{l_r}^{*}%
\end{array}
\right) . \ee%

We will employ Eqs. (\ref{alphaT}) and (\ref{betaT}) in the
following chapters.%

\chapter{Quantum mini black holes and the gravitational waves spectrum \label{mbh&gwspect}}

\markboth{CHAPTER \thechapter. QUANTUM MBHS AND GWS SPECTRUM}{}

In this chapter we consider in detail the power spectrum of relic
GWs assuming the Universe went through an expansion era dominated
by radiation and mini black holes (MBHs) intermediate between the
conventional inflationary (De Sitter) era and the radiation
dominated era \cite{MBHs}. The existence of that era is justified
in subsection \ref{MBH-scen}. In subsection \ref{three-stage} we
recall the derivation of the power spectrum of the conventional
three-stage scenario.

\section{Three-stage spatially flat FRW scenario \label{three-stage}}

Here we apply the fundamental equations found in the previous
chapter to a spatially flat FRW scenario consisting in an initial
De Sitter stage, followed by a radiation dominated era and a dust
era that includes the present time. The power spectrum predicted
for this scenario is well known in the literature, see Refs.
\cite{Maia93} and \cite{Allen}-\cite{Star}. The scale factor is
given by

\begin{equation}
a(\eta )=\left\{
\begin{array}{c}
-\frac{1}{H_{1}\eta }\qquad (-\infty <\eta <\eta _{1}<0), \\
\frac{1}{H_{1}\eta _{1}^{2}}(\eta -2\eta _{1})\qquad (\eta
_{1}<\eta <\eta
_{2}), \\
\frac{1}{4H_{1}\eta _{1}^{2}}\frac{(\eta +\eta _{2}-4\eta
_{1})^{2}}{\eta
_{2}-2\eta _{1}}{\qquad }(\eta _{2}<\eta <\eta _{0}).%
\end{array}%
\right.  \label{sclfac1}
\end{equation}%

The initial state is the vacuum associated with the modes of the
inflationary stage $\mu _{I}(\eta )$, which
are a solution to Lifshitz's equation (\ref{eqmu}) compatible with the condition (%
\ref{condcuan}). Taking into account that $l=-1$ in this era (De
Sitter), the modes read
\begin{equation}
\mu _{I}=(\sqrt{\pi }/2)e^{i\psi
_{I}}k^{-1/2}x^{1/2}H_{-3/2}^{(2)}(x), \label{muinfcuant}
\end{equation}%
where $x=k\eta$, $\psi _{I}$ is an arbitrary constant phase and $%
H_{-3/2}^{(2)}(x)$ is the Hankel function of order $-3/2$. The
proper modes of the radiation era ($l=1$) are%
\begin{equation}
\mu _{R}=(\sqrt{\pi }/2)e^{i\psi
_{R}}k^{-1/2}x_{R}^{1/2}H_{1/2}^{(2)}(x_{R}),  \label{murad}
\end{equation}%
where $x_{R}=k(\eta -2\eta _{1})$ and $\psi _{R}$ is again a
constant phase.

The ``parametric" amplification in this first transition was first
developed by Grishchuk \cite{grish74} and Starobinsky \cite{Star}.
Following the
quantum approach, the two families of modes are related by%
\begin{equation}
\mu _{I}(\eta )=\alpha _{1}\mu _{R}(\eta )+\beta _{1}\mu
_{R}^{\ast }(\eta ). \label{bogtr}
\end{equation}%
From equations (\ref{A-coef}) and (\ref{B-coef}), we obtain

\begin{equation}
\alpha _{1}=-1+\frac{i}{k\eta _{1}}+\frac{1}{2(k\eta
_{1})^{2}},\qquad \beta _{1}=\frac{1}{2(k\eta _{1})^{2}},
\label{coef1}
\end{equation}%
where we have neglected an irrelevant phase. Considering the
adiabatic quantum approximation, the modes whose frequency at the
transition are larger than the characteristic time scale
($H_{1}^{-1}$) get exponentially suppressed. Thus, the
coefficients are $\alpha _{1}=1$ and $\beta _{1}=0$ for GWs with
$k>2\pi a_{1}H_{1}$ and (\ref{coef1}) when $k<2\pi a_{1}H_{1}$.

In the dust era ($\eta >\eta _{2}$ and $l=2$) the solution for the modes is%
\begin{equation}
\mu _{D}=(\sqrt{\pi }/2)e^{i\psi
_{D}}k^{-1/2}x_{D}^{1/2}H_{3/2}^{(2)}(x_{D}),  \label{mudustcuan}
\end{equation}%
where $x_{D}=k\left( \eta +\eta _{2}-4\eta _{1}\right) $ and it is
related to the radiation ones by%
\begin{equation}
\mu _{R}(\eta )=\alpha _{2}\mu _{D}(\eta )+\beta _{2}\mu
_{D}^{\ast }(\eta ). \label{bogtrr_d}
\end{equation}%
Similarly one obtains%
\be
\alpha _{2}=-i\left( 1+\frac{i}{2k(\eta _{2}-2\eta
_{1})}-\frac{1}{8\left(
k(\eta _{2}-2\eta _{1})\right) ^{2}}\right) ,\qquad \beta _{2}=\frac{i}{%
8\left( k(\eta _{2}-2\eta _{1})\right) ^{2}}, \label{coef2}\ee%
for $k<2\pi a(\eta _{2})H_{2}$ and $\alpha _{2}=1$, $\beta _{2}=0$ for $%
k>2\pi a(\eta _{2})H_{2}$ where $H_{2}$ is the Hubble function
evaluated at the radiation-matter transition $\eta _{2}$.

In order to relate the modes of the inflationary era to the modes
of the dust era, we make use of the total Bogoliubov coefficients
$\alpha _{Tr2}$ and $\beta _{Tr2}$, given by equations
(\ref{alphaT}) and (\ref{betaT}). For $k>2\pi a_{1}H_{1}$, we find
that $\alpha _{Tr2}=1$ and $\beta _{Tr2}=0$; in the range $2\pi
a_{1}H_{1}>k>2\pi a_{2}H_{2}$, the coefficients are $\alpha
_{Tr2}=\alpha _{1}$ and $\beta _{Tr2}=\beta _{1}$, and finally for
$k<2\pi a(\eta _{2})H_{2}$
we obtain that%
\be%
\beta _{Tr2} \simeq -\frac{2(\eta _{2}-2\eta _{1})+\eta
_{1}}{8k^{3}\eta
_{1}^{2}(\eta _{2}-2\eta _{1})^{2}}. \ee%

Thus the number of GWs at the present time, $\eta _{0}$, created
from the initial vacuum state is $\left\langle N_{\omega
}\right\rangle =\left\vert \beta _{Tr2}\right\vert ^{2}\sim \omega
^{-6}(\eta _{0})$ for $\omega (\eta _{0})<2\pi
(a_{2}/a_{0})H_{2}$, $\omega ^{-4}(\eta _{0})$ for $2\pi
(a_{1}/a_{0})H_{1}>\omega (\eta _{0})>2\pi (a_{2}/a_{0})H_{2}$ and zero for $%
\omega (\eta _{0})>2\pi (a_{1}/a_{0})H_{1}$, where we have used
the present value of the frequency, $\omega (\eta _{0})=k/a_{0}$.

The present power spectrum of GWs predicted by this model is
\begin{equation}
P\left( \omega \right) \sim \left\{
\begin{array}{ll}
0 & \left( \omega (\eta _{0})>2\pi (a_{1}/a_{0})H_{1}\right) , \\
\omega ^{-1}(\eta _{0}) & \left( 2\pi (a_{2}/a_{0})H_{2}<\omega
(\eta _{0})<2\pi (a_{1}/a_{0})H_{1}\right) , \\
\omega ^{-3}(\eta _{0}) & \left( 2\pi H_{0}<\omega (\eta
_{0})<2\pi (a_{2}/a_{0})H_{2}\right) .%
\end{array}%
\right.  \label{espectr}
\end{equation}%

The spectrum (\ref{espectr}) is shown in figure \ref{figspec}
tagged as $l=1$ for different choices of $H_1$, $a_{1}/a_{0}$ and
$a_{2}/a_{0}$.

Below we compare the predictions of the three-stage model of
above, for the frequency ranges and powers, with the four-stage
model that includes the \textquotedblleft
MBHs+rad\textquotedblright\ era.

\markboth{CHAPTER \thechapter. QUANTUM MBHS AND GWS SPECTRUM}{GWS
IN A FRW UNIVERSE WITH AN ERA OF...}

\section{GWs in a FRW universe with an era of\ mini black holes and
radiation \label{MBH-scen}}%

\markboth{CHAPTER \thechapter. QUANTUM MBHS AND GWS
SPECTRUM}{\thesection.  GWS IN A FRW UNIVERSE WITH AN ERA OF...}

\subsection{Mini black holes in the very early Universe}
As is well known, MBHs can be created by quantum tunnelling from
the hot radiation with a rate of
nucleation given by \cite{gross}%
\be \Gamma
(T)=0.87T\left(\frac{\mu}{T}\right)^{\theta}\frac{1}{64\pi^3}e^{-1/16\pi
T^2},
\ee%
where $T$ is the temperature of the radiation, $\mu$ is a
parameter close to the Planck mass, and $\theta$ is a numerical
factor which depends on the number of spin fields accessible to
the system (i.e., $\theta=3.083$ for the standard model, $3.656$
for the supersymmetric standard model, $3.000$ for the
supersymmetric SU(5) and $0.283$ for the SU(5)).

Thus, the number of MBHs per unit comoving
volume at time $t$ is%
\be
n(t)\sim\frac{1}{a^3(t)}\int^t_{t^*}a^3(t^{\prime})%
\Gamma\left(T(t^{\prime})\right)dt^{\prime},
\ee%
where $t^*$ is the time of formation of a MBH which would have
just evaporated at time $t$. And the density of MBHs is%
\be%
\rho_{BH}=\frac{1}{a^3} \int^t_{t^*}a^3(t^{\prime})M\left(t,t^{\prime}\right)%
\Gamma\left(T(t^{\prime})\right)dt^{\prime},%
\ee where $M(t,t^{\prime})$ is the mass a MBH would have by time
$t$ if formed at time $t^{\prime}$.

The MBHs nucleated by quantum tunnelling will be strongly peaked
around an initial mass given by%
\be M_{BH}|_{t_{form}}=\left. \frac{1}{8\pi T}\right|_{t_{form}},%
\ee%
where $t_{form}$ is the instant of the MBH formation and $T$ is
the temperature of the radiation. Assuming that during the
formation of the MBHs the expansion of the Universe is dominated
by
the hot radiation%
\be%
p=\frac{\rho}{3},\qquad \rho=\frac{\pi^2}{30}g_*T^4,
\ee%
here $g_*$ is the number of relativistic particles species, and
using the Friedman equations, one has that $T\propto t^{-1/2}$.
Consequently, the initial mass of MBHs will increase as
$t^{1/2}_{form}$. The evolution of the mass of the MBH from this
point onward will depend on its interactions with the surrounding
radiation and with the other MBHs \cite{Hayward}.

If the total number of MBHs in a comoving volume is constant, the
relative velocity between two neighbor MBHs due to the
expansion is%
\be%
v_{exp}=n^{-1/3}H,
\ee%
where $n$ is the MBHs number density. The velocity required for a
MBH to escape the gravitational attraction of its neighbor is%
\be%
v_{esc}=\sqrt{4mn^{1/3}},
\ee%
where $m=\rho_{BH}/n$ is the mass of typical MBH. Given that
during this period $H=\sqrt{(8\pi/3)(\rho_{rad}+\rho_{BH})}$, it
is straightforward to conclude that $v_{exp}\geq v_{esc}$.
Therefore, the collisions between neighbor MBHs are not frequent
enough to significantly affect the mass spectrum of MBHs.

Even though the MBHs form at the same temperature as their
surroundings, one might imagine that cooling of the radiation due
to Hubble expansion would cause the MBH to begin to evaporate
freely immediately after formation. Therefore, their mass would
evolve according to \cite{Hawk}%
\be%
\left( dM/dt\right) _{free}=-g_{\ast \ast}/(3M^{2}),%
\label{freeev}
\ee%
where $g_{\ast \ast}$ is the number of particle species for the
black hole to evaporate into.

But the MBH evaporation can be delayed if it begins absorbing
radiation after its formation so that its mass increases.
Comparing the average time of interaction between the MBHs and the
surrounding radiation, $\tau_{int}=1/(36\pi m \rho_{BH})$, with
the characteristic time of expansion of the Universe (the inverse
of the Hubble factor, $t_{exp}=H^{-1}$), it is straightforward to
demonstrate that, although these interactions are not frequent
enough to consider an evolution different from (\ref{freeev}) for
all the particle models of interest, the energy density of the
MBHs can become comparable or even exceed that of the radiation
for sufficiently high temperatures \cite{Hayward}.

Further, it is also natural to assume that the MBHs are surrounded
by an atmosphere of particles in quasi thermal equilibrium with
them \cite{Thorne}. The MBH emits quanta in a perfectly thermal
manner and these quanta might create a thermal atmosphere
surrounding tightly the MBH. The vast majority of the quanta
emitted are prevented from escape the MBH gravitational potential
whether because they have a large angular momentum or  because,
having an adequately small angular momentum, they tend to have
such large wave lengths that when trying to escape they scatter of
the MBH spacetime curvature and are driven back towards the
horizon. Thus the absorption and emission of particles from the
atmosphere to the MBH and vice versa prevents the MBH to evaporate
freely. Therefore, we have that $\left| \left( dM/dt \right)_{atm}
\right|\ll \left| \left( dM/dt\right) _{free} \right|$.

Even assuming the MBHs begin to evaporate freely since they are
created, except in the SU(5) model, $\rho_{BH}$ will be comparable
to the radiation density in a time span of two to one hundred
times the Planck time from the instant the nucleation starts
\cite{Hayward, Barrow}. It is reasonable to expect that, at this
point, a steady state would be achieved where the total energy
density is shared between the black holes and the radiation whence
$\rho =\rho _{BH}+\rho _{R}$ and, consequently, the total pressure
is
\begin{equation}
p=p_{BH}+p_{R}=(\gamma -1)\rho ,  \label{eqstate}
\end{equation}%
where the constant $\gamma$ lies in the interval $1\leq \gamma
<4/3$. If the density of MBHs is large enough to dominate the
expansion of the Universe, then $\gamma \simeq 1 $. In the
opposite case, the Universe expansion is dominated by the
radiation, $\gamma \simeq 4/3$. From the Einstein equations and
(\ref{eqstate}) one finds that $1< l \leq 2$ during the
\textquotedblleft MBHs+rad\textquotedblright\ era. The MBHs
eventually evaporate in relativistic particles after some time
span which can be fairly large if MBHs have a thermal atmosphere.

Interestingly, the evaporation of MBHs when the age of the
Universe was about $100$ Planck times may explain why the cosmic
baryon-number to photon ratio is of the order of $10^{-9}$.
Lindley argued that that small figure can be obtained if the
Universe's expansion was dominated by black holes of a few hundred
Planck mass at the mentioned epoch \cite{lindley}. This looks
feasible in the scenario contemplated here.

\subsection{Power spectrum of the four-stage scenario}

In this subsection we obtain the power spectrum of the GWs
assuming the following eras in succession: an initial De Sitter
era, an era dominated by MBHs and radiation, the conventional
radiation-dominated era and the dust era.

The scale factor of this four-stage scenario is%
\be
a(\eta )=\left\{
\begin{array}{lcl}
-\frac{1}{H_{1}\eta } & (-\infty <\eta <\eta _{1}<0),
&\text{(De Sitter era)} \\
\frac{\left[ \eta _{BH}\right] ^{l}}{l^{l}H_{1}(-\eta _{1})^{l+1}}
& (\eta _{1}<\eta <\eta _{2}),& \text{(\textquotedblleft
MBHs+rad\textquotedblright\ era)} \\
\frac{\left( \eta _{R2}\right) ^{l-1}}{H_{1}(-\eta _{1})^{l+1}}\eta _{R} &%
(\eta _{2}<\eta <\eta _{3}),& \text{(radiation\ era)} \\
\frac{\left( \eta _{R2}\right) ^{l-1}}{2H_{1}(-\eta _{1})^{l+1}\eta _{D3}}%
\left[ \eta _{D}\right] ^{2}&(\eta _{3}<\eta <\eta _{0}),&%
\text{(dust era)}%
\end{array}%
\right. \ee%
where $\eta _{BH}=\eta -(l+1)\eta _{1}$, $\eta _{R}=\eta +\frac{(1-l)}{l}%
\eta _{2}-\frac{(l+1)}{l}\eta _{1}$, $\eta _{D}=\eta +\eta _{3}+2\frac{(1-l)%
}{l}\eta _{2}-2\frac{(l+1)}{l}\eta _{1}$, $\eta _{R2}=\left[ \eta
_{2}-(l+1)\eta _{1}\right] /l$ and $\eta _{D3}=2\left[ \eta _{3}+\frac{(1-l)%
}{l}\eta _{2}-\frac{(l+1)}{l}\eta _{1}\right] $. As in the
previous section, the sudden transition approximation is assumed.

The shape of $\mu (\eta )$ can be found by solving Lifshitz's
equation (\ref{eqmu}) in each era. For the De Sitter era, $\mu
(\eta )$ is given by (\ref{muinfcuant}) as above. For the
\textquotedblleft MBHs +rad \textquotedblright\ era the solution
of (\ref{eqmu}) is

\be \mu _{BH}=(\sqrt{\pi }/2)e^{i\psi
_{BH}}k^{-1/2}x_{BH}^{1/2}H_{l-1/2}^{(2)}(x_{BH}), \ee%
where $x_{BH}=k\,\eta _{BH},$ and it is related to the modes of
inflation by the Bogoliubov transformation (\ref{bogtr}) with $\mu
_{R}$ replaced by $\mu _{BH}$. By evaluating the Bogoliubov
coefficients from (\ref{A-coef}) and (\ref{B-coef}), we obtain

\be \alpha
_{1}^{l}=\frac{\pi}{4}l^{\frac{1}{2}}(k\eta_1)\left[%
H^{(2)}_{-\frac{3}{2}}(k\eta_1)H^{(1)}_{l+\frac{1}{2}}(-l
k\eta_1)-H^{(1)}_{l-\frac{1}{2}}(-l
k\eta_1)H^{(2)}_{-\frac{1}{2}}(k\eta_1)\right], \ee

\be \beta _{1}^{l}=-\frac{\pi}{4}l^{\frac{1}{2}}(k\eta_1)\left[%
H^{(2)}_{-\frac{3}{2}}(k\eta_1)H^{(2)}_{l+\frac{1}{2}}(-l
k\eta_1)-H^{(2)}_{l-\frac{1}{2}}(-l
k\eta_1)H^{(2)}_{-\frac{1}{2}}(k\eta_1)\right]. \ee

In the small argument limit  (i.e., $x\ll 1$) the Hankel functions
can be approximated by \cite{Abram} %

\be H_m^{(1),(2)}(x) \simeq \sqrt{\frac{2}{\pi x}}(i)^{\mp
(m+\frac{1}{2})}\frac{(2m-1)!}{\left(m-\frac{1}{2}\right)!}(\mp
2ix)^{-m+\frac{1}{2}}e^{\pm i x}. \label{hankf}\ee

Thus, the coefficients dominant term is \be \alpha
_{1}^{l},\text{\ }\beta _{1}^{l}\simeq \frac{l^{2}2^{l}}{\left(
-l\eta _{1}\right) ^{l+1}}k^{-(l+1)} \ee%
when $k<2\pi a_{1}H_{1}$ and $\alpha _{1}^{l}=1,$ $\beta _{1}^{l}=0$ when $%
k>2\pi a_{1}H_{1}$.

The solution for the radiation era is again (\ref{murad}) with $%
x_{R}=k\,\eta _{R}$. The coefficients that relate (\ref{murad})
with $\mu _{BH}$ are
\be
\alpha _{2}^{l}=-\frac{1}{2}\sqrt{\frac{\pi lx_{R2}}{2}}\left[ \left( \frac{1%
}{x_{R2}}-i\right) H_{l-\frac{1}{2}}^{(2)}(lx_{R2})-H_{l+\frac{1}{2}%
}^{(2)}(lx_{R2})\right] e^{ix_{R2}}, \ee%
\be
\beta _{2}^{l}=\frac{1}{2}\sqrt{\frac{\pi lx_{R2}}{2}}\left[ \left( \frac{1}{%
x_{R2}}+i\right) H_{l-\frac{1}{2}}^{(2)}(lx_{R2})-H_{l+\frac{1}{2}%
}^{(2)}(lx_{R2})\right] e^{-ix_{R2}}, \ee%
when $k<2\pi a_{2}H_{2}$ and $\alpha _{2}^{l}=1,$ $\beta _{2}^{l}=0$ when $%
k>2\pi a_{2}H_{2}.$

The modes of the dust era are given by (\ref{mudustcuan}), with $%
x_{D}=k\,\eta _{D}$, and they are related to the modes of the
radiation era by the
coefficients%
\be \alpha _{3}=-i\left(
1+\frac{i}{x_{D3}}-\frac{1}{2x_{D3}^{2}}\right) ,\qquad \beta
_{3}=i\frac{1}{2x_{D3}^{2}}, \ee%
when $k<2\pi a_{3}H_{3}$ and $\alpha _{3}=1,$ $\beta _{3}=0$ when
$k>2\pi a_{3}H_{3}.$

The total coefficients relating the initial vacuum state with the
modes of the radiation state can be evaluated from (\ref{alphaT})
and (\ref{betaT}). In this case, we get
\be
 \alpha _{Tr2}^{l}=\alpha
_{2}^{l}\alpha _{1}^{l}+\beta _{2}^{l\ast }\beta _{1}^{l},\qquad
\beta _{Tr2}^{l}=\beta _{2}^{l}\alpha _{1}^{l}+\alpha _{2}^{l\ast
}\beta _{1}^{l}, \ee%
and, consequently, the total coefficients dominant term is
\begin{equation}
\alpha _{Tr2}^{l},\text{ }\beta _{Tr2}^{l}\simeq \left\{
\begin{array}{c}
1,\text{ }0\text{\qquad }\left( k>2\pi a_{1}H_{1}\right) , \\
\alpha _{1}^{l},\text{ }\beta _{1}^{l}\text{\qquad }\left( 2\pi
a_{1}H_{1}>k>2\pi a_{2}H_{2}\right) , \\
\frac{l^{-l+2}(2l^{2}-3l+1)}{8\left( -l\eta _{1}\right)
^{l+1}\left( \eta
_{R2}\right) ^{l-1}}k^{-2l}\text{\qquad }\left( k<2\pi a_{2}H_{2}\right) ,%
\end{array}%
\right.  \label{btr2bh}
\end{equation}%
where we have made use of (\ref{hankf}) in order to approximate
the coefficients $\alpha_2^l$ and $\beta_2^l$.

Finally, the total coefficients relating the inflationary modes
with the modes of the dust era evaluated from%
\be \alpha _{Tr3}^{l}=\alpha _{3}\alpha _{Tr2}^{l}+\beta
_{3}^{\ast }\beta _{Tr2}^{l}\, ,\qquad \beta _{Tr3}^{l}=\beta
_{3}\alpha
_{Tr2}^{l}+\alpha _{3}^{\ast }\beta _{Tr2}^{l}\, , \ee%
are found to be
\begin{equation}
\alpha _{Tr3}^{l},\text{ }\beta _{Tr3}^{l}\simeq \left\{
\begin{array}{c}
\alpha _{Tr2}^{l},\text{ }\beta _{Tr2}^{l}\text{\qquad }\left(
k>2\pi
a_{3}H_{3}\right) , \\
\frac{l^{-l+2}(2l^{2}-3l+1)}{8\left( -l\eta _{1}\right)
^{l+1}\left( \eta _{R2}\right) ^{l-1}\eta
_{D3}}k^{-(2l+1)}\text{\qquad }\left( k<2\pi
a_{3}H_{3}\right) .%
\end{array}%
\right.  \label{btr3bh}
\end{equation}

We are now in position to calculate the current spectrum of GWs.
Taking into account that
\be
\begin{array}{l}
\eta _{1}=-(a_{1}H_{1})^{-1},\\ \eta
_{R2}=\left(\frac{a_{2}}{a_{1}}\right)^{1/l}(a_{1}H_{1})^{-1},\\
\eta
_{D3}=2\frac{a_{3}}{a_{2}}\left(\frac{a_{2}}{a_{1}}\right)^{1/l}(a_{1}H_{1})^{-1},
\end{array} \ee%
and $\omega =k/a_{0}$, the GWs power spectrum can be written as%
{\small
\begin{equation}
P(\omega) \simeq \left\{
\begin{array}{ll}
0 & \left( \omega >2\pi \frac{a_{1}}{a_{0}}H_{1}\right) , \\
\\
\frac{l^{2-2l}2^{2l}}{\pi ^{2}}\left( \frac{a_{1}}{a_{0}}\right)
^{2l+2}H_{1}^{2l+2}\omega ^{-(2l-1)}& \left( 2\pi \frac{a_{1}}{a_{0}}%
H_{1}>\omega >2\pi \frac{a_{2}}{a_{0}}H_{2}\right) , \\
\\
\frac{l^{2-4l}(2l^{2}-3l+1)^{2}}{64\pi ^{2}}\left( \frac{a_{1}}{a_{0}%
}\right) ^{4l}\left( \frac{a_{1}}{a_{2}}\right)
^{2-2/l}H_{1}^{4l}\omega
^{-(4l-3)} & \left( 2\pi \frac{a_{2}}{a_{0}}H_{2}>\omega >2\pi \frac{%
a_{3}}{a_{0}}H_{3}\right) , \\
\\
\frac{l^{2-4l}(2l^{2}-3l+1)^{2}}{16\pi ^{2}}\left( \frac{a_{1}}{a_{0}%
}\right) ^{4l+4}\left( \frac{a_{0}}{a_{3}}\right)
^{2}H_{1}^{4l+2}\omega ^{-(4l-1)} & \left( 2\pi
\frac{a_{3}}{a_{0}}H_{3}>\omega >2\pi
H_{0}\right) .%
\end{array}
\right.
\label{espectrbh}
\end{equation}
}

Comparing the power of $\omega $ in (\ref{espectr}) and
(\ref{espectrbh}) for $\omega <2\pi \frac{a_{1}}{a_{0}}H_{1}$, we
conclude that the four-stage scenario leads to a higher number of
GWs created at low frequencies than the three-stage scenario. This
fact can be explained intuitively with the classical amplification
approach. In the three-stage scenario the GWs are parametrically
amplified as long as $k^{2}<a^{\prime \prime }(\eta)/a(\eta )$.
For $\eta=\eta _{1}$, $a^{\prime \prime }(\eta )/a(\eta )$
vanishes and there is no further amplification. On the other hand,
in the four-stage scenario the GWs with $\omega <2\pi
\frac{a_{1}}{a_{0}}H_{1}$ are amplified until the instant
$\eta_{1}$ by identical term $a^{\prime \prime }(\eta )/a(\eta
)=2/\eta^2$ than in the three-stage scenario and from $\eta_{1}$
to $\eta_{2}$ by the term $l(l-1)/\eta_{BH}^2$. Consequently they
have a larger amplitude in the radiation era.

\begin{figure}[tbp]
\hspace{-1cm}
\includegraphics*[scale=0.6]{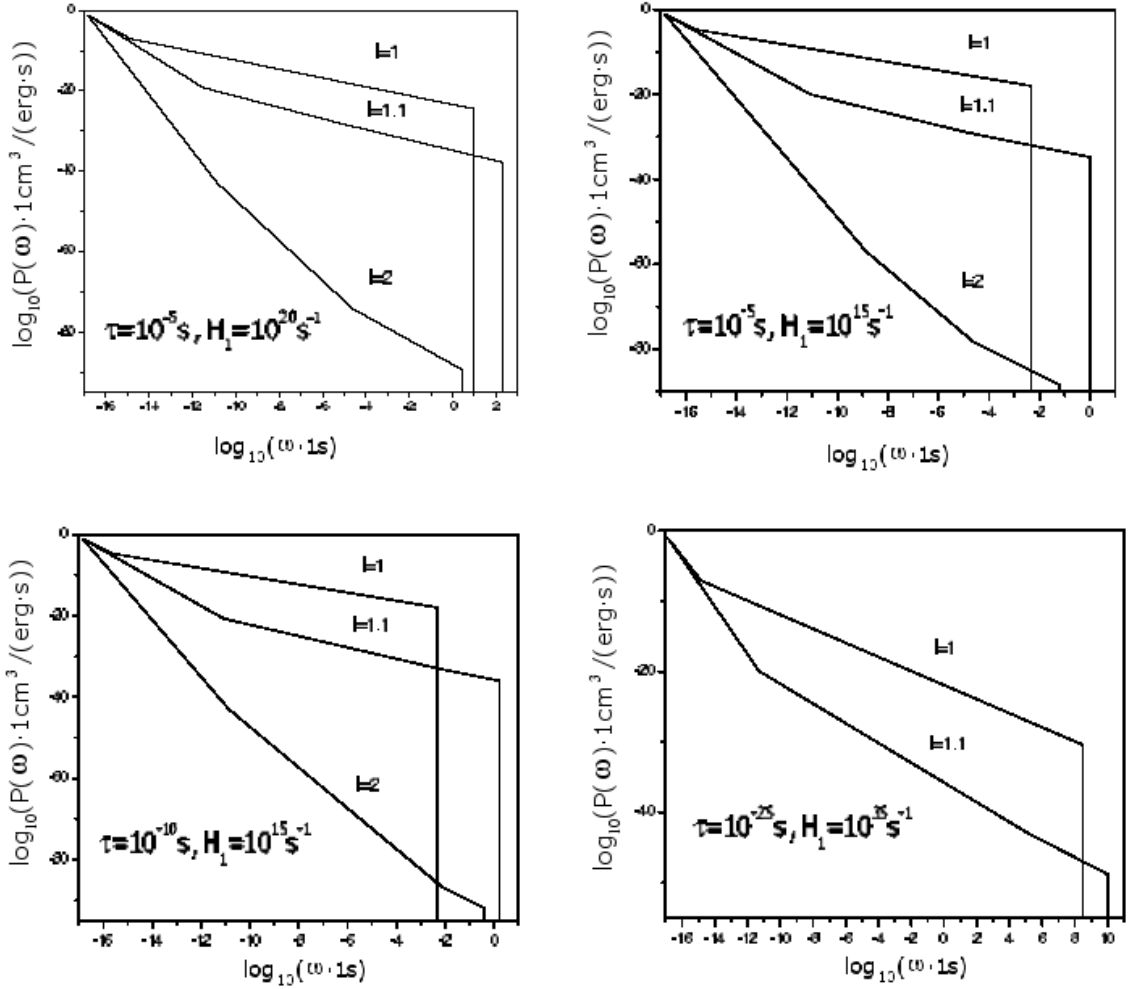}
\caption{GWs spectrum for an expanding universe with a ``MBHs+rad"
era for certain values of $l$, $\protect\tau $ and $H_{1}$. The
spectrum predicted for the three-stage model of the previous
section is plotted for comparison, $l=1$. Parameters $\tau$ and
$H_1$ are chosen assuming each spectrum has the maximum value
allowed by the CMB anisotropy data at the frequency $\omega=2\pi
H_0=2.24\times 10^{-18}s^{-1}$. In the bottom-right panel the
power spectrum with $l=2$ is rule out as it yields a CMB
anisotropy larger than observed.} \label{figspec}
\end{figure}

Figure \ref{figspec} shows the spectrum (\ref{espectrbh}) for
$l=2$ and $l=1.1$. As is apparent the four-stage scenario gives
rise to a much lower power spectrum than the three-stage scenario
assuming that in each case the spectrum has the maximum value
allowed by the CMB bound. The higher the MBHs contribution to the
energy density, the lower the final power spectrum.

Thus, if LISA fails to detect a spectrum at the level expected by
the three-stage model, rather than signaling that the recycle
model of Khoury \textit{et al.} \cite{Khoury} \footnote{Khoury
\textit{et al.} have proposed an alternative brane-world
cosmological scenario which addresses the cosmological horizon,
flatness and monopole problems an generates a nearly
scale-invariant spectrum of density perturbations without invoking
an inflationary period. In that model, the spectrum of GWs is
strongly blue in comparison with these of the standard big-bang
inflationary model.} should supersede the standard big-bang
inflationary model it may indicate a ``MBHs+radiation" era between
inflation and radiation dominance truly took place. Likewise, once
the spectrum is successfully measured we will be able to learn
from it the proportion of MBHs and radiation in the mixture phase.

In this four-stage cosmological scenario, the Hubble function
$H(\eta )$ decreases monotonically, while the energy density of
the GWs for $\eta
>\eta _{3}$ can be approximated by $\rho _{g}(\eta )\sim H^{-4l+2}(\eta )$
thereby it increases with expansion \cite{Maia93}. Obviously this
scenario will break down before $\rho _{g}(\eta )$ becomes
comparable to the energy density of matter and/or radiation since
from that moment on the linear approximation on which our approach
is based ceases to be valid. In the two next subsections
constraints on the parameters of the model are imposed so that
this does not happen.

\subsection{Free parameters of the four-stage scenario}

At this stage it is expedient to evaluate the parameters occurring in (\ref%
{espectrbh}). The redshift $\frac{a_{0}}{a_{3}}$, relating the
present value of the scale factor with the scale factor at the
transition radiation-dust, may be taken as $10^{4}$
\cite{Peebles}. The Hubble factor $H_{1}$ is connected to the
energy density at the inflationary era by%
\be \rho _{1}=\frac{c}{\hbar
}\frac{3m_{Pl}^{2}}{8\pi }H_{1}^{2}, \ee%
where we have restored momentarily the fundamental constants. In
any reasonable model the energy density at that time must be
larger than the nuclear density ($\sim 10^{35}erg/cm^{3}$) and
lower than the Planck density ($\sim 10^{115}erg/cm^{3}$)
\cite{grish93b}, therefore

\begin{equation}
10^{3}s^{-1}<H_{1}<10^{43}s^{-1}.  \label{H1}
\end{equation}

Using the expression for the scale factor at the ``MBHs+rad" era
in terms of the proper time, one obtains
\begin{equation}
\frac{a_{2}}{a_{1}}=\left( 1+H_{1}\frac{l+1}{l}\tau \right)
^{l/(l+1)}, \label{a2/a1}
\end{equation}%
where $\tau $ is the time span of the ``MBHs+rad" era which
depends on the evaporation history of the MBHs.

However, the ``MBHs+rad" era span $\tau$ should be longer than the
duration of the transition at $\eta _{1}$ (as the transition is
assumed instantaneous) in calculating the spectrum of GWs. To
evaluate the adiabatic vacuum cutoff for the frequency we have
considered that the transition between whatever two successive
stages has a duration of the same order as the inverse of the
Hubble factor. This places the additional constraint
\begin{equation}
\tau >H_{1}^{-1}.  \label{tau}
\end{equation}

Finally, $\frac{a_{1}}{a_{0}}$ can be evaluated from the evolution
of the Hubble factor until the present time
\be
H_{0}=\left( \frac{a_{0}}{a_{3}}\right) ^{1/2}\left( \frac{a_{2}}{a_{1}}%
\right) ^{(l-1)/l}\left( \frac{a_{1}}{a_{0}}\right) ^{2}H_{1}.
\ee%
The current value of the Hubble factor is estimated to be
$2.24\times 10^{-18}s^{-1}$ \cite{Spergel} and
\begin{equation}
\left( \frac{a_{1}}{a_{0}}\right) ^{2}=\left(
\frac{a_{3}}{a_{0}}\right)
^{1/2}\left( 1+H_{1}\frac{l+1}{l}\tau \right) ^{(1-l)/(l+1)}\frac{H_{0}}{%
H_{1}}.  \label{a1/a0}
\end{equation}

The only free parameters considered here are $l,$ $\tau $ and
$H_{1}$, with the restrictions $1<l\leq 2$,  (\ref{H1}) and
(\ref{tau}). The two first free parameters depend on the
assumption made on the MBHs, although it is possible to obtain
rigorous constraints on $H_{1}$ and $\tau $ from the density of
the GWs.

\subsection{Restrictions on the \textquotedblleft
MBHs+rad\textquotedblright\ era from the cosmic microwave
background}

\begin{figure}[tbp]
\includegraphics*[scale=0.6]{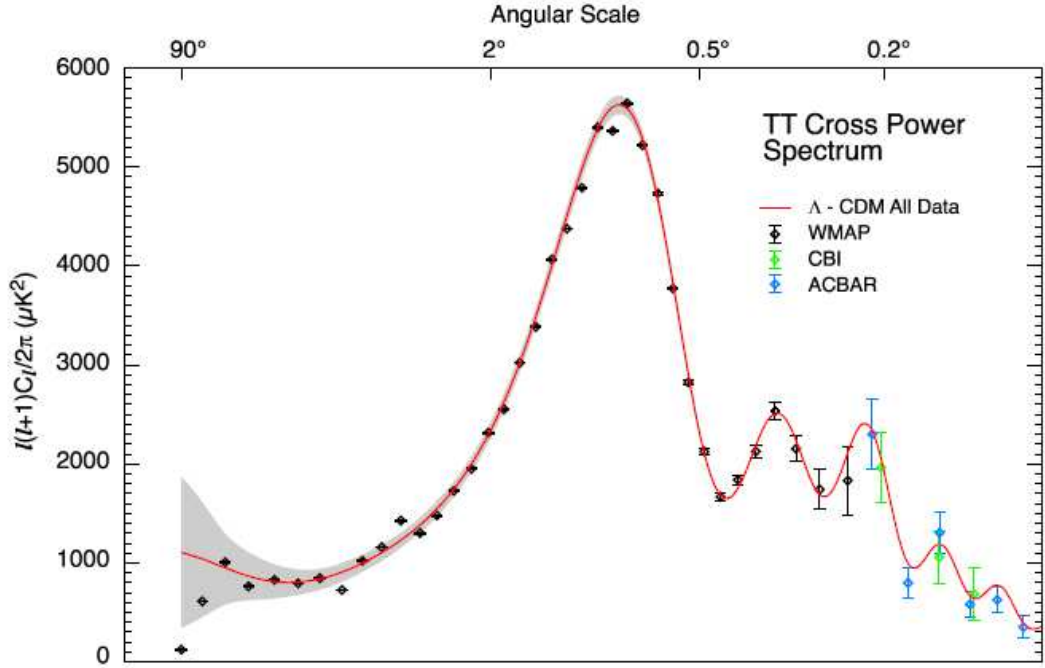}
\caption{The CMB power spectrum from the WMAP satellite
\cite{hinshaw}. The error bars on this plot are 1-$\sigma$ and the
solid line represents  the best-fit cosmological model
\cite{Spergel}.} \label{wmap1}
\end{figure}

It is obvious that $\rho _{g}$ cannot be arbitrarily large, in
fact the GWs are seen as linear perturbations of the metric. The
linear approximation holds only for $\rho _{g}(\eta )\ll \rho
(\eta )$, $\rho (\eta )$ being the total energy density of the
Universe. Several observational data place constraints on $\rho
_{g}$. The regularity of the pulses of stable millisecond pulsars
sets a constraint at frequencies of order $10^{-8}Hz$
\cite{Thors}. Likewise, there is a certain maximum value for $\rho
_{g}$ compatible with the primordial nucleosynthesis scenario. But
the most severe constraints come from the high isotropy degree of
the CMB. We will focus on the latter constraint. The observed
thermal fluctuations are usually analyzed by decomposing them into
spherical harmonics%
\be \frac{\delta T}{T}=\sum_{\ell ,m} a_{\ell m}Y_{\ell m}(\theta,\phi),\ee%
where $a_{\ell m}$ are expansion coefficients and $\theta$ and
$\phi$ are spherical polar angles on the sky. Defining the power
spectrum by $C_{\ell}=\langle |a_{\ell m}|^2 \rangle $, it is
conventional to plot $\ell (\ell +1)C_{\ell}$ versus $\ell$.
Figure \ref{wmap1} shows the first published results of the WMAP
experiment regarding the CMB temperature anisotropies as well as
the $\Lambda$CDM model (solid line) that fits the data rather
well.

Metric perturbations with frequencies between $%
10^{-16}$ and $10^{-18}$ $Hz$ at the last scattering surface can
produce thermal fluctuations in the CMB due to the Sachs-Wolfe
effect \cite{sachs}. These fluctuations cannot exceed the observed
value of $\delta T/T\sim 5\times 10^{-6}$ \cite{Peacock}.

A detailed analysis of the CMB bound yields \cite{allen96, Allen Koranda}%
\begin{equation}
\Omega _{g}h_{100}^{2}<7\times 10^{-11}\left( \frac{H_{0}}{f}\right) ^{2}{%
\qquad ({H}_{0}<f<30\times H}_{0}{)}  \label{condPCMB}
\end{equation}%
where $\Omega _{g}=fP(f)/\rho _{0},$ $\rho
_{0}=3cm_{Pl}^{2}H_{0}^{2}/(8\pi
\hbar )$ and $H_{0}=h_{100}^{{}}\times 100km/(s\times Mpc)$ with $%
h_{100}^{{}}=0.7$. The CMB bound for the spectrum
(\ref{espectrbh})
evaluated at $2\pi H_{0}\, $Hertz reads \footnote{%
It is necessary to multiply (\ref{espectrbh}) for $\hbar /c^{3}$
in order to obtain the right units.} \be
1>\left( 2\pi \right) ^{-4l}l^{2-4l}(2l^{2}-3l+1)^{2}\left( \frac{a_{1}}{%
a_{0}}\right) ^{4l+4}\left( \frac{a_{0}}{a_{3}}\right) ^{2}\left( \frac{H_{1}%
}{3.72\times 10^{19}s^{-1}}\right) ^{3}\left(
\frac{H_{1}}{H_{0}}\right) ^{4l-1}, \ee%
and consequently
\begin{equation}
f(l,H_{1},\tau )=-107.69+l\left( 28.10+2\log _{10}\left( \frac{H_{1}}{1s^{-1}%
}\right) \right) -(2l-2)\log _{10}\left( 1+\frac{l+1}{l}H_{1}\tau
\right) \label{eqH1taul}
\end{equation}%
\[
+(-4l+2)\log _{10}l+2\log _{10}(2l^{2}-3l+1)<0.
\]%
We next consider different values for $l$ and $\tau $.

\begin{figure}[tbp]
\hspace{-1.5cm}
\includegraphics*[angle=-90,scale=0.7,]{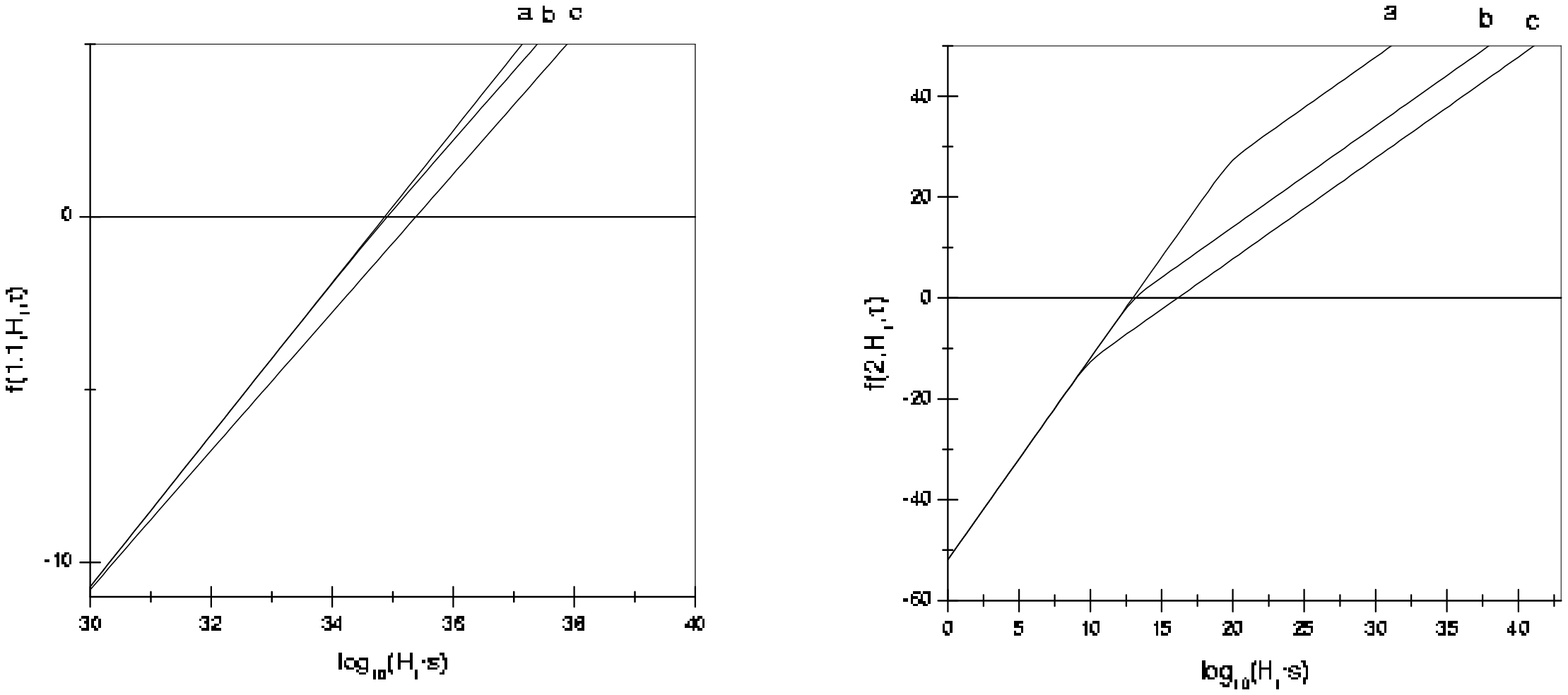}
\caption{{}The left panel depicts $f(1.1,H_{1}, \protect\tau) vs.
\log_{10}H_{1}$ for a) $\protect\tau =10^{-40}s$, b) $\protect\tau =\protect\tau %
_{c}^{l=1.1}=1.23\times 10^{-35}s$, and c) $\protect\tau
=10^{-30}s$. The right panel depicts $f(2,H_{1},\protect\tau )$ for a) $\protect\tau %
=10^{-20}s$, b) $\protect\tau =\protect\tau _{c1}^{l=2}=6.75\times 10^{-14}s$%
, and c) $\protect\tau =10^{-10}s$. Conditions
(\protect\ref{eqH1taul}) and (\protect\ref{tau}) are satisfied for
certain ranges of $\protect\tau $ and $H_{1}$ in each case. }
\label{graphf}
\end{figure}

(i) When $l=1.1$, the relation (\ref{eqH1taul}) reads%
\begin{equation}
f(1.1,H_{1},\tau )=-76.71+2.20\log _{10}\left(
\frac{H_{1}}{1s^{-1}}\right) -0.2\log _{10}\left(
1+\frac{l+1}{l}H_{1}\tau \right) <0, \label{eqH1taul=1.1}
\end{equation}%
see Fig. \ref{graphf}. From it we observe that:

\begin{enumerate}
\item For $\tau <\tau _{c}^{l=1.1}=1.22\times 10^{-35}s$, the condition (\ref%
{eqH1taul=1.1}) is satisfied if $H_{1}<8.13\times 10^{34}s^{-1}$
and
conflicts with (\ref{tau}), which in the most favorable case is $%
H_{1}=8.13\times 10^{34}s^{-1}$ for $\tau =\tau _{c}^{l=1.1}$. For
$l=1.1$, there is no compatibility with the observed CMB
anisotropy when $\tau <\tau _{c}^{l=1.1}$. Thus, this range of
$\tau $ is ruled out.

\item For $\tau >\tau _{c}^{l=1.1}$, one obtains
$H_{1}<H_{c}^{l=1.1}(\tau )$ from the condition
(\ref{eqH1taul=1.1}). $H_{c}^{l=1.1}(\tau )$ is
always larger than $\tau ^{-1}$ in the range considered, e.g. $%
H_{c}^{l=1.1}(\tau =10^{-30}s)=2.40\times 10^{35}s^{-1}$. Taking
into account (\ref{tau}) one obtains that the condition
(\ref{eqH1taul=1.1}) is satisfied for $\tau
^{-1}<H_{1}<H_{c}^{l=1.1}(\tau )$.
\end{enumerate}

\bigskip

(ii) When $l=2$ (the extreme case in which the expansion is
entirely dominated by the MBHs) we have
\begin{equation}
f(2,H_{1},\tau )=-51.89+4\log _{10}\left(
\frac{H_{1}}{1s^{-1}}\right) -2\log _{10}\left(
1+\frac{l+1}{l}H_{1}\tau \right) <0,  \label{eqH1taul=2}
\end{equation}%
see figure \ref{graphf}. Inspection of (\ref{eqH1taul=2}) and Fig. \ref%
{graphf} reveals that:

\begin{enumerate}
\item For $\tau <\tau _{c}^{l=2}=6.75\times 10^{-14}s$, one obtains $%
H_{1}<10^{13}s^{-1}$ which is totally incompatible with condition (\ref{tau}%
), $H_{1}>1.48\times 10^{13}s^{-1}$ for $\tau =\tau _{c}^{l=2}$ in
the most favorable case. Thus, the region $\tau <\tau _{c}^{l=2}$
is ruled out as predicts an excess of anisotropy in the CMB.

\item For $\tau >\tau _{c}^{l=2}$, one obtains
$H_{1}<H_{c}^{l=2}(\tau )$ from the condition (\ref{eqH1taul=2}).
$H_{c}^{l=2}(\tau )$ is always larger than $1.48\times
10^{13}s^{-1}$ for $\tau $ in the range considered, e.g.
$H_{c}^{l=2}(\tau =10^{-10}s)=1.35\times 10^{16}s^{-1}$.
Conditions (\ref{tau}) and (\ref{eqH1taul=2}) are both satisfied
in this range for $\tau ^{-1}<H_{1}<H_{c}^{l=2}(\tau )$.
\end{enumerate}

\bigskip

We may conclude by saying that the condition (\ref{eqH1taul})
leads to different allowed ranges for $H_{1}$ and $\tau $ for each
$l$ considered, although their interpretation is rather similar.
For $\tau <$ $\tau _{c}^{l}$ the condition of minimum duration of
the \ ``MBHs+rad" era (\ref{tau}) and the
upper bound given by the CMB anisotropy are incompatible. However, for $\tau >$ $%
\tau _{c}^{l}$ these two conditions are compatible for $\tau
^{-1}<H_{1}<H_{c}^{l}(\tau )$.

\section{Conclusions}

\markboth{CHAPTER \thechapter. QUANTUM MBHS AND GW SPECTRUM}{2.3
CONCLUSIONS}

We have calculated the power spectrum of GWs in a universe that
begins with an inflationary phase, followed by a phase dominated
by a mixture of MBHs and radiation, then a radiation dominated
phase (after the MBHs evaporated), and finally a dust dominated
phase. The spectrum depends just on three free parameters, namely
$H_{1}$ the Hubble factor at the transition inflation- ``MBHs+rad"
era, $\tau $, the cosmological time span of the \textquotedblleft
MBHs+rad\textquotedblright\ era, and the power $l$, being $a(\eta
)\propto\eta ^{l}$ the scale factor of the \textquotedblleft
MBHs+rad\textquotedblright\ era with $1<l\leq 2$.

The upper bound on the spectrum of GWs obtained from the CMB
anisotropy places severe constraints on $H_{1}$ and $\tau $. For
each value of $l$ considered, there is a minimum value of $\tau $,
$\tau _{c}^{l}$, compatible
with the CMB anisotropy. There is a range of $\tau $, $\tau >\tau _{c}^{l}$%
, for which $\tau ^{-1}<H_{1}<H_{c}^{l}(\tau )$ satisfies the CMB
upper bound.

The four-stage scenario predicts a much lower power spectrum of
GWs than the conventional three-stage scenario. If LISA fails to
detect the GWs spectrum at the level predicted by the three-stage
scenario, a possible explanation might be that the ``MBHs+rad" era
took place. Likewise, once the spectrum is successfully measured
we will be able to learn from it the proportion of MBHs and
radiation in the mixture phase as well as the time span of this
era.

%

\chapter{Gravitational waves and present accelerated expansion \label{presentaccel}}

\markboth{CHAPTER \thechapter. GW AND THE PRESENT
ACCELERATED...}{}
In this chapter we show how the power spectrum as well as the
dimensionless density parameter of the GWs created from the
initial vacuum state may help present (and future) observers
ascertain whether the expansion phase they are living in is
accelerated or not and if accelerated, which law follows the scale
factor \cite{RGWPAE}. The latter would facilitate enormously to
discriminate the nature of dark energy between a large variety of
proposed models (cosmological constant, quintessence fields,
interacting quintessence, tachyon fields, Chaplygin gas, etc)
\cite{iap}. To this end we calculate the power spectrum and energy
density of the GWs when the transitions to the dark energy era and
second dust era are considered. Obviously, the latter power
spectrum lies at the future and depending on the model under
consideration it may take very long for the Universe to enter the
second dust era.

\section{Accelerated expansion and decaying dark energy}
\markboth{CHAPTER \thechapter. GW AND THE PRESENT ACCELERATED...}{
\thesection. ACCELERATED EXPANSION AN DECAYING DE}

Nowadays the observational data regarding the apparent luminosity
of supernovae type Ia, together with the discovery of CMB angular
temperature fluctuations on degree scales and measurements of the
power spectrum of galaxy clustering, strongly suggests that the
Universe is nearly flat and that its expansion is accelerating at
present \cite{Spergel, snia}. In actual fact the debate now
focuses on when the acceleration did really commence, if it is
just a transient phenomenon or it is to last forever, and above
all the nature of the dark energy.

In Einstein gravity the accelerated expansion is commonly
associated to a sufficiently high negative pressure which might be
provided by a cosmological constant $\Lambda$ (vacuum energy),
whose equation of state is $p_{\Lambda}=-\rho_{\Lambda}$, with
$\rho_{\Lambda}=\Lambda/(8\pi)=\text{constant}$. The observational
data seems to indicate that the cosmological constant contributes
about the $70\%$ of the energy of the Universe, meanwhile the
remaining $30\%$ comes from non-relativistic matter (i.e., dust).
Thus, the question arises: ``Why are the vacuum and matter energy
densities of precisely the same order today?'', which is known in
the literature as the coincidence problem \cite{steinha}.

A possible answer to this question considers that the acceleration
of the Universe is associated to a sort of dynamical energy, the
so--called dark energy, that violates the strong energy condition
and clusters only at the largest accessible scales \cite{iap}. In
such a case the present state of the Universe would be dominated
by dark energy and since it redshifts more slowly with expansion
than dust, the contribution of the latter is bound to become
negligible at late times.

In an attempt to evade the particle horizon problem posed by an
everlasting accelerated expansion to string/M type theories
\cite{string/M}, some models propose dark energy potentials such
that the current accelerated phase would be just transitory and
sooner or later the expansion would revert to the Einstein--De
Sitter law $a(t) \propto t^{2/3}$, thereby slowing down (second
dust era) \cite{Alam}. The possibility that dark energy could be
unstable is in fact suggested by the remarkable qualitative
analogy between the presence of dark energy today and the
properties of a different type of dark energy - the inflaton field
- postulated in the inflationary scenario of the early Universe
-see e.g., \cite{Kolb}. This analogy has two main points. On one
hand, it makes natural that a form of matter with negative
pressure could have dominated the Universe in a distant past,
since a similar form of matter dominates the Universe today. On
the other hand, as the dark energy in the early Universe was
unstable and decayed aeons ago, one might be tempted to ask
whether the nature of dark energy observed today would be any
different.

For our purposes, we shall assume a simplified model of decaying
dark energy in which the usual dust dominated stage is followed by
an accelerated era where the adiabatic index $\gamma$ of the fluid
that dominates that era is a constant that lies in the range
$[0,2/3)$. Posteriorly, the dark energy decays in a time span much
lower than the duration of the accelerated era and the Universe
resumes the decelerated expansion dominated by the cold dark
matter. Models with  dark energy whose evolution mimics that of
dust can be also taken in consideration in our description as they
are formally equivalent to the model of above.

\markboth{CHAPTER \thechapter. GW AND THE PRESENT ACCELERATED...}
{BOGOLIUBOV COEFFICIENTS IN THE DECAYING...}%
\section{Bogoliubov coefficients in the decaying dark energy scenario}

\markboth{CHAPTER \thechapter. GW AND THE PRESENT ACCELERATED...}{
\thesection. BOGOLIUBOV COEFFICIENTS IN THE DECAYING...}

In this section we evaluate the coefficients of Bogoliubov in the
simplified model previously suggested, a spatially flat FRW
scenario initially De Sitter, then dominated by radiation,
followed by a dust dominated era, an accelerated expansion era
dominated by dark energy, and finally a second dust era.

The scale factor in terms of the conformal time reads%
{\small
\begin{equation}
a(\eta )=\left\{
\begin{array}{lcr}
-\frac{1}{H_{1}\eta }& (-\infty <\eta <\eta _{1}<0),& \text{%
De Sitter era} \\
\frac{1}{H_{1}\eta _{1}^{2}}(\eta -2\eta _{1})& (\eta _{1}<\eta
<\eta
_{2}),& \text{radiation era} \\
\frac{1}{4H_{1}\eta _{1}^{2}}\frac{(\eta +\eta _{2}-4\eta
_{1})^{2}}{\eta _{2}-2\eta _{1}}&(\eta _{2}<\eta <\eta _{3}),&
\text{first dust
era} \\
\left( \frac{l}{2}\right) ^{-l}\frac{(\eta _{3}+\eta _{2}-4\eta _{1})^{2-l}}{%
4H_{1}\eta _{1}^{2}(\eta _{2}-2\eta _{1})}\left( \eta _{l}\right)
^{l}&
(\eta _{3}<\eta <\eta _{4}),& \text{dark energy era} \\
\frac{a_{4}}{4}\left( a_{4}H_{4}\right) ^{2}\left( \eta -\eta _{4}+\frac{2}{%
a_{4}H_{4}}\right) ^{2}&(\eta _{4}<\eta ),& \text{second dust
era}%
\end{array}%
\right.  \label{sclfac3}
\end{equation}}%
where $l\leq -1$, $\eta _{l}=\eta +\frac{l}{2}\left[ \left(
-2/l+1\right) \eta _{3}+\eta_{2}-4\eta _{1}\right]$, the
subindexes $1,2,3,4$ correspond to sudden transitions from
inflation to radiation era, from radiation to first dust era, from
first dust era to dark energy era and from the latter to the
second dust era, respectively; $H_{i}$ is the Hubble factor at the
instant $\eta = \eta_{i}$. The present time $\eta _{0}$ lies in the range $%
\left[ \eta _{3},\eta _{4}\right] $, it is to say in the dark
energy dominated era.

\begin{figure}[tbp]
\includegraphics*[angle=-90,scale=0.7]{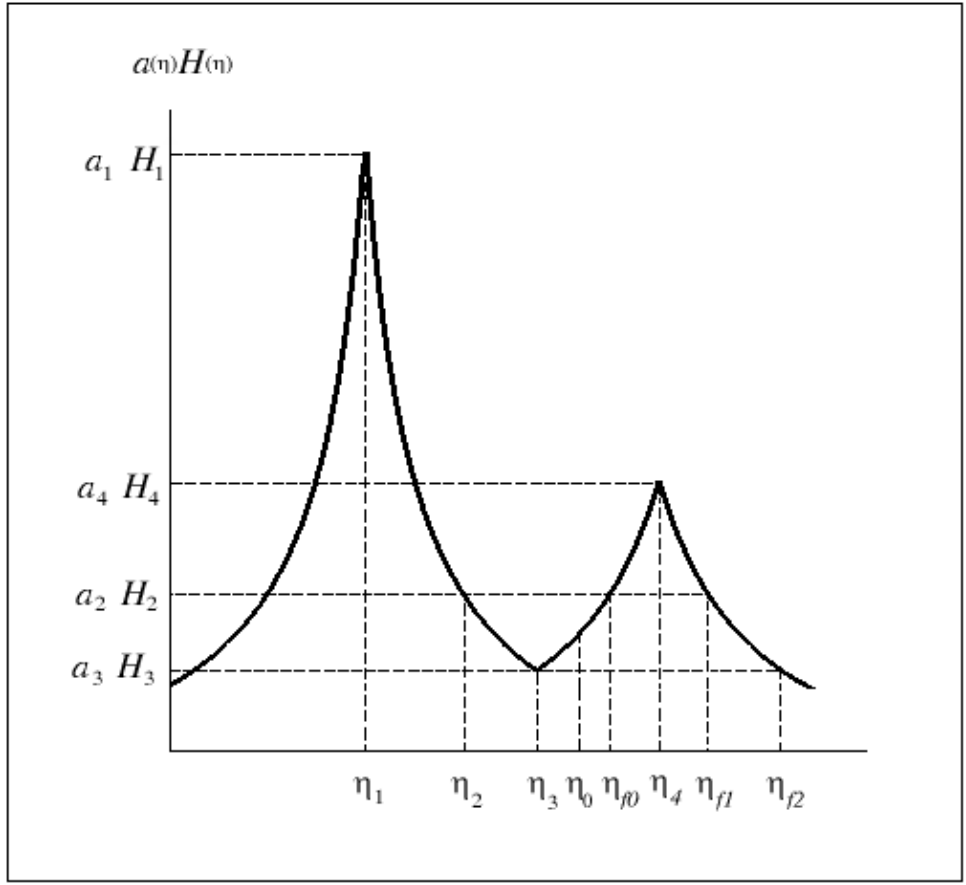}
\caption{{}Evolution of $a(\protect\eta )H(\protect\eta )$ in a
universe with scale factor given by Eq. (\protect\ref{sclfac3})
under the assumption $ a_{4}H_{4}>a_{2}H_{2}$. The quantity
$\protect\eta _{f0}$ is defined as the
instant in the dark energy era in which $a(\protect\eta _{f0})H(\protect\eta %
_{f0})=a_{2}H_{2}$, $\protect\eta _{f1}$ as the instant of the
second dust
era such that $a(\protect\eta _{f1})H(\protect\eta _{f1})=a_{2}H_{2},$ and $%
\protect\eta _{f2}$ as the instant of the second dust era in which $a(%
\protect\eta _{f0})H(\protect\eta _{f0})=a_{3}H_{3}$. } \label{aH}
\end{figure}

GWs which are inside the Hubble radius have a wave number $k$
lower than $a(\eta )H(\eta )$. Figure \ref{aH} sketches the
evolution of $a(\eta )H(\eta )$ in this scenario. During the
inflationary and dark energy eras $a(\eta )H(\eta )$ increases
with $\eta $, and decreases in the other eras. As a consequence,
$a_{4}H_{4}$ results higher than $a_{3}H_{3}$. Choosing $l$, $\eta
_{3}$ and $\eta _{4}$ in such a way that

\begin{equation}
\left( \frac{a_{4}}{a_{3}}\right) ^{-1/l}\left(
\frac{a_{2}}{a_{3}}\right) ^{1/2}>1,  \label{cond}
\end{equation}%
we have that $a_{4}H_{4}$ is also higher than $a_{2}H_{2}$. We assume that $%
a_{4}H_{4}$ is lower than $a_{1}H_{1}$ throughout.

The scale factor (\ref{sclfac3}) formally coincides with these of
the standard three-stage model given by (\ref{sclfac1}) until the
instant $\eta_3$. Thus, the Bogoliubov coefficients $\alpha_1$ and
$\beta_1$ are equal to those of equation (\ref{coef1}) in the
range $k<2\pi a_1 H_1$ and they are $1$ and $0$ for $k>a_1 H_1$,
respectively. Analogously, $\alpha_2$ and $\beta_2$ coincides with
those of (\ref{coef2}) for $k<2\pi a_2 H_2$ and are $1$ and $0$
for $k>a_2 H_2$.

In the dark energy era the solution to Lifshitz's equation
(\ref{eqmu}) reads \be \mu _{l}=(\sqrt{\pi }/2)e^{i\psi
_{l}}k^{-1/2}x_{l}^{1/2}H_{l-1/2}^{(2)}(x_{l}), \ee%
where $x_{l}=k\eta _{l}$. The Bogoliubov transformation%
\begin{equation}
\mu _{D}(\eta )=\alpha _{3}^{l}\mu _{l}(\eta )+\beta _{3}^{l}\mu
_{l}^{\ast} (\eta )  \label{bogtrr_l}
\end{equation}%
relates the dust and dark energy modes.

Using the well-known relation for Hankel functions \cite{Abram}%
\begin{equation}
H_{l-1/2}^{(2)}(x)\simeq e^{-ix}\sqrt{\frac{2}{\pi x}}\frac{(-2l)!}{(-l)!}%
(-2)^{l}x^{l}(1+ix),  \label{aproxhankl}
\end{equation}%
valid when $x\ll 1$ and $l<-1$, it follows that%
\begin{eqnarray*}
\alpha _{3} &=&\left( 6(-1)^{-l}l^{l}\frac{(-2l)!}{(-l)!}\right)
\eta
_{m3}^{l-2}k_{{}}^{l-2}+\mathcal{O}(k_{{}}^{l-1}), \\
\beta _{3} &=&-\left( 6(-1)^{-l}l^{l}\frac{(-2l)!}{(-l)!}\right)
\eta _{m3}^{l-2}k_{{}}^{l-2}+\mathcal{O}(k_{{}}^{l-1}),
\end{eqnarray*}%
for $k<2\pi a(\eta _{3})H_{3}$ and $\alpha _{3}=1$, $\beta _{3}=0$ for $%
k>2\pi a(\eta _{3})H_{3}$, where $H_{3}$ is the Hubble function evaluated at $%
\eta _{3}$ and $\eta _{m3}=\eta _{3}+\eta _{2}-4\eta _{1}$. The
transition between the first dust era to the dark energy era has a
time span of the order $H_{3}^{-1}$. This is much shorter than the
period of the waves we are considering and therefore it may be
assumed instantaneous when calculating the coefficients. As a
consequence, the time span from
this transition till today, $\tau =t_{0}-t_{3}$, must be larger than $%
H_{3}^{-1}$, otherwise the transition first dust era-dark energy
era would be too close to the present time for our formalism to
apply. This condition places a lower bound on the value of the
redshift $\frac{a_{0}}{a_{3}}$. When $l=-1$ the condition is
$\frac{a_{0}}{a_{3}}>2.72$, $\frac{a_{0}}{a_{3}}>2.25$ when
$l=-2$, $\frac{a_{0}}{a_{3}}>2.15$ when $l=-3$, and
$\frac{a_{0}}{a_{3}}>2$ when $l\rightarrow -\infty $. This bound
is compatible with the accepted values for $\frac{a_{0}}{a_{3}}\in
\lbrack 1.5,11]$ (see e.g., \cite{luca}). Henceforward we will
consider $\frac{a_{0}}{a_{3}}$ larger than $2.72$ and no larger
than $11$.

The solution to Lifshitz's equation in the second dust era (i.e.,
the one following the dark energy era) is \be \mu _{SD}=(\sqrt{\pi
}/2)e^{i\psi
_{SD}}k^{-1/2}x_{SD}^{1/2}H_{3/2}^{(2)}(x_{SD}), \ee%
where $x_{SD}=k\left( \eta -\eta _{4}+\frac{2}{a_{4}H_{4}}\right)
$.

The Bogoliubov coefficients relating the modes of the dark energy
era with
the modes of the second dust era are%
\be \mu _{l}(\eta )=\alpha _{4}^{l}\mu _{SD}(\eta )+\beta
_{4}^{l}\mu _{SD}^{\ast }(\eta ). \ee

Likewise, the continuity of $\mu $ at $\eta _{4}$ implies%
\begin{eqnarray*}
\alpha _{4}^{l} &=&\left( -1\right) ^{l}\frac{3}{8}2^{l}\frac{(-2l)!}{(-l)!}%
x_{l4}^{l-2}\left[ l^{2}+il(l-2)x_{l4}^{{}}+
\mathcal{O}(x_{l4}^{2})\right] ,
\\
\beta _{4}^{l} &=&\left( -1\right) ^{l}\frac{3}{8}2^{l}\frac{(-2l)!}{(-l)!}%
x_{l4}^{l-2}\left[ l^{2}+il(l+2)x_{l4}^{{}}+
\mathcal{O}(x_{l4}^{2})\right] ,
\end{eqnarray*}%
for $k<2\pi a_{4}H_{4}$, and $\alpha _{4}^{l}=1$, $\beta _{4}^{l}=0$ for $%
k>2\pi a_{4}H_{4}$, where $H_{4}$ is the Hubble function evaluated
at the transition time $\eta _{4}$.

\section{Power spectrums}

In this section we evaluate the total coefficients and  calculate
the current power spectrum as well as the power spectrum in the
second dust era. We find that the current power spectrum coincides
with those of the three-stage model of  section \ref{three-stage}
but it evolves differently. We also find two possible shapes for
the power spectrum in the second dust era, whether condition
(\ref{cond}) is fulfilled or not.

\subsection{Current power spectrum \label{CurrentPowerspec(DDE)}}

To evaluate the present power spectrum one must bear in mind that $%
a_{0}H_{0}>a_{3}H_{3}$ which implies the wave lengths of the
perturbations created at the transition dust era-dark energy era
are larger than the present Hubble radius $H_0^{-1}$ \cite{Chiba}.
One must also consider the possibility that
$a_{0}H_{0}>a_{2}H_{2}$, it is to say

\be \left( \frac{a_{0}}{a_{3}}\right) ^{-1/l}\left(
\frac{a_{2}}{a_{3}}\right) ^{1/2}>1. \ee%

For $l=-1$ and assuming $\frac{a_{0}}{a_{2}}\sim 10^{4}$
\cite{Peebles} this condition implies $\frac{a_{0}}{a_{3}}>21.5$,
$\frac{a_{0}}{a_{3}}>100$
when $l=-2,$ and $\frac{a_{0}}{a_{3}}>251.2$ when $l=-3$. The values for $%
\frac{a_{0}}{a_{3}}$ considered by us are larger than $2.72$ and no lower than $%
11$, consequently we can safely assume $a_{0}H_{0}<a_{2}H_{2}$.

For $k>2\pi a_{1}H_{1}$, we find that $\alpha _{Tr2}=1$, $\beta
_{Tr2}=0$;
in the range $2\pi a_{1}H_{1}>k>2\pi a_{2}H_{2}$, the coefficients are $%
\alpha _{Tr2}=\alpha _{1}$ and $\beta _{Tr2}=\beta _{1}$, and finally for $%
k<2\pi a(\eta _{2})H_{2}$ we obtain \cite{Allen, Maia93}

\begin{equation}
\beta _{Tr2}\simeq -\frac{1}{8k^{3}\eta _{1}^{2}\eta _{R2}}.
\label{coeft2}
\end{equation}

Thus, the number of GWs at the present time $\eta _{0}$ created
from the initial vacuum state is $\left\langle N_{\omega
}\right\rangle =\left\vert \beta _{Tr2}\right\vert ^{2}\sim \omega
^{-6}(\eta _{0})$ for $\omega (\eta _{0})<2\pi
(a_{2}/a_{0})H_{2}$, $\omega ^{-4}(\eta _{0})$ for $2\pi
(a_{1}/a_{0})H_{1}>\omega (\eta _{0})>2\pi (a_{2}/a_{0})H_{2}$,
and zero for $\omega (\eta _{0})>2\pi (a_{1}/a_{0})H_{1}$, where
we have used the present value of the frequency, $\omega (\eta
_{0})=k/a_{0}$.

In summary, the current power spectrum of GWs in this scenario is
\begin{equation}
P(\omega )\simeq \left\{
\begin{array}{l}
0\text{\qquad }\left( \omega (\eta _{0})>2\pi (a_{1}/a_{0})H_{1}\right) , \\
\\
\frac{1}{4\pi ^{2}}\left( \frac{a_{1}}{a_{0}}\right)
^{4}H_{1}^{4}\omega ^{-1}\text{\qquad }\left( 2\pi
(a_{2}/a_{0})H_{2}<\omega
(\eta _{0})<2\pi (a_{1}/a_{0})H_{1}\right) , \\
\\
\frac{1}{16\pi ^{2}}\left( \frac{a_{0}}{a_{2}}\right) ^{2}\left(
\frac{a_{1}}{a_{0}}\right) ^{8}H_{1}^{6}\omega ^{-3}\text{\qquad
}\left(
2\pi H_{0}<\omega (\eta _{0})<2\pi (a_{2}/a_{0})H_{2}\right) .%
\end{array}%
\right. \label{espectrde}
\end{equation}

While this power spectrum is not at variance with the power
spectrum of the conventional three-stage scenario evaluated at
section \ref{three-stage}, it evolves differently. The power
spectrum in the dark
energy scenario at $\eta =\eta _{3}$ formally coincides with Eq. (\ref%
{espectrde}) but with $a_{3}$ substituted by $a_{0}$ throughout,
and from then up to now waves with $2\pi a_{3}H_{3}<k<$ $2\pi
a_{0}H_{0}$ cease to contribute to the spectrum as soon as their
wave length exceeds the Hubble radius. By contrast, in the
three-stage scenario waves are continuously being added to the
spectrum. As we shall see in section \ref{EDGW}, this implies that
the evolution of the energy density of the gravitational waves in
the three-stage scenario differs from the scenario in which the
Universe expansion is dominated by dark energy.

\subsection{Power spectrum in the second dust era}

Here we evaluate the power spectrum at some future time $\eta $ larger than $%
\eta _{f2}$ for which the waves created at the transition dust
era-dark energy era ($\eta =\eta _{3}$) are considered in the
spectrum by the first time (see Figure \ref{aH}). Let $\alpha
_{Tr4}$ and $\beta _{Tr4}$ be the Bogoliubov coefficients relating
the modes of the inflationary era to the modes of the second dust
era. Because of condition (\ref{cond}) we must consider two
possibilities with two different power spectrums.

$(i)$ If condition (\ref{cond}) is not fulfilled, then $%
a_{1}H_{1}>a_{2}H_{2}>a_{4}H_{4}>a_{3}H_{3}$ and the power
spectrum can be obtained from the following total coefficients. In
the range $k>2\pi
a_{1}H_{1}$ the total coefficients are $\alpha _{Tr4}=1$ and $\beta _{Tr4}=0$%
. For $2\pi a_{1}H_{1}>k>2\pi a_{2}H_{2}$, the coefficients are
$\alpha
_{Tr4}=\alpha _{1}$ and $\beta _{Tr4}=\beta _{1}$, where $\alpha _{1}$ and $%
\beta _{1}$ are defined in Eq. (\ref{coef1}). For $2\pi
a_{2}H_{2}>k>2\pi a_{4}H_{4}$, $\alpha _{Tr4}=\alpha _{Tr2}$ and
$\beta _{Tr4}=\beta _{Tr2}$ where $\beta _{Tr2}$ is defined in Eq.
(\ref{coeft2}) and the dominant term of $\alpha _{Tr2}$ coincides
with $\beta_ {Tr2}$.

When $2\pi a_{4}H_{4}>k>2\pi a_{3}H_{3}$, except for $\alpha _{3}$ and $%
\beta _{3}$, all the coefficients obtained in the previous section
must be
considered when evaluating $\alpha _{Tr4}$ and $\beta _{Tr4}$, therefore%
\begin{eqnarray}
\alpha _{Tr4}^{l} &=&\alpha _{4}^{l}\alpha _{Tr2}^{{}}+\beta
_{4}^{l\ast }\beta _{Tr2}^{{}}\simeq \left[ \frac{3}{32}\left(
-1\right) ^{-l-1}l^{2}2^{l}\frac{(-2l)!}{(-l)!}\right]
\frac{2}{\eta _{1}^{2}\eta
_{R2}\eta _{l4}^{-l+2}}k^{l-5},\qquad  \label{coeft4a} \\
\beta _{Tr4}^{l} &=&\beta _{4}^{l}\alpha _{Tr2}^{{}}+\alpha
_{4}^{l\ast }\beta _{Tr2}^{{}}\simeq \left[ \frac{3}{32}\left(
-1\right) ^{-l-1}l^{2}2^{l}\frac{(-2l)!}{(-l)!}\right]
\frac{2}{\eta _{1}^{2}\eta _{R2}\eta _{l4}^{-l+2}}k^{l-5}.
\end{eqnarray}%
Finally, for $2\pi a_{3}H_{3}>k>2\pi H(\eta )$, we get%
\begin{equation}
\beta _{Tr4}^{l}\simeq -\left[ \frac{9}{16}l^{l+2}2^{l}\left( \frac{(-2l)!}{%
(-l)!}\right) ^{2}\right] \frac{2l}{\eta _{1}^{2}\eta _{R2}\eta
_{m3}^{-l}\eta _{l4}^{-l+2}}k^{2l-5}.  \label{coeft4b}
\end{equation}%
Accordingly, the power spectrum reads{\small \be P(\omega )\simeq
\left\{
\begin{array}{l}
0\text{\qquad }\left( \omega (\eta )>2\pi (a_{1}/a(\eta ))H_{1}\right) , \\
\\
\frac{1}{4\pi ^{2}}\left( \frac{a_{1}}{a(\eta )}\right)
^{4}H_{1}^{4}\omega ^{-1}\text{\qquad }\left( 2\pi (a_{2}/a(\eta
))H_{2}<\omega (\eta )<2\pi (a_{1}/a(\eta ))H_{1}\right) , \\
\\
\frac{1}{16\pi ^{2}}\left( \frac{a_{0}}{a_{2}}\right) ^{2}\left(
\frac{a_{1}}{a_{0}}\right) ^{8}\left( \frac{a_{0}}{a(\eta
)}\right) ^{6}H_{1}^{6}\omega ^{-3}\text{\qquad }\left( 2\pi
(a_{4}/a(\eta
))H_{4}<\omega (\eta )<2\pi (a_{2}/a(\eta ))H_{2}\right) \\
\\
\frac{9}{\pi ^{2}}l^{2l}2^{2l-8}\left( \frac{(-2l)!}{(-l)!}%
\right) ^{2}\left( \frac{a_{1}}{a_{0}}\right) ^{16-4l}\left( \frac{a_{2}}{%
a_{0}}\right) ^{l-4}\left( \frac{a_{3}}{a_{0}}\right) ^{l-4+4/l}\left( \frac{%
a_{4}}{a_{0}}\right) ^{-8+2l-4/l} \\
\text{\qquad }\times \left( \frac{a(\eta )}{a_{4}}\right)
^{2l-10}H_{1}^{10-2l}\omega ^{2l-7}\text{\qquad }\left( 2\pi
(a_{3}/a(\eta
))H_{3}<\omega (\eta )<2\pi (a_{4}/a(\eta ))H_{4}\right) , \\
\\
\frac{27}{\pi ^{2}}l^{4l}2^{4l-6}\left( \frac{(-2l)!}{(-l)!}%
\right) ^{4}\left( \frac{a_{1}}{a_{0}}\right) ^{16-8l}\left( \frac{a_{2}}{%
a_{0}}\right) ^{2l-4}\left( \frac{a_{3}}{a_{0}}\right)
^{2l-4+4/l}\left(
\frac{a_{4}}{a_{0}}\right) ^{4l-8-4/l} \\
\text{\qquad }\times \left( \frac{a(\eta )}{a_{4}}\right)
^{4l-10}H_{1}^{10-4l}\omega ^{4l-7}\text{\qquad }\left( 2\pi
H(\eta )<\omega
(\eta )<2\pi (a_{3}/a(\eta ))H_{3}\right) .%
\end{array}%
\right. \ee%
}$(ii)$ If condition (\ref{cond}) is fulfilled, then $%
a_{1}H_{1}>a_{4}H_{4}>a_{2}H_{2}>a_{3}H_{3}$. As in the previous
case, in the range $k>2\pi a_{1}H_{1}$ the total coefficients are
$\alpha _{Tr4}=1$ and $\beta _{Tr4}=0$. For $2\pi
a_{1}H_{1}>k>2\pi a_{4}H_{4}$, the coefficients are $\alpha
_{Tr4}=\alpha _{1}$ and $\beta _{Tr4}=\beta _{1}$.
In the range $2\pi a_{4}H_{4}>k>2\pi a_{2}H_{2}$, we obtain%
\begin{eqnarray*}
\alpha _{Tr4}^{l} &=&\alpha _{4}^{l}\alpha _{1}+\beta _{4}^{l\ast
}\beta
_{1}\simeq i\frac{3}{8}\left( -1\right) ^{-l}l^{2}2^{l}\frac{(-2l)!}{(-l)!}%
\frac{1}{\eta _{1}^{2}\eta _{l4}^{-l+1}}k^{l-3},\qquad \\
\beta _{Tr4}^{l} &=&\beta _{4}^{l}\alpha _{1}+\alpha _{4}^{l\ast
}\beta
_{1}\simeq i\frac{3}{8}\left( -1\right) ^{-l}l^{2}2^{l}\frac{(-2l)!}{(-l)!}%
\frac{1}{\eta _{1}^{2}\eta _{l4}^{-l+1}}k^{l-3}.
\end{eqnarray*}%
Again, for $2\pi a_{2}H_{2}>k>2\pi a_{3}H_{3}$, the total
coefficients are given by Eq. (\ref{coeft4a}) while for $2\pi
a_{3}H_{3}>k>2\pi H(\eta )$ they obey Eq. (\ref{coeft4b}). The
power spectrum in this case is{\small
\begin{equation}
P(\omega )\simeq \left\{
\begin{array}{l}
0\text{\qquad }\left( \omega (\eta )>2\pi (a_{1}/a(\eta ))H_{1}\right) , \\
\\
\frac{1}{4\pi ^{2}}\left( \frac{a_{1}}{a(\eta )}\right)
^{4}H_{1}^{4}\omega ^{-1}\text{\qquad }\left( 2\pi (a_{4}/a(\eta
))H_{4}<\omega (\eta )<2\pi (a_{1}/a(\eta ))H_{1}\right) , \\
\\
\frac{9}{\pi ^{2}}l^{2l+2}2^{2l-6}\left( \frac{(-2l)!}{(-l)!}%
\right) ^{2}\left( \frac{a_{1}}{a_{0}}\right) ^{8-4l}\left( \frac{a_{2}}{%
a_{0}}\right) ^{l-1}\left( \frac{a_{3}}{a_{0}}\right) ^{l-3+2/l}\left( \frac{%
a_{4}}{a_{0}}\right) ^{2l-4-4/l} \\
\text{\qquad } \times \left( \frac{a(\eta )}{a_{4}}\right)
^{2l-6}H_{1}^{6-2l}\omega ^{2l-3}\text{\qquad }\left( 2\pi
(a_{2}/a(\eta
))H_{2}<\omega (\eta )<2\pi (a_{4}/a(\eta ))H_{4}\right) , \\
\\
\frac{9}{\pi ^{2}}l^{2l}2^{2l-8}\left( \frac{(-2l)!}{(-l)!}%
\right) ^{2}\left( \frac{a_{1}}{a_{0}}\right) ^{16-4l}\left( \frac{a_{2}}{%
a_{0}}\right) ^{l-4}\left( \frac{a_{3}}{a_{0}}\right) ^{l-4+4/l}\left( \frac{%
a_{4}}{a_{0}}\right) ^{2l-8-4/l} \\
\text{\qquad } \times \left( \frac{a(\eta )}{a_{4}}\right)
^{2l-10}H_{1}^{10-2l}\omega ^{2l-7}\text{\qquad }\left( 2\pi
(a_{3}/a(\eta
))H_{3}<\omega (\eta )<2\pi (a_{2}/a(\eta ))H_{2}\right) , \\
\\
\frac{27 }{\pi ^{2}}l^{4l}2^{4l-6}\left( \frac{(-2l)!}{(-l)!}%
\right) ^{4}\left( \frac{a_{1}}{a_{0}}\right) ^{16-8l}\left( \frac{a_{2}}{%
a_{0}}\right) ^{2l-4}\left( \frac{a_{3}}{a_{0}}\right)
^{2l-4+4/l}\left(
\frac{a_{4}}{a_{0}}\right) ^{4l-8-4/l} \\
\text{\qquad }\times \left( \frac{a(\eta )}{a_{4}}\right)
^{4l-10}H_{1}^{10-4l}\omega ^{4l-7}\text{\qquad }\left( 2\pi
H(\eta )<\omega
(\eta )<2\pi (a_{3}/a(\eta ))H_{3}\right) .%
\end{array}%
\right.  \label{Ps}
\end{equation}%
} The power spectrum governed by Eq.(\ref{Ps}) is plotted in Fig. \ref%
{spec} for different choices of the free parameters $l$,
$\frac{a_{0}}{a_{3}} $, $\frac{a_{4}}{a_{0}}$ and
$\frac{a(\eta)}{a_{0}}$ as well as the power spectrum assuming the
three-stage model, i.e., non-accelerated phase and no second dust
era.

The shape of the power spectrum given by (\ref{Ps}) in the range
$2\pi H(\eta )<\omega (\eta )<2\pi (a_{3}/a(\eta ))H_{3}$ is the
same in both cases as in this range all the coefficients are
present in the evaluation of the total coefficients. It is
interesting to see how markedly this spectrum differs from the one
arising in the three-stage model (dot-dashed line) at low
frequencies.

\begin{figure}[tbp]
\hspace{-1.5cm}
\includegraphics*[scale=0.7]{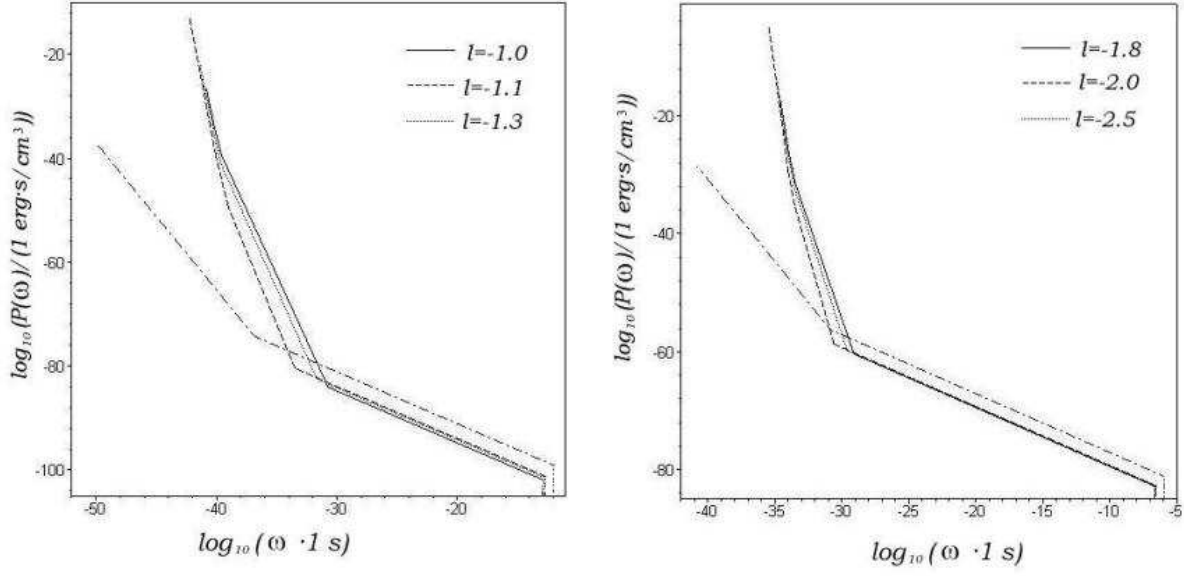}
\caption{Power spectrum given by Eq. (\protect\ref{Ps}) for
different values
of $l$. The left panel depicts the power spectrum at the instant $\protect%
\eta $ for which $\frac{a\left(\protect\eta \right)}{a_{0}}=10^{22}$ (when $%
\Omega_{g}(\protect\eta)$ is near unity), with
$\frac{a_{0}}{a_{3}}=11$, and $\frac{a_{4}}{a_{0}}=10^{6}$ (to
fulfill condition (\ref{cond})). The right panel shows the power
spectrum at the instant $\protect\eta $ for which
$\frac{a\left(\protect\eta \right) }{a_{0}}=10^{16}$ (again when
$\Omega_{g}(\protect\eta)$ is near unity), with
$\frac{a_{0}}{a_{3}}=11$, and $\frac{a_{4}}{a_{0}}=10^{6}$. For
the sake of comparison the power spectrum of the three-stage
scenario (De Sitter inflation, radiation, dust with $l = 2$) is
also shown in both panels (dot-dashed line). Notice the difference
in slopes at the dust era.} \label{spec}
\end{figure}

\subsubsection{Topological defects}

Up to now we assumed that at the end of the dark energy era the
Universe will evolve as if it became dominated by dust once again.
Nevertheless if the expansion achieved in the accelerated phase
were large enough, either cosmic strings, or domain walls, or a
cosmological constant will take over instead. We will not
consider, however, cosmic strings (whose equation of
state is $p_{K}=-\frac{1}{3}\rho _{K}$) for, as pointed out by Maia \cite%
{Maia93}, it seems problematic to define an adiabatic vacuum in an
era dominated by these topological defects since the
creation--annihilation operators, $A_{(k)}$, $A_{(k)}^{\dag }$,
fail to satisfy the commutation relations (i.e., condition
(\ref{condcuan})) in the range of frequencies where one should
expect GWs amplification. In short, our approach, as it is, does
not apply to this case.

As for domain walls (topological stable defects of second order
with equation of state $p_{dw}=-\frac{2}{3}\rho _{dw}$ and energy
density that varies as $a^{-1}(\eta)$ -see e.g.,
\cite{Kolb,shellard}), once the dark energy evolved as
pressureless matter at $\eta =\eta _{4}$ the scale factor may be
approximated by \be a(\eta >\eta _{4})=4a_{4}\left(
a_{4}H_{4}\right) ^{-2}\left( -\eta +\eta
_{4}+\frac{2}{a_{4}H_{4}}\right) ^{-2}, \ee%
so long as $a^{-1}\gg a^{-3}$. That is to say, for $\eta >\eta
_{4}$ the expansion of the Universe is again accelerated whereby
$a(\eta )H(\eta )$ resumes growing. The GWs will be leaving the
Hubble radius as soon as $H^{-1}$ becomes smaller than their wave
length, and eventually none of them will contribute to the
spectrum.

Finally, we consider the existence of a positive cosmological constant $%
\Lambda $. (Recall that $\rho _{\Lambda }=\Lambda /\left( 8\pi \right) $ and $%
p_{\Lambda }=-\rho _{\Lambda }$). Once the dark energy dynamically
mimicked dust the Universe will become dominated by a very tiny
cosmological constant. The corresponding scale factor is \be
a(\eta >\eta _{4})=H_{4}^{-1}\left( -\eta +\eta _{4}+\frac{1}{a_{4}H_{4}}%
\right) ^{-1}.
\ee%
Once again, the expansion is accelerated and GWs will leave the
Hubble radius and, in the long run, none of them will contribute
to the spectrum.

\section{Energy density of the gravitational waves \label{EDGW}}

Now we are in position to evaluate the energy density of the GWs
in terms of the conformal time by integrating the power spectrum
$P(\omega)$ obtained in the previous section. As we shall see, the
evolution of the energy density strongly depends  on the free
parameters of the model.

Its current value, evaluated from Eq. (\ref{espectrde}), can
be approximated by \cite{Allen}%
\begin{equation}
\rho _{g}(\eta _{0})\simeq \frac{\hbar }{32\pi ^{2}c^{3}}\left( \frac{a_{0}}{%
a_{2}}\right) ^{2}\left( \frac{a_{1}}{a_{0}}\right)
^{8}H_{1}^{6}(2\pi H_{0})^{-2}.
\label{dens0}
\end{equation}%
where, in this section, we return to conventional units.

To study the evolution of $\rho _{g}(\eta )$ from this point
onward we first consider that $a_{4}H_{4}>a_{2}H_{2}$. In this
case, $\rho _{g}(\eta )$ evolves as (\ref{dens0}) with $H(\eta )$
and $a(\eta )$ substituted by $H_{0}$ and $a_{0}$, respectively,
till some instant $\eta _{f0} $ in the range $\eta _{0}<\eta
_{f0}<\eta _{4}$. When $\eta \geqslant \eta _{f0}$ the GWs with
$\omega (\eta _{f0})<2\pi \left( a_{2}/a(\eta _{f0})\right) H_{2}$
must no longer be considered in evaluating $\rho _{g}$ as their
wave length exceed the Hubble radius. Consequently
\begin{equation}
\rho _{g}(\eta _{f0}<\eta <\eta _{4})\simeq \frac{\hbar }{4\pi ^{2}c^{3}}%
\left( \frac{a_{1}}{a(\eta )}\right) ^{4}H_{1}^{4}\ln \left( \frac{a_{1}H_{1}%
}{a(\eta )H(\eta )}\right) .  \label{dens4}
\end{equation}%
For $\eta =\eta _{4}$, the gravitational waves created at the
transition dark energy era-second dust era begin contributing to
$\rho _{g}$ thereby, {\small
\begin{eqnarray}
\rho _{g}(\eta _{4} &<&\eta <\eta _{f1})\simeq \rho _{g}(\eta
_{4})\left(
\frac{a_{4}}{a\left( \eta \right) }\right) ^{4}+\frac{9\hbar }{\pi ^{2}c^{3}}%
l^{2l+2}2^{2l-6}\left( \frac{(-2l)!}{(-l)!}\right) ^{2}\left( \frac{a_{1}}{%
a_{0}}\right) ^{8-4l}  \label{densf1} \\
&& \times \left( \frac{a_{2}}{a_{0}}\right) ^{l-1}\left( \frac{a_{3}}{a_{0}}%
\right) ^{l-3+2/l}\left( \frac{a_{4}}{a_{0}}\right) ^{2l-4-4/l}\left( \frac{%
a(\eta )}{a_{4}}\right) ^{2l-6}H_{1}^{6-2l}\frac{(2\pi H(\eta ))^{2l-2}}{%
-2l+2},  \nonumber
\end{eqnarray}%
}where $\rho _{g}(\eta _{4})$ corresponds to Eq.(\ref{dens4}) evaluated at $%
\eta =\eta _{4}$. For $\eta =\eta _{f1}>\eta _{4}$ where $\eta
_{f1}$ is
defined by the condition $a(\eta _{f1})H(\eta _{f1})=a_{2}H_{2}$ (see Figure %
\ref{aH}), the gravitational waves which left the Hubble radius at
$\eta _{f0}$ reenter it, therefore{\small
\begin{eqnarray}
\rho _{g}(\eta _{f1} &<&\eta <\eta _{f2})\simeq \rho _{g}(\eta
_{f1})\left(
\frac{a_{f1}}{a\left( \eta \right) }\right) ^{4}+\frac{9\hbar }{\pi ^{2}c^{3}%
}l^{2l}2^{2l-8}\left( \frac{(-2l)!}{(-l)!}\right) ^{2}\left( \frac{a_{1}}{%
a_{0}}\right) ^{16-4l}  \label{densf2} \\
&& \times \left( \frac{a_{2}}{a_{0}}\right) ^{l-4}\left( \frac{a_{3}}{a_{0}}%
\right) ^{l-4+4/l}\left( \frac{a_{4}}{a_{0}}\right) ^{2l-8-4/l}\left( \frac{%
a(\eta )}{a_{4}}\right) ^{2l-10}H_{1}^{10-2l}\frac{(2\pi H(\eta ))^{2l-6}}{%
-2l+6},  \nonumber
\end{eqnarray}%
}where $\rho _{g}(\eta _{f1})$ corresponds to Eq.(\ref{densf1})
evaluated at
$\eta =\eta _{f1}$. For $\eta =\eta _{f2}$, with $\eta _{f2}$ defined by $%
a(\eta _{f2})H(\eta _{f2})=a_{3}H_{3}$ (see Figure \ref{aH}), the
gravitational waves created at the transition first dust era-dark
energy era have wave lengths shorter than the Hubble radius for
first time, and from that point on the density of gravitational
waves can be approximated by{\small
\begin{eqnarray}
\rho _{g}(\eta &>&\eta _{f2})\simeq \rho _{g}(\eta _{f2})\left( \frac{a_{f2}%
}{a\left( \eta \right) }\right) ^{4}+\frac{27\hbar }{\pi ^{2}c^{3}}%
l^{4l}2^{4l-6}\left( \frac{(-2l)!}{(-l)!}\right) ^{4}\left( \frac{a_{1}}{%
a_{0}}\right) ^{16-8l}  \label{densf} \\
&& \times \left( \frac{a_{2}}{a_{0}}\right) ^{2l-4}\left( \frac{a_{3}}{a_{0}}%
\right) ^{2l-4+4/l}\left( \frac{a_{4}}{a_{0}}\right) ^{4l-8-4/l}\left( \frac{%
a(\eta )}{a_{4}}\right) ^{4l-10}H_{1}^{10-4l}\frac{(2\pi H(\eta ))^{4l-6}}{%
-4l+6},  \nonumber
\end{eqnarray}%
} where $\rho _{g}(\eta _{f2})$ corresponds to Eq.(\ref{densf2})
evaluated at $ \eta =\eta _{f2}$.

In the case that $a_{4}H_{4}>a_{2}H_{2}$, Eq. (\ref{dens0})
dictates the
evolution of $\rho _{g}$ between $\eta _{0}$ till $\eta _{4}$. Then, from $%
\eta _{4}$ till $\eta _{f2}$, $\rho _{g}$ obeys Eq. (\ref{densf2})
(note that $\eta _{f1}$ cannot be defined in this case) and from
$\eta _{f2}$ onwards $\rho _{g}$ obeys Eq. (\ref{densf}).

A natural restriction on $\rho _{g}(\eta )$ is that it must be
lower than the total energy density of the flat FRW universe \be
\rho (\eta )=\frac{c}{\hbar }\frac{3m_{Pl}^{2}}{8\pi }H^{2}(\eta
), \ee%
where $m_{Pl}=\sqrt{\hbar c/G}$ stands for the Planck mass.

It seems reasonable to consider $H_{1}\simeq 10^{38}s^{-1}$ as it
corresponds to the grand unification energy scale of inflationary
models \cite{allen96, Kolb, alan}. The redshift $%
\frac{a_{0}}{a_{2}}$, relating the present value of the scale
factor with the scale factor at the transition radiation era-first
dust era, may be taken as $10^{4}$ \cite{Peebles}, and the current
value of the Hubble factor $H_{0}$ is estimated to be $2.24\times
10^{-18}s^{-1}$ \cite{Spergel, freed}. It
then follows%
\begin{eqnarray*}
\frac{a_{1}}{a_{0}} &=&1.50\times 10^{-29}\left(
\frac{a_{0}}{a_{3}}\right)
^{-\frac{1}{4}+\frac{1}{2l}},\text{\qquad }H_{2}=2.24\times
10^{-12}\left(
\frac{a_{0}}{a_{3}}\right) ^{-\frac{1}{2}+\frac{1}{l}}s^{-1}, \\
H_{3} &=&H_{0}\left( \frac{a_{0}}{a_{3}}\right) ^{1+\frac{1}{l}}s^{-1},\text{%
\qquad }H_{4}=H_{0}\left( \frac{a_{4}}{a_{0}}\right)
^{-1-\frac{1}{l}}s^{-1},
\\
H_{f1} &=&2.24\times 10^{-12}\left( \frac{a_{0}}{a_{3}}\right) ^{-\frac{3}{2}%
+\frac{3}{l}}\left( \frac{a_{4}}{a_{0}}\right) ^{-1+\frac{2}{l}}s^{-1}, \\
H_{f2} &=&2.24\times 10^{-18}\left( \frac{a_{0}}{a_{3}}\right) ^{\frac{3}{l}%
}\left( \frac{a_{4}}{a_{0}}\right) ^{-1+\frac{2}{l}}s^{-1},
\end{eqnarray*}%
where we have used the relation%
\begin{equation}
H(\eta >\eta _{4})=\left( \frac{a_{4}}{a(\eta )}\right)
^{3/2}H_{4}, \label{Hsde}
\end{equation}%
valid in the second dust era (see Eq. (\ref{sclfac3})) with $%
a_{f1}H_{f1}=a_{2}H_{2}$ and $a_{f2}H_{f2}=a_{3}H_{3}$ to evaluate
$H_{f1}$ (only defined if $a_{4}H_{4}>a_{2}H_{2}$) and $H_{f2}$,
respectively. In our
model, there are only three parameters, namely $l\leq -1$, $2.72<\frac{a_{0}%
}{a_{3}}<11$, and $\frac{a_{4}}{a_{0}}$.

We are now in position to evaluate the evolution of the
dimensionless density parameter $ \Omega _{g}(\eta)\equiv \rho
_{g}(\eta )/\rho (\eta )$. Its current value
is\ \cite{Allen}%
\be
\Omega _{g0}\simeq \frac{1}{48\pi ^{3}}\frac{G\hbar }{c^{5}}\left( \frac{%
a_{0}}{a_{2}}\right) ^{2}\left( \frac{a_{1}}{a_{0}}\right)
^{8}H_{1}^{6}H_{0}^{-4}, \ee%
which in our description boils down to
\begin{equation}
\Omega _{g0}\simeq 2.00\times 10^{-13}\left( \frac{a_{0}}{a_{3}}\right) ^{-2+%
\frac{4}{l}}. \label{Og0}
\end{equation}
$\Omega _{g0}$ is much lower than unity for any choice of $\frac{a_{0}}{a_{3}%
}$ and $l$ in the above intervals. At later times $\Omega _{g}$
evolves as
\begin{equation}
\Omega _{g}(\eta >\eta _{0})=\Omega _{g0}\left( \frac{a(\eta )}{a_{0}}%
\right) ^{-2+\frac{4}{l}},  \label{omf0}
\end{equation}%
where we have used $H(\eta _{4}>\eta >\eta _{0})=\left( \frac{a(\eta )}{a_{0}%
}\right) ^{-1-\frac{1}{l}}H_{0}$. It is obvious that $\Omega
_{g}(\eta )$ is a decreasing function of $\eta $. For $\eta >\eta
_{0}$ we shall distinguish the two cases mentioned in the previous
section.

When condition (\ref{cond}) is fulfilled, the evolution of $\Omega
_{g}$ is given by Eq. (\ref{omf0}) till $\eta =\eta _{f0}$. Then,
$\rho _{g}$ changes in shape from $\eta =\eta _{f0}$ till $\eta
=\eta _{4}$, as we have seen. Consequently
\begin{eqnarray}
\Omega _{g}(\eta _{4} &>&\eta >\eta _{f0})\simeq \frac{2}{3\pi
}\frac{G\hbar
}{c^{5}}\left( \frac{a_{0}}{a_{3}}\right) ^{-1+\frac{2}{l}}\frac{a_{2}}{a_{0}%
}\left( \frac{a(\eta )}{a_{0}}\right) ^{-2+\frac{2}{l}}H_{1}^{2}
\label{om4}
\\
&& \times \ln \left[ \left( \frac{a_{2}}{a_{0}}\right)
^{\frac{1}{4}}\left(
\frac{a_{0}}{a_{3}}\right) ^{\frac{1}{4}-\frac{1}{2l}}\left( \frac{a(\eta )}{%
a_{0}}\right)
^{\frac{1}{l}}H_{1}^{\frac{3}{2}}H_{0}^{-\frac{1}{2}}\right] ,
\nonumber
\end{eqnarray}%
is a decreasing function of $\eta $. Finally, from $\eta _{4}$ on,
$\Omega _{g}$ evolves as {\small
\begin{eqnarray}
\Omega _{g}(\eta _{f1} >\eta >\eta _{4})\simeq \Omega _{g}(\eta
_{4})\left( \frac{a_{4}}{a(\eta )}\right) +\frac{24}{\pi }\frac{G\hbar }{%
c^{5}}l^{2l+2}2^{2l-6}\left( \frac{(-2l)!}{(-l)!}\right) ^{2}\left( \frac{%
a_{1}}{a_{0}}\right) ^{8-4l}  \left( \frac{a_{2}}{a_{0}}\right) ^{l-1}\label{omf1} \\
 \;\;\times\left( \frac{a_{3}}{a_{0}}%
\right) ^{l-3+2/l}\left( \frac{a_{4}}{a_{0}}\right) ^{2l-4-4/l}\left( \frac{%
a(\eta )}{a_{4}}\right) ^{2l-6}H_{1}^{6-2l}\frac{(2\pi )^{2l-2}}{-2l+2}%
\left( H(\eta )\right) ^{2l-4},  \nonumber
\end{eqnarray}

\begin{eqnarray}
\Omega _{g}(\eta _{f2} >\eta >\eta _{f1})\simeq \Omega _{g}(\eta
_{f1})\left( \frac{a_{f1}}{a(\eta )}\right) +\frac{24}{\pi }\frac{G\hbar }{%
c^{5}}l^{2l}2^{2l-8}\left( \frac{(-2l)!}{(-l)!}\right) ^{2}\left( \frac{a_{1}%
}{a_{0}}\right) ^{16-4l} \left( \frac{a_{2}}{a_{0}}\right) ^{l-4} \label{omf2} \\
 \times \left( \frac{a_{3}}{a_{0}}%
\right) ^{l-4+4/l}\left( \frac{a_{4}}{a_{0}}\right) ^{2l-8-4/l}\left( \frac{%
a(\eta )}{a_{4}}\right) ^{2l-10}H_{1}^{10-2l}\frac{(2\pi )^{2l-6}}{-2l+6}%
\left( H(\eta )\right) ^{2l-8},  \nonumber
\end{eqnarray}

\begin{eqnarray}
\Omega _{g}(\eta >\eta _{f2})\simeq \Omega _{g}(\eta _{f2})\left( \frac{%
a_{f2}}{a(\eta )}\right) +\frac{72}{\pi }\frac{G\hbar }{c^{5}}%
l^{4l}2^{4l-6}\left( \frac{(-2l)!}{(-l)!}\right) ^{4}\left( \frac{a_{1}}{%
a_{0}}\right) ^{16-8l} \left( \frac{a_{2}}{a_{0}}\right) ^{2l-4}  \label{omf} \\
 \times\left( \frac{a_{3}}{a_{0}}%
\right) ^{2l-4+4/l}\left( \frac{a_{4}}{a_{0}}\right) ^{4l-8-4/l}\left( \frac{%
a(\eta )}{a_{4}}\right) ^{4l-10}H_{1}^{10-4l}\frac{(2\pi )^{4l-6}}{-4l+6}%
\left( H(\eta )\right) ^{4l-8}.  \nonumber
\end{eqnarray}%
} As follows from (\ref{Hsde}) and (\ref{omf1})-(\ref{omf}), in
each case, the first term redshifts with expansion while the
second term of $\Omega _{g}$ grows with $\eta $ (recall that $l
\leq -1$). Conditions $\Omega_{g}(\eta _{f1}>\eta >\eta _{4})\ll
1$ and $\Omega _{g}(\eta _{f2}>\eta >\eta _{f1})\ll 1$ become just
$\Omega _{g}(\eta =\eta _{f1})\ll 1$ and $\Omega _{g}(\eta =\eta
_{f2})\ll 1$, which are true in either case whatever the choice of
parameters. Finally, from Eq. (\ref{omf}) we may conclude that at
some future time larger than $\eta _{f2}$ the condition $\Omega
_{g}\ll 1$ will no longer be fulfilled and the linear
approximation in which our approach rests will cease to apply.

In the opposite case, when condition (\ref{cond}) is not fulfilled, $%
\Omega_{g}$ also grows following Eq. (\ref{omf0}) till $\eta =\eta
_{4}$, in the interval $\eta _{4}<\eta <\eta _{f2}$ it grows according to Eq. (\ref%
{omf2}), and according to Eq. (\ref{omf}) from $\eta _{f2}$
onwards. Our conclusions of the first case regarding the evolution
of $\Omega _{g}$ during these time intervals still hold true.

\begin{figure}[tbp]
\includegraphics*[scale=0.5,angle=-90]{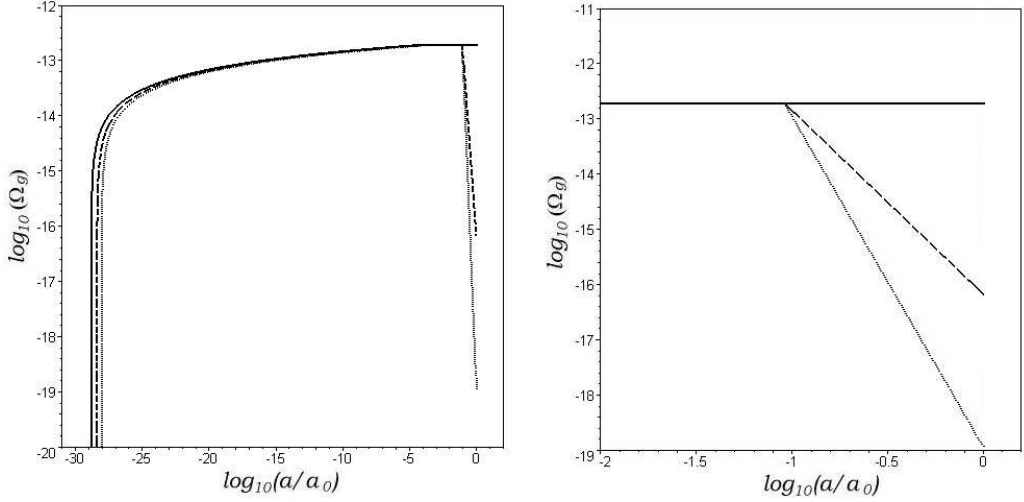}
\caption{Left panel: Evolution of the density parameter $\Omega_g$
with the scale factor from the beginning of the radiation era till
the present time. The solid line shows the density parameter
predicted by the three-stage model (De Sitter inflation,
radiation, dust). The dashed and dotted lines depict the density
parameter predicted by a four-stage model having a dark energy era
right after the dust era with $l=-3,\, -1$, respectively, and
$a_{0}/a_{3}=11$. Note that the four-stage model predicts a lower
present value for $\Omega_g$ than the three-stage one as
$\Omega_g$ decreases, in a $l$ dependent fashion, during the dark
energy era. The right panel is a blow up of the region in which
the four-stage models ($l=-3,\, -1$),  notably differ from the
three-stage one.} \label{Omegag1}
\end{figure}

The behavior of the density parameter $\Omega_{g}$ differs from
one scenario to the other. In the three-stage model (inflation,
radiation, dust) $\Omega_{g}$ remains constant during the dust era
\cite{Allen}. In a four-stage model, with a dark energy era right
after the conventional dust era, $ \Omega_{g}$ sharply decreases
in a $l$ dependent way during the fourth (accelerated) era since
long wave lengths are continuously exiting the Hubble sphere
\cite{Chiba}, see Fig. \ref{Omegag1}.

\begin{figure}[tbp]
\hspace{1cm}
\includegraphics*[scale=0.8]{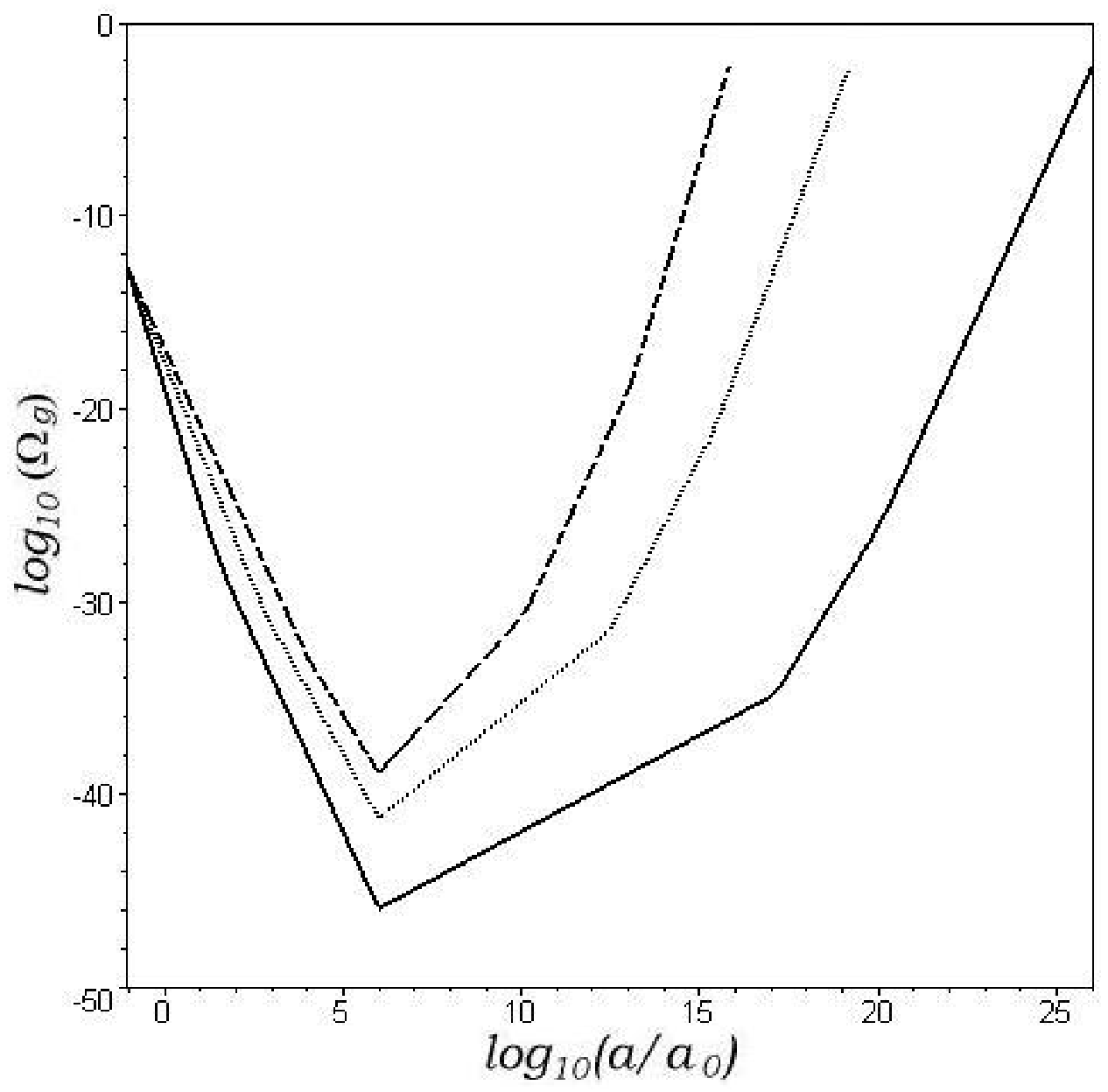}
\caption{Evolution of the density parameter $\Omega_g$ given by
Eqs. (\ref{Og0})-(\ref{omf}) with the scale factor in a five-stage
scenario (De Sitter inflation-radiation-dust-dark energy-second
dust era) from the beginning of the dark energy era. The second
dust era starts at the time in which the graphs attain their
minimum. The solid, dotted and dashed lines correspond to $l=-1$,
$l=-1.5$ and $l=-2$, respectively. The density parameter decreases
in the dark energy era and increases in the second dust era till
its  value is comparable to unity. From that point on our approach
(being linear) ceases to apply. We have chosen a dark energy era
which begins when $a_{3}/a_{0}=1/11$ and ends when
$a_{4}/a_{0}=10^{6}$.} \label{Omegag2}
\end{figure}

By contrast, if a second dust era followed the accelerated (dark
energy) era, $\Omega_{g}$ would grow in this second dust era
because long wave lengths would continuously be entering that
sphere, see Fig. \ref{Omegag2}. This immediately suggests a
criterion to be used by future observers to ascertain whether the
era they are living in is still our accelerated, dark
energy-dominated, era or a subsequent non-accelerated era. By measuring $%
\Omega_{g}$ at conveniently spaced instants they shall be able to
tell. Further, if that era were the accelerated one and the
$\Omega_{g}$ measurements were accurate enough they will be able
to find out the value of the parameter $l$ occurring in the
expansion law given by Eq. (\ref{sclfac3}). The lower $l$, the
higher the slope of $\Omega_{g}$ in the second dust era.

One may argue, however, that these observers may know more easily
from supernova data. Nonetheless, if this epoch lies in the
faraway future it may well happen that by then the ability of
galaxies to generate stars (and hence enough supernovae) has
seriously gone down and as a result this prime method might be
unavailable or severely impaired. At any rate, even if there were
plenty of supernova, the simple gravitational wave method just
outlined could still play a complementary role.

\section{Conclusions}

In this chapter we studied the power spectrum and the energy
density evolution of the relic gravitational waves generated at
the big bang by considering the transitions between successive
stages of the Universe expansion. In particular, we considered the
effect of the present phase of accelerated expansion as well as a
hypothetical second dust phase that may come right after the
present one. As it turns out, the power spectrum at the current
accelerated era, Eq. (\ref{espectrde}), formally coincides with
the power spectrum of the conventional three-stage scenario. As a
consequence, measurements of $P(\omega)$ will not directly tell us
if the Universe expansion is currently accelerated (as we know
from the high redshift supernove data) or non-accelerated.
However, the density parameter of the gravitational waves evolves
differently during these two phases: it stays constant in the
decelerated one and  goes down in the accelerated era in a $l$
dependent manner. Therefore, the present value of $\Omega_{g}$ may
not only confirm us the current acceleration but also may help
determine the value the parameter $l$ -see Fig. \ref{Omegag1}- and
hence give us invaluable information about the nature of dark
energy.

In the faraway future measurements of $P(\omega)$, if sufficiently
accurate, will be able to tell if the Universe is still under
accelerated expansion (driven by dark energy) or has entered a
hypothetical decelerated
phase (second dust era) suggested by different authors \cite%
{Alam}. This may also be ascertained by measuring the density
parameter of the gravitational waves at different instants to see
whether it decreases or increases.%


\chapter{Gravitational waves entropy and the generalized second law \label{GWsentropy}}

\markboth{CHAPTER \thechapter. GWS ENTROPY AND THE GSL}{}

In this chapter we study the evolution of the entropy associated
to the GWs as well as the generalized second law (GSL) of
gravitational thermodynamics in the present era of accelerated
expansion of the Universe. In spite of the fact that the entropy
of matter and relic gravitational waves inside the event horizon
diminish, the mentioned law is fulfilled provided the expression
for the entropy density of the gravitational waves satisfies a
certain condition \cite{RGWsGSL}. Section \ref{secGSL1} gives the
power spectrum of the GWs at the beginning of the present era of
accelerated expansion. In section \ref{secGSL2}, the entropy of
the GWs is defined and its evolution during that era is found.
Finally, in section \ref{secGSL3} an upper bound on the entropy
density of the GWs is obtained by straightforward application of
the GSL.

\section{Gravitational waves in the dark energy era \label{secGSL1}}

\begin{figure}[tbp]
\hspace{2cm}
\includegraphics*[angle=0,scale=0.9]{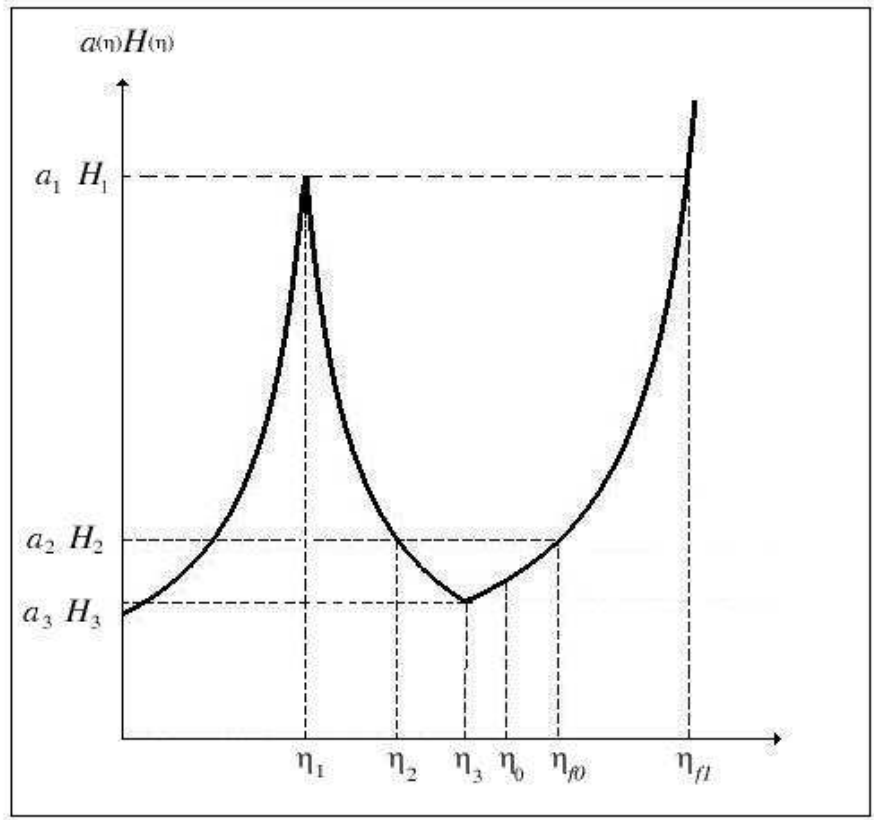}
\caption{{}Evolution of $a(\protect\eta )H(\protect\eta )$ in a
universe with scale factor given by Eq.
(\protect\ref{sclfacGSL}).} \label{aH2}
\end{figure}

As explained in the previous chapter, the current era of cosmic
acceleration is believed to be dominated by the dark energy. For
our purposes in this chapter, we shall assume the Universe is
currently dominated by a form of everlasting dark energy of
constant $\gamma$ lying in the range $(0,2/3)$ (i.e., phantom
energy and cosmological constant dominated universes are
excluded). The dependence of the scale factor on the conformal
time is formally equal to the scale factor in (\ref{sclfac3}),
without the second dust era, i.e.,
\\
\begin{equation}
a(\eta )=\left\{
\begin{array}{lcr}
-\frac{1}{H_{1}\eta }& (-\infty <\eta <\eta _{1}<0),& \text{De
Sitter era} \\
\frac{1}{H_{1}\eta _{1}^{2}}(\eta -2\eta _{1})& (\eta _{1}<\eta
<\eta
_{2}),& \text{radiation era} \\
\frac{1}{4H_{1}\eta _{1}^{2}}\frac{(\eta +\eta _{2}-4\eta
_{1})^{2}}{\eta _{2}-2\eta _{1}}&(\eta _{2}<\eta <\eta _{3}),&
\text{dust era}
\\
\left( \frac{l}{2}\right) ^{-l}\frac{(\eta _{3}+\eta _{2}-4\eta _{1})^{2-l}}{%
4H_{1}\eta _{1}^{2}(\eta _{2}-2\eta _{1})}\left( \eta _{l}\right) ^{l}&
(\eta _{3}<\eta ),& \text{dark energy era}%
\end{array}%
\right.  \label{sclfacGSL}
\end{equation}%
where $l < -1$, $\eta _{l}=\eta +\frac{l}{2}\left[
-\textstyle{2\over{l+1}} \, \eta_{3}+\eta _{2}-4\eta _{1}\right]
$. Again, the subindexes $1,2,3$ correspond to sudden transitions
from De Sitter era to radiation era, from radiation to dust era,
and from dust era to dark energy era, respectively, and $H_{i}$ is
the Hubble factor at the instant $\eta =\eta _{i}$. The Hubble
factor during the current dark energy era obeys
\\
\begin{equation}
H(\eta )=\left( \frac{a_{3}}{a(\eta )}\right)
^{1+\frac{1}{l}}H_{3}. \label{h}
\end{equation}
\\
The evolution of the quantity $aH$ in terms of the conformal time
$\eta$ is sketched in Fig. \ref{aH2}.
\\

The modes solution to the gravitational wave equation during the
De Sitter era are related with those of the final dark energy era
by a Bogoliubov transformation with coefficients $\alpha _{Tr2}$
and $\beta _{Tr2}$, which are formally equal to those found in
subsection \ref{CurrentPowerspec(DDE)}. At the beginning of the
dark energy era, $\eta = \eta_{3}$, the power spectrum was
\\
\begin{equation}
P(\omega )\simeq \left\{
\begin{array}{ll}
0&\left( \omega (\eta_3 )>2\pi (a_{1}/a_3)H_{1}\right) , \\
\\
\frac{1}{4\pi ^{2}}\left( \frac{a_{1}}{a_3}\right)
^{4}H_{1}^{4}\omega ^{-1}&\left( 2\pi (a_{2}/a_3)
H_{2}<\omega (\eta_3 )<2\pi (a_{1}/a_3)H_{1}\right) , \\
\\
\frac{1}{16\pi ^{2}}\left( \frac{a_3}{a_{2}}\right) ^{2}\left(
\frac{a_{1}}{a_3}\right) ^{8}H_{1}^{6}\omega ^{-3}&
\left( 2\pi H_3<\omega (\eta_3 )<2\pi (a_{2}/a_3)H_{2}\right) .%
\end{array}%
\right.  \label{espectr1}
\end{equation}%

During the radiation and dust eras $a(\eta )H(\eta )$ decreases
with $\eta $ and increases during the De Sitter and dark energy
eras. Consequently, as we explained before, GWs are continuously
leaving the Hubble radius during the accelerated dark energy era
\cite{RGWPAE, Chiba}. At some instant $\eta_{f0}$, defined by
$a(\eta _{f0})H(\eta _{f0})=a_{2}H_{2}$, the third term in
(\ref{espectr1}) ceases to contribute to the power spectrum since
the wave lengths of the corresponding GWs exceed the size of the
horizon. Finally, at $\eta _{f1}$, defined by $a(\eta _{f1})H(\eta
_{f1})=a_{1}H_{1}$, all GWs have their wave length longer than the
Hubble radius and the power spectrum vanishes altogether.

\section{GWs entropy and its evolution in the dark energy era \label{secGSL2}}

\markboth{CHAPTER \thechapter. GWS ENTROPY AND THE GSL}{GWS
ENTROPY AND ITS EVOLUTION IN THE DE ERA}

There are different expressions in the literature for the entropy
density of gravitational waves -see e.g., \cite{gasperini,
brandenberger, nesteruk}. All of them are based on the assumption
that the gravitational entropy is associated with the amount of
GWs inside the horizon. In this section, we review the expressions
discussed in \cite{brandenberger}. Next, we adopt the proposal of
Nesteruk and Ottewill  \cite{nesteruk} to find the evolution to
the GWs entropy during the dark energy era.

\subsection{Entropy of the GWs}
Branderberger \textit{et al.} defined the nonequilibrium entropy
of cosmological perturbations in two ways \cite{brandenberger}.
One of them is based in the microcanonical ensemble
\cite{microensam} while the other is a formula for the entropy
that can be associated with the stochastic distribution describing
the state of the classical gravitational field. Both descriptions
are in agreement.

We shall now summarize the second approach concerning GWs. When
the metric perturbations are sufficiently large, the GWs field can
be treated as a classical field. The entropy source in this case
is the lack of information about the exact field configuration. If
a classical field $\varphi (t, \textbf{x})$ and its canonical
momentum $\pi(t,\textbf{x})$ are known at all points $\textbf{x}$
at the instant $t$, the state of the system is completely
specified and consequently its entropy is zero. But, if all we
know about the system is the probability of finding it in a state
$\varphi (t, \textbf{x})$, $\pi(t,\textbf{x})$, i.e., $P(\varphi
(t, \textbf{x}), \pi(t,\textbf{x}))$, the entropy can be expressed
as
\be%
S=-\int P\left(\varphi (\textbf{x}), \pi(\textbf{x})\right)\; \ln
P\left(\varphi (\textbf{x}), \pi(\textbf{x})\right)\; D\varphi
(\textbf{x})D\pi(\textbf{x}),
\ee%
where $D\varphi (\textbf{x})D\pi(\textbf{x})$ denotes the
functional integral measure for a scalar field.

When the initial state of the system is Gaussian and the time
evolution preserves its Gaussian character, the probability
distribution $P(\varphi (t, \textbf{x}), \pi(t,\textbf{x}))$ can
be expressed in terms of the two point correlations functions
$\langle\varphi (t, \textbf{x})\varphi (t, \textbf{y})\rangle$,
$\langle\pi(t,\textbf{x}))\pi(t,\textbf{y})\rangle$ and
$\langle\varphi (t, \textbf{x})\pi(t,\textbf{y})\rangle$, where
$\langle q \rangle$ stands for the ensemble average of $q$ (wich
coincides with the space average of $q$ for a spatially
homogeneous stochastic process). In this the case, the entropy per
unit volume can be expressed as
\be%
s=\int
d^3\textbf{k}\ln\left(\langle|\varphi_k|^2\rangle\langle|\pi_k|^2
\rangle-\langle|\varphi_k|^2|\pi_k|^2\rangle\right),
\ee%
where $\varphi_k$ and $\pi_k$ are the Fourier transformed of
$\varphi (t, \textbf{x})$ and $\pi (t, \textbf{x})$ respectively.

The correlation functions of the GWs field can be expressed in
terms of the Bogoliubov coefficients (see \cite{brandenberger} for
more details). Finally, the entropy
density reads  %
\be%
s=\int d^3k \ln N (k),
\ee%
where $N(k)=|\beta_{Tr2}|^2$ is the number of GWs created from the
initial vacuum with a given wave number $k$. The above expression
is valid in the classical limit, i.e., $N(k) \gg 1$.

In \cite{nesteruk}, Nesteruk and Ottewill proposed a definition of
the entropy of GWs based on the idea that the number of generated
GWs describes the capacity of the gravitational field to create
matter and is associated with the gravitational entropy. Their
proposal assume the entropy density is proportional to the GWs
number density, i.e.,
\\
\begin{equation}
s_{g} = A\, n_{g}\, , \label{ng}
\end{equation}
\\
where $n_{g}$ is the number density of gravitational waves, and
$A$ is an unknown positive--definite constant of proportionality.
We shall now make use of this definition to evaluate the GWs
entropy and to study its evolution in the scenario of the previous
section.

\subsection{GWs entropy in the dark energy era}

We are interested in the evolution of $n_{g}$ during the dark
energy era. The number density of GWs created from the initial
vacuum state is
\\
\begin{equation}
dn_{g}(\eta )=\left[ \frac{\omega ^{2}(\eta )}{2\pi ^{2}}d\omega (\eta )%
\right] |\beta_{Tr2}|^2 =\frac{P(\omega (\eta ))}{%
2\omega (\eta )}\, d\omega (\eta ), \label{dng}
\end{equation}
\\
where the term in square brackets is the density of states and, as
we have seen, $|\beta_{Tr2}|^2 $ is the number of GWs created. We
can obtain $n_{g}(\eta )$ by inserting Eq. (\ref{espectr1}) into
Eq. (\ref{dng}) and integrating over the frequency.

At the beginning of the dark energy era the GWs number density is
\\
\begin{equation}
n_{g}(\eta _{3})=n_{g}(\eta _{2})\left( \frac{a_{2}}{a_{3}}\right) ^{3}+%
\frac{1}{768\pi ^{5}}\left( \frac{a_{1}}{a_{2}}\right) ^{2}\left( \frac{%
a_{2}}{a_{3}}\right) ^{3}H_{1}^{3}\left[ \left( \frac{a_{3}}{a_{2}}\right) ^{%
\frac{3}{2}}-1\right] ,
\end{equation}
\\
where
\\
\begin{equation}
n_{g}(\eta _{2})=\frac{1}{16\pi ^{3}}\left(
\frac{a_{1}}{a_{2}}\right) ^{3}H_{1}^{3}\left[
\frac{a_{2}}{a_{1}}-1\right]
\end{equation}
\\
is the number density at the transition radiation era--dust era.

For $\eta <\eta _{f0}$, i.e. $\frac{a(\eta )}{a_{3}}<\left( \frac{a_{3}}{%
a_{2}}\right) ^{-\frac{l}{2}}$, one has%
\begin{eqnarray}
n_{g}(\eta _{3}<\eta <\eta _{f0})=n_{g}(\eta _{2})\left(
\frac{a_{2}}{a(\eta )}\right) ^{3}+\frac{1}{768\pi ^{5}}\left(
\frac{a_{1}}{a_{2}}\right) ^{2} \left( \frac{a_{2}}{a_{3}}\right)
^{3}H_{1}^{3} \nonumber \\ \times \left( \frac{a_{3}}{a(\eta
)}\right) ^{3}
\left[ \left( \frac{a_{3}}{a(\eta )}\right) ^{-\frac{3}{%
l}}\left( \frac{a_{3}}{a_{2}}\right) ^{\frac{3}{2}}-1\right]
.\qquad \label{dens1}
\end{eqnarray}

The density number given by Eq. (\ref{dens1}) decreases with the
scale factor because of two effects: $(i)$ the evolution of the
volume considered, which grows as $a^{3}$, and $(ii)$ the exit of
those waves whose wave length becomes longer than the Hubble
radius, which appears in the term in square brackets.
As $\frac{a(\eta )}{a_{3}}$ approaches $\left( \frac{a_{3}}{a_{2}}\right) ^{-%
\frac{l}{2}}$, this term tends to zero. From this instant (defined
as $\eta _{f0}$) on, the number density reads
\\
\begin{eqnarray}
n_{g}(\eta _{f0}<\eta <\eta _{f1})=\frac{1}{16\pi ^{3}}\left( \frac{%
a_{1}}{a_{3}}\right) ^{3}H_{1}^{3}\left( \frac{a_{3}}{a(\eta )}\right) ^{3}%
\nonumber \\ \times
\left[ \left( \frac{a_{2}}{a_{1}}\right) ^{\frac{1}{2}}\left( \frac{a_{3}}{%
a_{1}}\right) ^{\frac{1}{2}}\left( \frac{a_{3}}{a(\eta )}\right) ^{-\frac{1}{%
l}}-1\right] . \label{dens2}
\end{eqnarray}
\\
As time goes on, the term in brackets tends to zero and at the
instant $\eta_{f1}$, $n_{g}$ vanishes.

Consequently, the entropy density, proportional to the number
density, decreases during the dark energy era not just because of
the variation of the volume considered but also because of the
disappearance of the GWs from the Hubble volume.

The GWs entropy inside the event horizon is
\\
\begin{equation}
S_{g}=\textstyle{4 \pi\over{3}} R_{H}^{3}s_{g}\, ,
\end{equation}
\\
where $R_{H} = a(t) \, \int_{t} ^{\infty}{ dt'/a(t')} \, $ is the
radius of the event horizon and $t$ the cosmic time. For the
horizon to exist $R_{H}$ must not diverge. For expansions of the
general form  $a(\eta) \propto \eta^{l}$ where $l < -1$, the
horizon exists and can be expressed as $R_{H} = -l H^{-1}(\eta)$.
Since it is of the order of magnitude of the Hubble radius, $
H^{-1}$, we will neglect $-l$ and use both terms interchangeably.

Making use of Eq. (\ref{h}) and Eqs. (\ref{dens1})-(\ref{dens2}) ,
we obtain%
\\
{\small
\begin{eqnarray}
S_{g}(\eta _{3} &<&\eta <\eta
_{f0})=\frac{4\pi}{3}H_{3}^{-3}A\left( \frac{a_{3}}{a(\eta
)}\right) ^{-3/l} \left\{ n_{g}(\eta _{2})
\left( \frac{a_{2}}{a_{3}}\right) ^{3}\right. \nonumber \\
&&\left.+\frac{1}{768\pi ^{5}}\left( \frac{a_{1}}{a_{2}}\right) ^{2}\left( \frac{%
a_{2}}{a_{3}}\right) ^{3}H_{1}^{3}\left[ \left( \frac{a_{3}}{a(\eta )}%
\right) ^{-\frac{3}{l}}\left( \frac{a_{3}}{a_{2}}\right) ^{\frac{3}{2}}-1%
\right] \right\} ,
\end{eqnarray}
}
\\
and
\\
\begin{eqnarray}
S_{g}(\eta _{f0} &<&\eta <\eta
_{f1})=\frac{4\pi}{3}H_{3}^{-3}A\left(
\frac{a_{3}}{a(\eta )}\right) ^{-3/l}\frac{1}{16\pi ^{3}}\left( \frac{%
a_{1}}{a_{3}}\right) ^{3}H_{1}^{3} \nonumber \\
&&\times \left[ \left( \frac{a_{2}}{a_{1}}\right) ^{\frac{1}{2}}\left( \frac{%
a_{3}}{a_{1}}\right) ^{\frac{1}{2}}\left( \frac{a_{3}}{a(\eta )}\right) ^{-%
\frac{1}{l}}-1\right] .
\end{eqnarray}

The GWs entropy is a decreasing function of the scale factor and
consequently of conformal time. The next section explores whether
this entropy descent can be compensated by an increase of the
entropy of the other contributors, namely, matter and horizon.

\section{The generalized second law \label{secGSL3}}

In this section, we study the implications the GSL of
thermodynamics has over the GWs entropy during the dark energy
era.

By extending the GSL of black hole spacetimes \cite{bhth} to
cosmological settings, several authors have considered the
interplay between ordinary entropy and the entropy associated to
cosmic event horizons \cite{paul,gsl}. According the GSL of
gravitational thermodynamics the entropy of the horizon plus its
surroundings (in our case, the entropy in the volume enclosed by
the horizon) cannot decrease. Consequently, we must evaluate the
total entropy to see its evolution during the present dark energy
era.

\begin{figure}[tbp]
\includegraphics*[angle=0,scale=0.7]{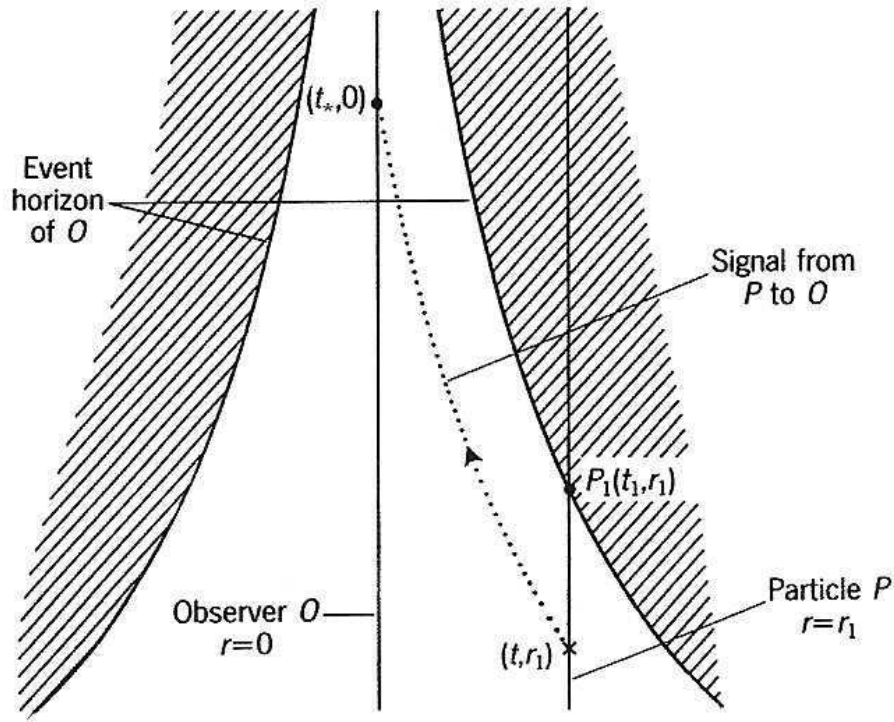}
\caption{The event horizon of an observer in an accelerated FRW
universe (a De Sitter universe in this case). The signal emitted
by a particle $P$ at the instant $t<t_1$ arrives to the observer
at the finite instant $t_*$. Therefore, as $t \rightarrow t_1$, it
follows that $t_* \rightarrow \infty$ and the light signal takes
longer and longer to arrive from the particle to the observer.
Signals emitted once the particle is beyond the horizon will never
reach the observers position. (Adapted from Fig. $23.5$ of Ref.
\cite{dInverno}).} \label{eventhor}
\end{figure}

The cosmological event horizon has an associated entropy that may
be interpreted analogously to the entropy of the black hole
\cite{GH}. An observer living in an accelerated FRW universe has a
lack of information about the regions outside its event horizon
(see figure \ref{eventhor}). The area of the horizon, ${\cal
A}=4\pi\, H^{-2}$, represents a measure of this lack of
information. As in the case of the black hole horizon, we  define
the entropy of the cosmological horizon as $S_H={\cal A}/4$.

Thus, the entropy of the horizon in our case reads
\\
\begin{equation}
S_{H}=\pi H^{-2}(\eta )=\pi H_{3}^{-2}\left( \frac{a_{3}}{a(\eta
)}\right) ^{-2-\frac{2}{l}},
\end{equation}
\\
and increases with expansion. (Bear in mind that $l < -1$, i.e.,
the dark energy behind the acceleration is not of ``phantom" type
\cite{phantom}).

Here we assume that the entropy of the dark energy field
responsible for the acceleration does not vary.

The non-relativistic matter fluid entropy must be also considered.
If the latter consists in particles of mass $m$ and  each of them
contributes $k_{B}=1$ to the matter entropy, we get
\\
\begin{equation}
S_{m}=\frac{\rho _{m}}{m}\frac{4\pi}{3}H^{-3}=\frac{H_{3}^{-1}}{%
2m}\left( \frac{a_{3}}{a(\eta )}\right) ^{-\frac{3}{l}},
\label{sm}
\end{equation}
\\
for the entropy of the non-relativistic fluid. Here, we made use
of the conservation equation
$\rho _{m}(\eta )=\rho _{m}(\eta_{3})\left( \frac{a_{3}}{a(\eta )}\right) ^{3}=\frac{3}{8\pi}%
H_{3}^{2}\left( \frac{a_{3}}{a(\eta )}\right) ^{3}$   with
$\rho_{m}$ the energy density of matter. From (\ref{sm}), it is
apparent that $S_{m}$ decreases with expansion.

In virtue of the above equations, the GSL, $ S ^{\prime}_{m} + S
^{\prime}_{g}+S ^{\prime}_{H}\geqslant 0$, where the prime
indicates derivation with respect to $\eta$, can be written as
\\
\[
\frac{3}{l}\left[ \frac{H_{3}^{-1}}{2m}+\frac{4\pi}{3}%
AH_{3}^{-3}\left( \frac{a_{2}}{a_{3}}\right) ^{3}n_{g}(\eta
_{2})\right]
+\left( 2+\frac{2}{l}\right)\pi%
H_{3}^{-2}\left( \frac{a_{3}}{a(\eta )}\right) ^{-2+\frac{1}{l}}
\]
\begin{equation}
+\frac{3}{l}\frac{A}{576\pi ^{4}}\left( \frac{a_{1}}{a_{2}}\right)
^{2}\left( \frac{a_{2}}{a_{3}}\right)
^{3}H_{1}^{3}H_{3}^{-3}\left[ 2\left(
\frac{a_{3}}{a(\eta )}\right) ^{-\frac{3}{l}}\left( \frac{a_{3}}{a_{2}}%
\right) ^{\frac{3}{2}}-1\right] \geqslant 0, \label{cond1}
\end{equation}
\\
for $\frac{a(\eta )}{a_{3}}<\left( \frac{a_{3}}{a_{2}}\right)
^{-\frac{l}{2}}$, and
\\
\[
\frac{3}{l}\frac{H_{3}^{-1}}{2m}+\left( 2+\frac{2}{l}\right)
\pi H_{3}^{-2}\left( \frac{a_{3}}{a(\eta )}%
\right) ^{-2+\frac{1}{l}}
\]
\begin{equation}
+\frac{3}{l}\frac{A}{12\pi ^{2}}\left( \frac{a_{1}}{a_{3}}\right)
^{3}H_{1}^{3}H_{3}^{-3}\left[ \frac{4}{3}\left( \frac{a_{2}}{a_{1}}\right) ^{%
\frac{1}{2}}\left( \frac{a_{3}}{a_{1}}\right) ^{\frac{1}{2}}\left( \frac{%
a_{3}}{a(\eta )}\right) ^{-\frac{1}{l}}-1\right] \geqslant 0,
\label{cond2}
\end{equation}
\\
for $\frac{a(\eta )}{a_{3}}>\left( \frac{a_{3}}{a_{2}}\right)
^{-\frac{l}{2}}$.

For $l<-1$, both conditions are of the type $f\left( \frac{a(\eta
)}{a_{3}}\right) \geqslant 0$, $\, \,  f$ being an increasing
function of $a(\eta)$. Therefore,  if the condition holds true at
the beginning of the dark energy era, $\eta = \eta _{3}$, it will
hold for $\eta > \eta_{3}$.

By setting $a(\eta ) = a_{3}$ in Eq. (\ref{cond1}), a restriction
over the unknown constant $A$ of proportionality  follows
\\
\begin{equation}
A\leqslant \frac{-\left( 2l+2\right) \frac{\pi }{3}%
H_{3}^{{}}-\frac{H_{3}^{2}}{2m}}{\frac{4\pi}{3}\left( \frac{%
a_{2}}{a_{3}}\right) ^{3}n_{g}(\eta _{2})+\frac{1}{576\pi ^{4}}\left( \frac{%
a_{1}}{a_{2}}\right) ^{2}\left( \frac{a_{2}}{a_{3}}\right) ^{3}H_{1}^{3}%
\left[ 2\left( \frac{a_{3}}{a_{2}}\right) ^{\frac{3}{2}}-1\right]
}\, , \label{bound}
\end{equation}
\\
implying that for the GSL to be satisfied the above upper bound
must be met. In this case the event horizon soon comes to dominate
the total entropy and steadily augments with expansion. So, even
though the entropy of matter and GWs within the horizon decrease
during the present dark energy era, the GSL is preserved provided
Eq. (\ref{bound}) holds. Note that since the authors of Ref.
\cite{nesteruk} left the constant $A$ unspecified restriction
(\ref{bound}) turns to be all the more important: it is the only
knowledge we have about how large $A$ may be.

Obviously, our conclusions hang on the expression adopted for the
entropy density of the gravitational waves. Here we have chosen
(\ref{ng}) since, on the one hand, it is the simplest one based on
particle production in curved spacetimes \cite{Birrell}, and on
the other hand, $s_{g}$ cannot fail to be an increasing function
of $n_{g}$. We believe, that any sensible expression for $s_{g}$
should not run into conflict with the GSL, and that the latter may
impose restrictions on the parameters entering the former.

As mentioned above, we have left aside models of late acceleration
driven by dark energy of ``phantom type" (i.e., $ -1 < l <0$)
\cite{phantom}. In this case, owing to the fact that the dominant
energy condition is violated, the event horizon decreases with
expansion.  We shall focus our attention on this issue in the next
chapter.

\section{Conclusions}

The number of gravitational waves inside the cosmological event
horizon in a four-stage model with scale factor given by
(\ref{sclfacGSL}) is a decreasing function of time in the present
dark energy era. The GWs entropy in the horizon is associated with
the amount of GWs inside it. Consequently, the GWs entropy in the
horizon decreases with time.

In order to study the GSL in this scenario, we must consider all
the possible contributions to the total entropy: the entropy of
the horizon itself (which is proportional to the horizon area and,
consequently, it increases with time), the entropy of the dark
energy field (which is zero as we assume the field is in a
specified state) and the entropy of ordinary matter (which is
proportional to the number of particles of ordinary matter present
inside the horizon and is a decreasing function of time). If we
assume the relation between the GWs entropy and the number of GWs
is (\ref{ng}), the GSL imposes a bound over the constant $A$ of
proportionality. %
\chapter{The generalized second law in universes dominated by dark energy \label{GSLDE}}

\markboth{CHAPTER \thechapter. THE GSL IN UNIVERSES DOMINATED BY
DE}{}

In this chapter we leave the main subject of the thesis, the
spectrum of GWs, to explore the thermodynamics of dark energy
taking into account the existence of the observer's event horizon
in accelerated universes. Here we shall not consider GWs as we
will focus our attention on one single stage of expansion. Thus,
there is no transitions between different stages and no GWs
creation.

As we shall see, phantom and non–-phantom dark energies satisfy
the generalized second law of gravitational thermodynamics whether
the equation of state parameter is constant or not. Phantom energy
transcends the holographic principle, non--phantom energy complies
with it \cite{PHlightGSL}.

\section{Phantom dark energy}

As mentioned before, nowadays there is an ample consensus on the
observational side that the Universe is undergoing an accelerated
expansion likely driven by some unknown fluid, the DE, with the
distinguishing feature of violating the strong energy condition,
$\rho+3p>0$ \cite{Spergel,iap,snia,consensus}. The strength of
this acceleration is presently a matter of debate mainly because
it depends on the theoretical model employed when interpreting the
data. While a model independent analysis suggests it to be below
the De Sitter value \cite{moncy} it is certainly true that the
body of observational data allows for a wide parameter space
compatible with an acceleration value larger than De Sitter's
\cite{phantom}. If eventually this proves to be the case, the
fluid driving the expansion would violate not only the strong
energy condition but the dominant energy condition, $\rho+p>0$, as
well. In general, fluids of such characteristics, dubbed "phantom"
fluids, face some theoretical difficulties on their own such as
quantum instabilities \cite{carroll}. Nevertheless, phantom models
have attracted much  interest, among other things because in their
barest formulation they predict the Universe to end from an
infinite expansion in a finite time ("big rip"), rather than in a
big crunch, preceded by the ripping apart of all bound systems,
from galaxy clusters down to atomic nuclei \cite{phantom}. While
this scenario might look weird, given our incomplete understanding
of the physics below $\rho+p=0$ and the scarcity of reliable
observational data, it should not be discarded right away; on the
contrary we believe it warrants some further consideration.

Recently, it has been demonstrated that if the expansion of the
Universe is dominated by phantom fluid, black holes will decrease
their mass and eventually disappear altogether \cite{babichev}. At
first sight this means a threat for the second law of
thermodynamics as these collapsed objects are the most entropic
entities of our world. This short consideration spurs us to
explore the thermodynamic consequences of phantom–-dominated
universes. In doing so one must take into account that ever
accelerating universes have a future event horizon (or
cosmological horizon). Since the horizon implies a classically
unsurmountable barrier to our ability to see what lies beyond it,
it is natural to attach an entropy to the horizon size (i.e., to
its area) for it is a measure of our ignorance about what is going
on in the other side. However, this has been proved in a rigorous
manner for the De Sitter horizon only \cite{GH} which, in
addition, has a temperature proportional to the Hubble expansion
rate. Nevertheless, following previous authors
\cite{paul,gsl,brustein} here we conjecture that this also holds
true for non-stationary event horizons.

Phantom expansions of pole-like type%

\be
a(t) \propto \frac{1}{(t_{*} - t)^{n}} \qquad (t \leq t_{*},
\; \;  0 < n = \mbox{constant})\, , \label{phexpansions}
\ee%
as proposed by Caldwell \cite{phantom}, arise when the equation of
state parameter $w \equiv p/\rho = \gamma -1 = \mbox{constant} <
-1$. Current cosmological observations hint that  $w$ may be as
low as $-1.5$ \cite{Spergel, snia}. In what follows it will be
useful to bear in mind that $ n = - 2/[3(1+w)] > 0$.

From the above equation for the scale factor it is seen that the
Hubble expansion rate augments
\\
\be H(t) \equiv \frac{\dot{a}}{a} = \frac{n}{t_{*} - t}\,  ,
\label{hubble1} \ee
\\
whereby the radius of the observer's event horizon
\\
\be R_{H} = a(t)\int_{t}^{\infty}{\frac{dt'}{a(t')}} =
a(t)\int_{t}^{t_{*}}{\frac{dt'}{a(t')}} = \frac{t_{*} - t}{1+n} =
\frac{n}{1+n} H^{-1}  \qquad (c = 1)\, \label{event1} \ee
\\
decreases with time, i.e., $\dot{R}_{H}< 0$, and vanishes
altogether at the big rip time $t_{*}$. Consequently the horizon
entropy,
\\
\be S_{H} = \frac{{\cal A}}{4}= \pi R_H^2, \label{hentropy} \ee
\\
diminishes with time, $\dot{S}_{H} < 0$. This is only natural
since for spatially flat FRW phantom-dominated universes one has
\\
\be \dot{H} = -4\pi\, (\rho+ p) > 0,  \ee
\\
as these fluids violate the dominant energy condition. Then the
question arises, ``will the generalized second law (GSL) of
gravitational thermodynamics, $\dot{S}\, + \, \dot{S}_{H} \geq 0$,
be respected?" According to the GSL the entropy of matter plus
fields inside the horizon plus the entropy of the event horizon
cannot decrease with time. For the interplay between ordinary
matter (and radiation) and the cosmological event horizon, see
Refs. \cite{RGWsGSL,paul,gsl,brustein}.

\markboth{CHAPTER \thechapter. THE GSL IN UNIVERSES DOMINATED BY
DE}{PH AND NON-PH DE WITH CONSTANT $W$}

\section{Phantom and non-phantom dark energy with constant $w$ \label{phcntw}}

\markboth{CHAPTER \thechapter. THE GSL IN UNIVERSES DOMINATED BY
DE}{\thesection. PH AND NON-PH DE WITH CONSTANT $W$}

In this section we study the evolution of the phantom DE with
constant $w$, next we compare it with the entropy of the
non--phantom DE with constant w and, finally, we examine the GSL
in both models.

In the previous chapter we assumed that the dark energy has no
entropy associated with it, which would be a reasonable choice if
the dark energy is given by a scalar field in a pure state. But
this is far from being the more natural assumption. Henceforward,
we shall see both phantom and non--phantom fluids as
phenomenological representations of a mixture of fields, each of
which may or may not be in a pure state but the overall (or
effective) "field" is certainly in a mixed state and therefore
entitled to have an entropy. This is the case, for instance, of
assisted inflation \cite{assisted}.

The entropy of the phantom fluid inside the cosmological event
horizon of a comoving observer can be related to its energy and
pressure in the horizon by Gibbs equation

\be T \, dS = dE\, + \, p \, d \left(\textstyle{4\over{3}}\pi
R_{H}^{3}\right)\, , \label{gibbs1} \ee%
where owing to the fact that the number of "phantom particles"
inside the horizon is not conserved we have set the chemical
potential to zero.From the relation $E = (4\pi/3) \rho R_{H}^{3}$,
together with the Friedmann equation $\rho = (3/8\pi) H^{2}$, Eq.
(\ref{event1}), and the equation of state $p= w \rho$  (with $w <
-1$), we get
\\
\be T \, dS = - \frac{n}{1+n}\, dR_{H}. \label{gibbs2} \ee
\\
Since $dR_{H} <0$ the phantom entropy increases with expansion (so
long as $T >0$).

To proceed further, we must specify the temperature of the phantom
fluid. The only temperature scale we have at our disposal is the
temperature of the event horizon, which we assume to be given by
its De Sitter expression \cite{GH}
\\
\be T_{H} =  \frac{n}{1+n} \frac{1}{2 \pi\,R_{H}} \, , \label{TH1}
\ee
\\
though in our case $\dot{H} \neq 0$. Thus, it is natural to
suppose that $T \propto T_{H}$ and then figure out the
proportionality constant. As we shall see below, this choice is
backed by the realization that it is in keeping with the
holographic principle \cite{holographic}. The simplest choice is
to take the proportionality constant as unity which means thermal
equilibrium with the event horizon, $T = T_{H}$. Otherwise energy
would spontaneously flow between the horizon and the fluid (or
viceversa), something at variance with the FRW geometry. Note
that, explicitly or implicitly,  every FRW model (including
inflation models) assumes thermal equilibrium.

After integration of Eq. (\ref{gibbs2}), bearing in mind that $S%
\rightarrow 0$ as $t \rightarrow t_{*}$, the entropy of the
phantom fluid can be written as $S = - {\cal A}/4$.

Some consequences follow: $(i)$ The phantom entropy is a negative
quantity, something already noted by other authors
\cite{tnegative}, and equals to minus the entropy of a black hole
of radius $R_{H}$. $(ii)$ It bears no explicit dependence on $w$
and, but for the sign, it exactly coincides with the entropy of
the cosmological event horizon. $(iii)$ Since  $ -S \propto E^{2}$
it is not an extensive quantity. Note that $S$ does not scale with
the volume of the horizon but with its area. One may argue that
the  first consequence was to be expected since (for $T> 0$) it
readily follows from  Euler's equation $T \, s = \rho +P$, $s$
being the entropy density. However, it is very doubtful that
Euler's equation holds for phantom fluids since it is based on the
extensive character of the  entropy of the system under
consideration \cite{callen} and we have just argued that for
phantom fluids this is not the case. The fact that $S <0$ tells us
that the entropy of the phantom fluid is not to be understood as a
measure of the number of microstates associated to the macroscopic
state of the fluid and that, in general, the current formulation
of statistical mechanics might not apply to systems that violate
the dominant energy condition. Before closing this paragraph, it
is noteworthy that negative entropies are not so new after all; in
principle monoatomic ideal gases may exhibit such a feature for
sufficiently low temperatures \cite{appendixD}.

Two further consequences are: $(iv)$ The sum $S+ S_{H}$ vanishes
at any time,  therefore the GSL is not violated -the increase of
the (negative) phantom entropy exactly offsets the entropy decline
of the event horizon. $(v)$ $\mid S \mid$ saturates the bound
imposed by the holographic principle \cite{holographic}. The
latter asserts that the total entropy of a system enclosed by a
spherical surface cannot be larger than a quarter of the area of
that surface measured in Planck units. (For papers dealing with
the holographic principle in relation with dark energy, see
\cite{cohen} and references therein). In this connection, it is
interesting to see that if the equation of state were such the
entropy obeyed $S^{*}= \varkappa \,  S$, with $\varkappa $ a
positive--definite constant, then the GSL would impose $1 \leq
\varkappa$. This leads us to conjecture  that the entropy of
phantom energy is not bounded from below but from above, being $-
S_{H}$ its upper limit.

Nevertheless, in a less idealized cosmology one should consider
the presence of other forms of energy, in particular of black
holes and the decrease in entropy of these objects
\cite{babichev}. It is unclear whether in such scenario the GSL
would still hold its ground. Nonetheless, one may take the view
that the GSL may impose an upper bound to the entropy stored in
the form of black holes. At any rate, the calculation would be
much more involved and we leave it to a future research. Among
other things, one should take into account that the scale factor
would not obey such a simple expansion law as (\ref{phexpansions})
and that the black holes would be evaporating via Hawking
radiation \cite{Hawk} which would also modify the expansion rate
and, accordingly, the horizon size.

A straightforward and parallel study for a non--phantom dark
energy-dominated universe with constant parameter of state (lying
in the range $-1 < w <-1/3$), shows that in this case too
$\dot{S}+\dot{S}_{H} = 0$. There are two main differences,
however; on the one hand the entropy of the fluid  decreases while
the area of event horizon augments, and on the other hand $S \geq
0$. The latter comes from adopting the view that the dark energy
must vanish for $T \rightarrow 0$ (Planck's statement of the third
law of thermodynamics) and realizing that this happens for $t
\rightarrow \infty$. Again, $S$ saturates the bound imposed by the
holographic principle.

Another important difference between phantom and non--phantom
dark--energy expansions is that while the former gets hotter and
hotter the latter gets colder and colder. This implies the
following. As the phantom fluid heats up its ability to produce
black holes via quantum tunnelling may augment and the probability
for this to occur becomes high near the Planck limit \cite{gross}.
This is a very strongly irreversible process since the variation
of entropy is enormous going from negative values (phantom fluid)
to positive values (black holes). Obviously, the black hole
production would severely alter the dynamics of the last stage of
expansion. Two competing effects come into play, namely: the
tendency to produce black holes via tunnelling, and the tendency
to destroy these black holes via the Babichev effect
\cite{babichev} (also the tendency of the black holes to evaporate
may play a non--negligible role). Depending on which one of these
eventually comes to dominate two rather different scenarios arise:
$(i)$ If the  Babichev effect dominates, the play will be
rehearsed again and again, probably never reaching the big rip ($t
= t_{*}$) but the Universe will always be accelerating. $(ii)$ If
the quantum tunnelling process takes the upper hand, the Universe
will exit the phantom regime and come back to the old Einstein-De
Sitter expansion during the time taken by the black hole to
evaporate into radiation. This achieved, the universe will start a
new radiation dominated era. As a consequence the analysis of
above will be found inconsistent as a whole because if the
Universe is to resume a non--accelerating regime no cosmological
event horizon would have ever existed.

Coming back to the phantom energy discussion, one may adopt the
point of view that the phantom temperature is negative
\cite{pedro}. But, as mentioned above, this would destroy the FRW
geometry. On the other hand, negative temperatures are linked to
condensed matter systems whose energy spectrum is bounded from
above and may therefore exhibit the phenomenon of population
inversion, i.e.,  their upper energy states can be found more
populated than their lower energy states \cite{populated}.
However, all models of phantom energy  proposed so far consist in
some or other type of scalar field with no upper bound on their
energy spectrum. In addition, while population inversion is a
rather transient phenomenon the phantom regime is supposed to last
for cosmological times. Moreover, bearing in mind that $w< -1$ and
$dR_{H} < 0$, it follows from Eq. (\ref{gibbs2}) that if $T$ were
negative, then the phantom entropy would decrease with expansion
and the GSL would  be violated.

\markboth{CHAPTER \thechapter. THE GSL IN UNIVERSES DOMINATED BY
DE}{PH AND NON-PH DE WITH VARIABLE $W$}

\section{Phantom and non--phantom dark energy with variable $w$}

\markboth{CHAPTER \thechapter. THE GSL IN UNIVERSES DOMINATED BY
DE}{\thesection. PH AND NON-PH DE WITH VARIABLE $W$}

One may argue that the result $\dot{S}+\dot{S}_{H} = 0$ critically
depends on the particular choice of the equation of state
parameter $w$ to the point that if  it were not constant, then the
GSL could be violated. Here we shall explore this issue taking the
phantom model with $w$ dependent on time of Sami and Toporensky
\cite{sami}.

In this model a scalar field, $\phi(t)$, with negative kinetic
energy, minimally coupled to gravity sources with energy density
and pressure given by%
\begin{equation*}
\rho =-\frac{{\overset{.}{\phi }}^{2}(t)}{2}+V(\phi ),\text{\qquad }%
p =-\frac{{\overset{.}{\phi }}^{2}(t)}{2}-V(\phi ),
\end{equation*}%
respectively, is adopted. The equation of state parameter
\begin{equation*}
w=\frac{p}{\rho}=\frac{\frac{\overset{.}{\phi }^{2}(t)}{2}%
+V(\phi )}{\frac{\overset{.}{\phi }^{2}(t)}{2}-V(\phi )},
\end{equation*}%
is lower than $-1$ as long as $\rho$ is positive and depends on
time.

\begin{figure}[tbp]
\includegraphics*[scale=0.7,angle=0]{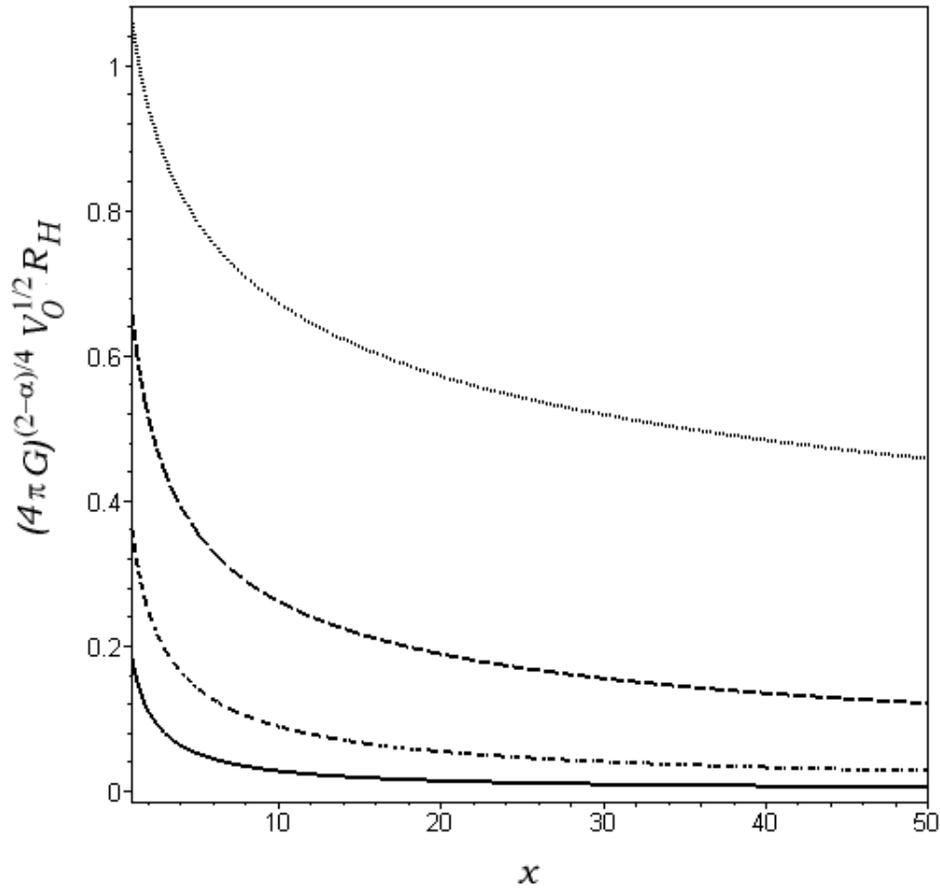}
\caption{Radius of the event horizon {\it vs} $x$ for $\alpha= $
$1$, $2$, $3$ and $4$ from top to bottom. In all the cases $R_{H}
\rightarrow 0$ as $t \rightarrow \infty$. For convenience sake we
have set $x_{i}=1$.} \label{horvsa}
\end{figure}

When the kinetic energy term is subdominant (``slow climb"
\cite{sami}) the evolution equation of the phantom field
simplifies to
\\
\begin{equation}
3H\overset{.}{\phi }(t)\simeq \frac{dV(\phi )}{d\phi }.%
\text{\qquad } \label{fielev1}
\end{equation}%
\\
Assuming also the power law potential $V(\phi) = V_{0}\,
\phi^{\alpha}$ with $\alpha $ a constant, restricted to the
interval $0<\alpha \leq 4$ to evade the big rip, the equation of
state parameter reduces to $w \simeq - 1 - \frac{\alpha}{6x}$,
where $x = 2\pi\, \phi^{2}$ is a dimensionless variable, and the
field is an ever increasing function of time, namely,
\\
\begin{eqnarray}
\phi(t)&=& \left[ \phi _{i}^{\frac{4-\alpha }{2}}+\sqrt{\frac{V_{0}}{24\pi}}%
\, \frac{\left( 4-\alpha \right) \alpha }{2}\left( t-t_{i}\right) \right] ^{%
\frac{2}{4-\alpha }} \quad (0 < \alpha <4), \\
\phi(t)&=& \phi_{i} \, \exp\left[4 \sqrt{\frac{V_{0}}{24\pi}}\,
(t-t_{i})\right] \qquad (\alpha = 4), \label{field} \end{eqnarray}
\\
where the subscript $i$ stands for the time at which the phantom
energy begins to dominate the expansion. Thus $w$ increases with
expansion up to the asymptotic value $-1$. Likewise the scale
function obeys
\\
\be a(\phi) = a_{i}\, \exp \left[ \frac{4\pi}{\alpha}\, (\phi^{2}
- \phi_{i}^{2}) \right]\, . \label{scalefactor} \ee As it should,
the Hubble factor
\\
\be H = \sqrt{\frac{2 V_{0}}{3}} \,
\left(4\pi\right)^{\frac{1}{2}-\frac{\alpha}{4}} \alpha^{\alpha/4}
\, x^{\alpha/4} \label{hfactor} \ee
\\
augments with expansion while the cosmological event horizon
\\
\be R_{H} = x^{\alpha/4}\, \Gamma \left(1- \frac{\alpha}{4},
x\right) \, \frac{e^{x}}{H} \label{event2} \ee
\\
decreases monotonically to vanish asymptotically -see Figure
\ref{horvsa}. Here $\Gamma \left( \frac{4-\alpha }{4},x\right) $
is the incomplete Gamma function \cite{Abram}.
\\

As in the preceding section, the entropy of this phantom fluid can
be obtained by using Gibbs's equation and $T = T_{H} =H/2\pi$.
Again, we get a negative quantity that increases with time
\\
\be S = - \pi\,  C^{2}\, \int_{x}^{\infty}{} \Gamma^{2}\left( 1-
\frac{\alpha}{4},\, x\right)\, \frac{\alpha}{2x}\, e^{2x}\, dx \,
, \label{sphantom2} \ee
\\
here $C= \sqrt{3/(2V_{0})}\,
(4\pi)^{-\frac{1}{2}+\frac{\alpha}{4}}\, \alpha^{-\alpha/4}$.

From the above expressions, it follows that
\\
\be \dot{S} + \dot{S}_{H} = \pi\,C \, \left[\Gamma \left(
\frac{4-\alpha }{4},\, x\right)\, e^{x} (2 + \frac{\alpha}{2x})- 2
x^{-\alpha/4} \right]\, R_{H}\, \dot{x} \, . \label{gsl2} \ee
\\
Since the quantity in square brackets is positive--definite for
any finite $x$ and $\dot{x}>0 $, the GSL in the form $\dot{S} +
\dot{S}_{H} \geq 0 $ is satisfied. The equality sign occurs just
for $t \rightarrow \infty $, i.e., when $R_{H}$ vanishes.

\begin{figure}[tbp]
\includegraphics*[scale=0.9,angle=0]{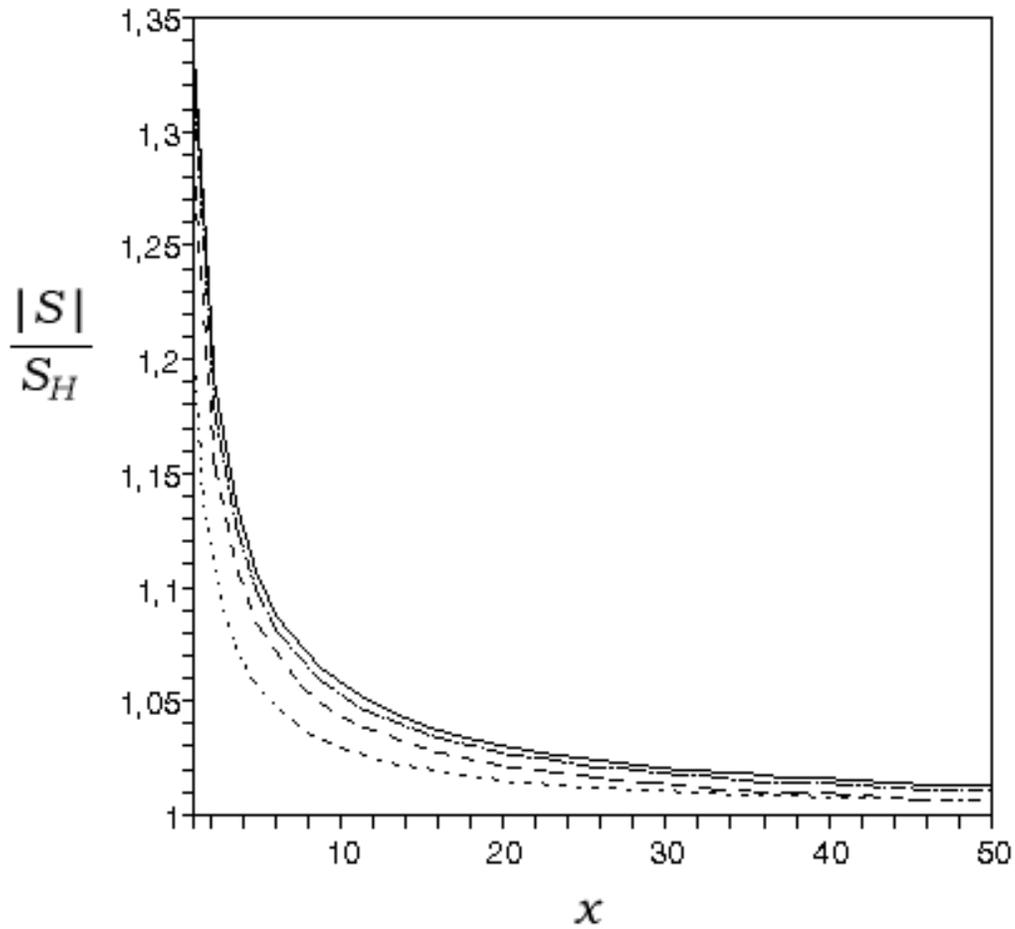}
\caption{Evolution of the ratio between the absolute value of the
phantom entropy and the horizon entropy for the model of Ref.
\cite{sami}. The graphs correspond successively to $\alpha$ values
$1$, $2$, $3$, and $4$ from bottom to top. For convenience sake we
have set $x_{i} = 1$ in drawing the graphs. All the curves
asymptote to $x$ = 1.} \label{sami-holographic}
\end{figure}

As Fig. \ref{sami-holographic} illustrates, the holographic
principle is again transcended in the sense that $-\infty < S \leq
-S_{H}$.
\\

As an example of non-phantom dark energy with variable $w$ we
consider the Chaplygin gas \cite{chaplygin}. This fluid is
characterized by the equation of state $p = - A/\rho$. Integration
of the energy conservation equation yields $\rho =
\sqrt{A+(B/a^{6})}$, whereby $w = -A \, a^{6}/(A\, a^{6}\,+B)$.
Here $a$ is the scale factor normalized to its present value and
$A$ and $B$ denote positive--definite constants. Accordingly, this
fluid interpolates between cold matter, for large redshift, and a
cosmological constant, for small redshift. This has led to propose
it as a candidate to unify dark matter and dark energy
\cite{unify}.

\begin{figure}[tbp]
\includegraphics*[scale=0.7,angle=0]{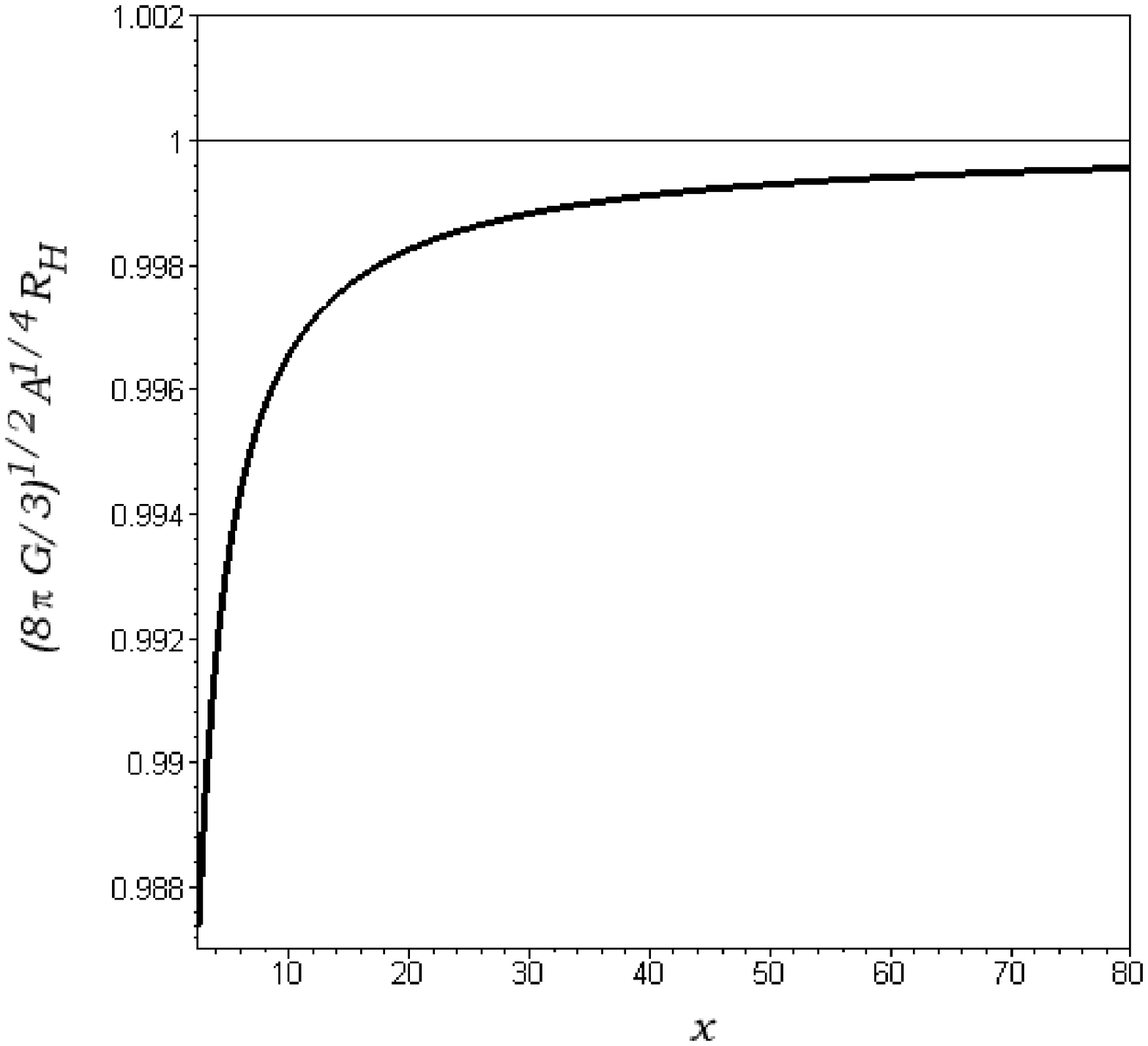}
\caption{Evolution of the adimensionalized radius of the event
horizon in the Chaplygin gas model. It tends to unity for $x
\rightarrow \infty$.} \label{rh-chaplygin}
\end{figure}

The radius of the cosmological event horizon reads
\\
\be R_{H} = \sqrt{\frac{8\pi}{3}} \, \frac{2}{A^{1/4}}\, x^{1/6}
\, \left[ 1.4712 - x^{1/12} \, \, _{2}F_{1}\left(\frac{1}{12},
\frac{1}{4}, \frac{13}{12}; -x\right) \right]\, , \label{event3}
\ee
\\
where $_{2}F_{1}$ is the hypergeometric function and $x = A\,
a^{6}/B$ is a dimensionless variable. As figure \ref{rh-chaplygin}
illustrates, $\mbox{Lim}_{\, x \rightarrow \infty} \,R_{H} =
\sqrt{3/(8\pi)}\, A^{-1/4}$ which is nothing but the value taken
by $H^{-1}$ in that limit. This was to be expected: for large
scale factor the  Chaplygin gas expansion goes over De Sitter's.
\\

The horizon temperature is
\\
\be T_{H} =  \sqrt{\frac{8\pi}{3}} \,\frac{A^{1/4}}{2 \pi} \,
\left(\frac{1+x}{x}\right)^{1/4}, \label{TH2} \ee
\\
while the entropy of the fluid and the entropy of the horizon
evolve as
\\
\be dS= - \frac{3}{4\, A^{1/2}} \, \frac{1}{x^{2/3}(1+x)}
\left[1.4712 - x^{1/2}\,\, _{2}F_{1}\left(\frac{1}{12},
\frac{1}{4}, \frac{13}{12}; -x\right) \right]^{2} \, dx \, , \ee
\label{schaplygin}
\\
and
\\
\begin{eqnarray} dS_{H} = \frac{1}{4\,
A^{1/2}} \, x^{1/6}\,
\left[1.4712 - x^{1/2}\, \,  _{2}F_{1}\left(\frac{1}{12}, \frac{1}{4}, \frac{13}{12}; -x\right) \right] \\
\nonumber \times \left\{2x^{-5/6} \left[1.4712  - x^{1/2}\, \,
_{2}F_{1}\left(\frac{1}{12}, \frac{1}{4}, \frac{13}{12};
-x\right)\right] - \frac{1}{x^{3/4}\, (1+x)^{1/4}} \right\} \, ,
\label{sh3} \end{eqnarray}
\\
respectively. The two latter equation imply that $\dot{S}\, + \,
\dot{S}_{H}  \geq 0$ for $x \geq 2.509$ or  equivalently, for
$\rho_{f(i)}\leq 1.18 A^{1/2}$. Thus there is an early period in
which the GSL seems to fail. However, this may be seen not as a
failure of the GSL but as an upper bound on the initial value of
the energy density of the Chaplygin gas. This is somewhat similar
to the bound imposed on the initial entropy of relic gravitational
waves in a universe dominated by non--phantom dark energy
\cite{RGWsGSL}.

Notice that, $S$ is a positive, ever--decreasing function of the
scale factor that complies with the holographic bound $S \leq
S_{H}$.

\markboth{CHAPTER \thechapter. THE GSL IN UNIVERSES DOMINATED BY
DE}{QUASI-DUALITY BETWEEN PH AND NON-PH THERM}

\section{Quasi-duality between phantom and non-phantom thermodynamics}

\markboth{CHAPTER \thechapter. THE GSL IN UNIVERSES DOMINATED BY
DE}{\thesection. QUASI-DUALITY BETWEEN PH AND NON-PH THERMO}

\begin{figure}[tbp]
\includegraphics*[scale=0.5,angle=-90]{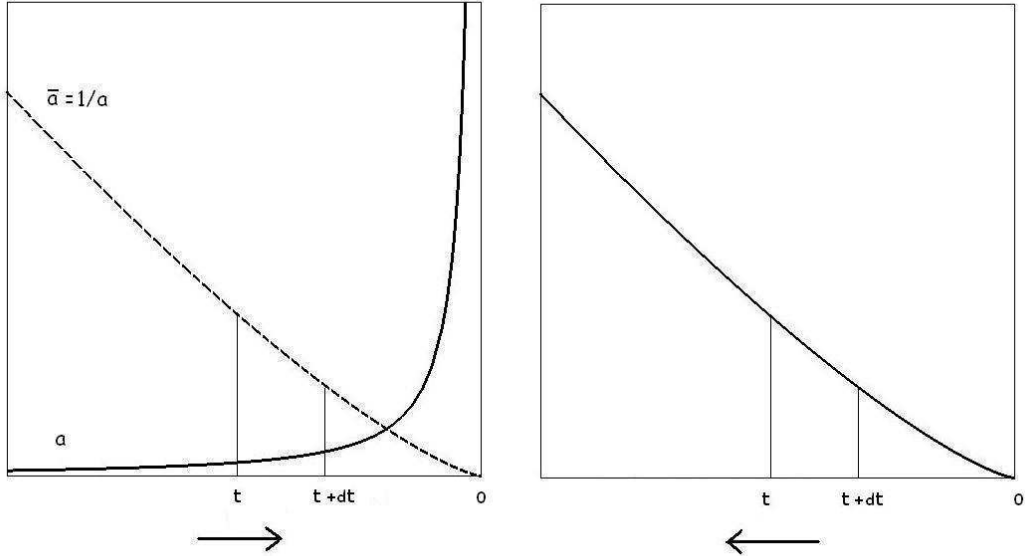}
\caption{The left panel depicts the scale factor of the original
phantom dominated universe (solid line). The dashed line shows the
evolution of the scale factor dual of this phantom DE model, a
contracting universe. In both models (the phantom and its dual)
the time grows from left to right as tends to $t=0^-$ and the
instant $t+dt$ lies at the future of the instant $t$. In the right
panel, the time direction is reversed ($t+dt$ lies at the past of
$t$). Under this operation, the contracting model of the left
panel becomes the non-phantom DE dominated universe of the right
panel if $-5/3<w<-1$.} \label{duality}
\end{figure}

The duality transformation

\be \left\{
\begin{array}{ll}
H&\rightarrow\overline{H}=-H\\
\rho+p&\rightarrow \overline{\rho}+\overline{p}=-(\rho+p)
\end{array}
\right.\ee%
leaves the Einstein's equations%
\be
\begin{array}{ll}
H^2=\frac{8\pi}{3}\rho,&\\
\dot{\rho}+3H(\rho+p)=0,&\\
\dot{H}=-4\pi(\rho+p),& \label{einsteinequations}
\end{array}
\ee%
of spatial FRW universes invariant and the scale factor transforms
as $a\rightarrow\overline{a}=1/a$ \cite{duality}. Thus, a phantom
DE dominated universe with scale factor $a=1/(-t)^n$ (for
simplicity we assume a phantom scale factor as Eq.
(\ref{phexpansions}) with $t_*=0$ and $t<0$ that tends to $0^-$)
becomes a contracting universe whose scale factor obeys
$\overline{a}=(-t)^n$. This universe begins contracting from an
infinite expansion at the infinite past and collapses to a
vanishing scale factor for $t=0^-$ (left panel of Fig.
\ref{duality}).

Reversing the direction of time (i.e., the future of $t$, $t +
dt$, transforms into $t -dt$), the contracting universe is mapped
into an expanding one whose scale factor grows with no bound as
$t$ tends to $-\infty$. Notice that the first transformation takes
the universe from expanding to contracting and the second
transformation makes the universe expand again, i.e., the final
cosmology has $\overline{H} > 0$. It should be remarked that the
latter transformation also preserve Eqs.
(\ref{einsteinequations}).

Under this two successive operations, the final and the initial
equation of states parameters are related by $w \rightarrow
\overline{w}=-(w+2)$. Consequently, phantom DE dominated universes
with $n> 1$, (i.e., $-5/3 < w < -1$) are mapped into non-phantom
DE universes (see Fig. \ref{duality}). Therefore, both of them,
the original and the transformed universes, have an event horizon
and entropies whose relation we derive below.

The radius of the event horizon of the original universe is $R_H =
(-t)/(1+ n)$ meanwhile the radius of the event horizon of the
transformed universe is $\overline{R}_H = (-t)/(n - 1) = R_H[(1 +
n)/(1-n)]$. The transformations preserve the horizon temperature
as the Hubble factor does not change. From the definition of the
horizon, it is readily seen that%
\be%
\frac{dR_H} {dt} = H R_H-1. \ee%
In virtue of the above expression together with the Gibbs equation
one obtains

\be%
T \frac{d S} {dt} = -4\pi(\rho + p)R_H^2 , \ee%
whereby, the entropy transforms as \cite{PHlightGSL}%
\be%
S\rightarrow \overline{S}= -[(1 + 3w)/(5 + 3w)]^2 S.\ee%
Therefore, the entropy of the final non-phantom dark energy
dominated universe is proportional, but of the opposite the sign,
to the entropy of the original phantom universe. We might say that
the duality transformation ``quasi'' preserves the thermodynamics
of dark energy. This result is in keeping with the findings of
Section \ref{phcntw}.

\section{Conclusions}
We explored some thermodynamical consequences for the dark energy
when the entropy of the cosmological event horizon is not ignored.
Irrespective of if $w$ is constant or not, we have found that
phantom fluids possess negative entropy, transcend the holographic
bound and their temperature and entropy increase as the Universe
expands. By contrast, non-phantom fluids have positive entropy,
satisfy the holographic bound and their temperature and entropy
decrease with expansion. In all the cases, the GSL is fulfilled.

It goes without saying that negative entropies are hard to
assimilate; especially because the Einstein-Boltzmann
interpretation of entropy as a measure of the probability breaks
down under such circumstances. Systems of negative entropy appear
to lie outside the province of statistical mechanics as is
currently formulated \cite{populated, mazenko}. Nevertheless, the
laws of thermodynamics alone do not entail that S must be a
positive quantity. The latter will be found to be positive or
negative only after combining these laws with the equation (or
equations) of state of the system under consideration. What we
believe to be at the core of thermodynamics is the law that
forbids the entropy of isolated systems to decrease, this law
contains the GSL for gravitating systems with a horizon.

As pointed out by Nojiri and Odintsov, it may well happen that in
the last stages of phantom dominated universes the scalar
curvature grows enough for quantum effects to play a
non-negligible role before the big rip. If so, the latter may be
evaded or at least softened \cite{nojiri}. While our study does
not incorporate such effects they should not essentially alter our
conclusions so long as the cosmic horizon persists.

It is noteworthy that the duality transformation $\rho + p
\rightarrow -(\rho + p)$ and $H\rightarrow-H$ along with reversing
the direction of time (i.e., $dt\rightarrow -dt$) leaves
Einstein's equations invariant and maps phantom cosmologies, with
both $H$ and $\dot{H}$ positive, into non-phantom cosmologies with
$H >0$ but $\dot{H} <0$ -see, e.g., \cite{duality}. However, this
duality does not necessarily extend to the thermodynamics of the
respective universes since future event horizons exist only when
$\ddot{a} > 0$. Nevertheless, in the particular case of pole-like
expansions, Eq. (\ref{phexpansions}), the transformation preserves
the temperature while the entropy transforms as $S \rightarrow%
-[(1+3w)/(5+3w)]^{2}\, S$ with the constraint $-5/3 < w <-1$.
\newpage
\renewcommand{\chaptername}{}
\renewcommand{\thechapter}{}
\chapter{Summary \label{summary}}

\markboth{}{}

Cosmology has for a long time been a rather speculative science.
Hubble's discovery that the Universe is expanding, and -more
recently-  the realization that at present this expansion is
accelerated, the measured abundance of light elements, the mass
distribution of galaxies and clusters thereof, and the discovery
and posterior measurements of the anisotropies of the CMB have
changed this picture. Hopefully, measurements of GWs will soon be
added to this short list. At any rate, now we can  speak
confidently of physical cosmology as a fully--fledged branch of
Science.

The relic GWs constitute a privileged window to determine the
evolution of the Universe. Little is known from the early
evolution of the Universe and the predictions for their spectrum
depend on the model considered. According to these predictions, a
spectrum of relic GWs is generated making feasible its detection
with the technology currently being developed.

In this thesis, using the adiabatic vacuum approximation, we have
reviewed how the expansion of the Universe amplifies the quantum
vacuum fluctuations, and how the relic GWs spectrum is related
with the scale factor.

We have later evaluated the spectrum in a four-stage model (which
consist on a De Sitter stage, a stage dominated by a mixture of
MBHs and radiation, a radiation dominated stage and finally a
non-relativistic matter (dust) dominated stage). We have
demonstrated that the spectrum in this scenario is much lower than
the predicted by three--stage model (De Sitter-radiation era-dust
era). We have also shown how the bound over the GWs spectrum from
the measured CMB anisotropies places severe constraints over the
free parameters of the four-stage model.

We have also considered a scenario featuring an accelerated
expanding era dominated by dark energy, right after the dust era
of the three--stage model. We have found that the current power
spectrum of this four--stage scenario exactly coincides with that
of the three--stage, but it evolves in a different fashion. We
have considered as well the possibility that the dark energy
decays in non--relativistic matter leading to a second dust era in
the far future and obtained the power spectrum of the GWs as well
as the evolution of the density parameter.

We have applied the generalized second law of thermodynamics to
the four--stage model of above. Assuming the GWs entropy
proportional to the number of GWs, we have found the GSL is
fulfilled provided a certain proportionality constant does not
exceed a given upper bound.

Finally, we have extended the GSL study to a single stage universe
model dominated by  dark energy (either phantom or not), and found
that the GSL is satisfied and that the entropy of the phantom
fluid is negative. Likewise, we have found a transformation
between phantom and non--phantom scenarios preserving the Einstein
field equations that entails a ``quasi" duality between the
thermodynamics of both scenarios.

\appendix
\chapter{Sudden transition approximation} \label{st}
Let us assume a spatially flat FRW universe which gradually
evolves from an initial stage (whose expansion is dominated by a
barotropic fluid of index $\gamma_{(r-1)}$) to a next stage
dominated by another perfect fluid  of barotropic index
$\gamma_{r}$. If the transition between both stages  begins at the
instant $\eta^{-} = \eta_{i} - \Delta \eta_{i}$ and ends at
$\eta^{+} = \eta_{i} + \Delta \eta_{i}$, the scale factor as a
function of time in this scenario reads%
\be%
a(\eta)=\left\{
\begin{array}{lr}
a_{i-1}\left(\frac{a_{i-1}H_{i-1}}{l_{(r-1)}}\right)^{l_{(r-1)}}
\left(\eta-\eta_{i-1}+\frac{l_{(r-1)}}{a_{i-1}H_{i-1}}\right)^{l_{(r-1)}},
& \left(\eta_{i-1}<\eta<\eta^-\right),\\%
a_{tr}(\eta),& \left(\eta^-<\eta<\eta^+,\right),
\\ %
a_+\left(\frac{a_+H_+}{l}\right)^{l_r} \left(\eta-\eta^+
+\frac{l_r}{a_+H_+}\right)^{l_r}, &
\left(\eta^+<\eta \right),%
\end{array} \right.
\ee%
where the subindex $i-1$ denotes the instant when the expansions
begins and the subindex $+$ denotes the quantities evaluated at
$\eta_+$. The function $a_{tr}(\eta)$ describes the smooth
transition from one stage to the other. Consequently, in this
scenario $a(\eta)$ and $a^\prime (\eta)$ are continuous functions
of time at the points $\eta_-$ and $\eta_+$.

The solution to equation (\ref{eqmu}) is
\be%
\mu_{l_{(r-1)}}(\eta)=\left\{
\begin{array}{lr}
(\sqrt{\pi }/2)e^{i\psi
_{l_{(r-1)}}}k^{-1/2}x_{l_{(r-1)}}^{1/2}H_{l_{(r-1)}-\frac{1}{2}}^{(2)}(x_{l_{(r-1)}}),
& \left(\eta_{i-1}<\eta<\eta^-\right),\\%
\mu_{tr}(\eta),& \left(\eta^-<\eta<\eta^+,\right),
\\ %
\alpha_i \mu _{l_r}(\eta)+\beta_i \mu _{l_r}^{\ast }(\eta), &
\left(\eta^+<\eta \right),%
\end{array} \right.
\ee%

but also

\be%
\mu_{l_{r}}(\eta)=\left\{
\begin{array}{lr}
\alpha_i^* \mu _{l_{(r-1)}}(\eta)-\beta_i \mu _{l_{(r-1)}}^{\ast
}(\eta),
& \left(\eta_{i-1}<\eta<\eta^-\right),\\%
\overline{\mu}_{tr}(\eta),& \left(\eta^-<\eta<\eta^+,\right),
\\ %
(\sqrt{\pi }/2)e^{i\psi
_{l_{r}}}k^{-1/2}x_{l_r}^{1/2}H_{l_r-\frac{1}{2}}^{(2)}(x_{l_r}),
&
\left(\eta^+<\eta \right),%
\end{array} \right.
\ee%
where the solutions in the transition region $\mu_{tr}$ and
$\overline{\mu}_{tr}$ are some unknown functions.

By imposing the continuity of $\mu_{l_{(r-1)}}$ and $\mu_{l_r}$
and its first derivatives at $\eta^-$ and $\eta^+$, we get

\be%
 \begin{array}{l} \alpha_i=i\left[ \mu _{tr}^{\prime
}(\eta^+)\mu _{l_r}^{*}(\eta^+)-\mu _{tr}(\eta^+){\mu
_{l_r}^{*}}^{\prime
}(\eta^+)\right]\\
\;\;\;\;\,=i\left[ \mu _{l_{r-1}}^{\prime }(\eta^-)\overline{\mu}
_{tr}^{*}(\eta^-)-\mu
_{l_{r-1}}(\eta^-){\overline{\mu}_{tr}^{*}}^{\prime
}(\eta^-)\right],\end{array} \label{A4}
\ee%
\be%
\begin{array}{l} \beta_i=i\left[ \mu _{tr}(\eta^+)\mu _{l_r}^{\prime
}(\eta^+)-\mu _{tr}^{\prime }(\eta^+)\mu _{l_r}(\eta^+)\right]\\
\;\;\;\;\,=i\left[ \mu _{l_{r-1}}(\eta^-)\overline{\mu}
_{tr}^{\prime}(\eta^-)-\mu _{l_{r-1}}^{\prime
}(\eta^-)\overline{\mu}_{tr}(\eta^-)\right].\end{array} \label{A5}
\ee%

For modes whose period is much larger than the duration of the
transition era (i.e., those modes allowed by the adiabatic vacuum
bound $\omega(\eta)<\omega_i(\eta)$), the transition can be
assumed instantaneous. This is the so-called ``sudden transition
approximation". In this limit, we can safely let
$\eta^-\rightarrow\eta^+$. Relabelling the instant $\eta^+$ as
$\eta_i$, we have that $\mu_{tr}(\eta_i)=\mu_{l_{(r-1)}}(\eta_i)$,
$\overline{\mu}_{tr}(\eta_i)=\mu_{l_{r}}(\eta_i)$, and  a similar
relation for the derivatives evaluated at $\eta_i$. Equations
(\ref{A-coef}) and (\ref{B-coef}) readily follow from (\ref{A4})
and (\ref{A5}) in this approximation.

\end{document}